\documentclass{aa}  

\makeatletter
\renewcommand\paragraph{\@startsection{paragraph}{4}{\z@}%
                                     {-3.25ex\@plus -1ex \@minus -.2ex}%
                                     {1.5ex \@plus .2ex}%
                                     {\normalfont\normalsize\bfseries}}
\makeatother

\usepackage{graphicx}
\usepackage{lscape}
\usepackage[version=4]{mhchem}

\usepackage{txfonts}
\usepackage{natbib}
\bibpunct{(}{)}{;}{a}{}{,}
\newcommand{\cmcms}{cm\textsuperscript{2}/s}
\newcommand{\cmss}{cm/s\textsuperscript{2}}

\begin{document}

\title{Evidence for disequilibrium chemistry from vertical mixing in hot Jupiter atmospheres}
\subtitle{A comprehensive survey of transiting close-in gas giant exoplanets with warm-Spitzer/IRAC} 

   \author{Claire Baxter \inst{1} \and
          Jean-Michel D\'esert \inst{1} \and
          Shang-Min Tsai \inst{2}\and
          Kamen O. Todorov \inst{1} \and
          Jacob L. Bean \inst{3} \and 
          Drake Deming \inst{4} \and 
          Vivien Parmentier \inst{2} \and
          Jonathan J. Fortney \inst{6} \and
          Michael Line \inst{5} \and
          Daniel Thorngren \inst{7}\and
          Raymond T. Pierrehumbert \inst{2} \and
          Adam Burrows \inst{8} \and 
          Adam P. Showman \inst{9,10} }

   \institute{Anton Pannekoek Institute for Astronomy, University of Amsterdam, Science Park 904, 1098 XH Amsterdam, The Netherlands \\ \email{c.j.baxter@uva.nl} \and
            Atmospheric, Ocean, and Planetary Physics, Department of Physics, University of Oxford, Oxford, OX1 3PU, United Kingdom \and 
            Department of Astronomy \& Astrophysics, University of Chicago, 5640 S. Ellis Avenue, Chicago, IL 60637, USA \and 
            Department of Astronomy University of Maryland at College Park, College Park, MD 20742, USA \and
            School of Earth \& Space Exploration, Arizona State University, Tempe AZ 85287, USA \and
            Department of Astronomy and Astrophysics, University of California, Santa Cruz, CA 95064, USA \and
            Institut de Recherche sur les Exoplanètes, Université de Montréal, Canada \and
            Department of Astrophysical Sciences, Princeton University, 4 Ivy Lane, Princeton, NJ 08544, USA \and
            Department of Atmospheric and Oceanic Sciences, Peking University, Beijing, China \and
            Department of Planetary Sciences and Lunar and Planetary Laboratory, The University of Arizona, 1629 University Blvd., Tucson, AZ 85721 USA
             }

   \date{Accepted February 9, 2021}

 
  \abstract
   {}
  {We present a large atmospheric study of 49 gas giant exoplanets using infrared transmission photometry with Spitzer/IRAC at 3.6 and 4.5~$\mu$m.}
   {We uniformly analyze 70 photometric light curves of 33 transiting planets using our custom pipeline, which implements pixel level decorrelation. Augmenting our sample with 16 previously published exoplanets leads to a total of 49.
   We use this survey to understand how infrared photometry traces changes in atmospheric chemical properties as a function of planetary temperature. We compare our measurements to a grid of 1D radiative-convective equilibrium forward atmospheric models which include disequilibrium chemistry. We explore various strengths of vertical mixing ($K_{zz} = 0$ - $10^{12}$~\cmcms) as well as two chemical compositions (1x and 30x solar).}
   {We find that, on average, Spitzer probes a difference of 0.5 atmospheric scale heights between 3.6 and 4.5~$\mu$m, which is measured at $7.5~\sigma$ level of significance.
   Changes in the opacities in the two Spitzer bandpasses are expected with increasing temperature due to the transition from methane-dominated to carbon-monoxide-dominated atmospheres at chemical equilibrium.
   Comparing the data with our model grids, we find that the coolest planets show a lack of methane compared to expectations, which has also been reported by previous studies of individual objects. We show that the sample of coolest planets rule out 1x solar composition with >$3~\sigma$ confidence while supporting low vertical mixing ($K_{zz} = 10^8$~\cmcms). On the other hand, we find that the hot planets are best explained by models
with 1x solar metallicity and  high vertical mixing  ($K_{zz}= 10^{12}$~\cmcms). We interpret this as the lofting of \ce{CH4} to the upper atmospheric layers. Changing the interior temperature changes the expectation for equilibrium chemistry in deep layers, hence the expectation of disequilibrium chemistry higher up. We also find a significant scatter in the transmission signatures of the mid-temperate and ultra-hot planets, likely due to increased atmospheric diversity, without the need to invoke higher metallicities. Additionally, we compare Spitzer transmission with emission in the same bandpasses for the same planets and find no evidence for any correlation.
   Although more advanced modelling would test our conclusions further, our simple generic model grid points towards different amounts of vertical mixing occurring across the temperature range of hot Jupiters. This finding also agrees with the observed scatter with increasing planetary magnitude seen in Spitzer/IRAC color-magnitude diagrams for planets and brown dwarfs.}
   {}


\maketitle
   

%

\section{Introduction}

\label{P1:sec:introduction}

Studying exoplanets is critical in order to gain insight into the dominant composition and physical atmospheric processes and for understanding the theory of planet formation and evolution \citep{Seager2010, Crossfield2015, Deming2017}. Hot Jupiters with large scale heights are ideal targets for detecting molecular signatures in their atmospheres via transmission spectroscopy \citep{Seager2000, Brown2001}. The atmospheres of such planets have been studied across a large range of wavelengths with a myriad of different instruments. Given the number of exoplanet atmospheres already observed, we now enter the era of statistical study of exoplanet atmospheres \citep[e.g.,][]{Triaud2014a, Beatty2014, Gao2020, Keating2019, Garhart2020, Baxter2020, Fu2017, Tsiaras2018, Wallack2019}. 

Wavelength-dependent transit depths are in principle primarily sensitive to the atmospheric composition \citep{Seager2000}, but in practice these observations have often been plagued by the presence of clouds/hazes dampening the expected molecular signals \citep[e.g.,][]{Fortney2005, Sing2016, Barstow2017}. Nevertheless, Cloud-free hot Jupiter atmospheres in chemical equilibrium are predicted to exhibit traces of water, carbon monoxide, and methane \citep{Seager2000, Fortney2005, Fortney2010}. Studies are conducted to demonstrate whether such elements are statistically and systematically observed in exoplanets \citep{Tsai2018}. However, non-equilibrium chemistry and clouds are predicted to be present in close-in giant exoplanet atmospheres, and will impact their observations \citep[e.g.,][]{Agundez2012, Drummond2016, Steinrueck2019}. \citet{Sing2016} performed a mini-survey of the transmission spectra of ten hot Jupiters, which they characterize in terms of a cloud index, and found a transition between cloudy and cloud-free atmospheres. They note that a temperature-pressure profile crossing a condensation curve is not solely responsible for the resulting dampened spectra, but rather it is likely that nonequilibrium effects such as atmospheric circulation and vertical mixing play a role.  

There are several important atmospheric processes to consider that can drive atmospheres away from cloud-free chemical equilibrium. \citet{Zhang2018a} showed that atmospheric transport can move atmospheric abundances away from chemical equilibrium and greatly alter the expected spectroscopic observations. They develop a 1D framework to capture these complex atmospheric processes and parameterize it with an eddy diffusion co-efficient ($K_{zz}$). For hot Jupiters, $K_{zz}$ ranges from $10^8$ to $10^{12}$~\cmcms\  according to estimations of the mean vertical wind in global circulation models \citep[GCMS;][]{Moses2011, Parmentier2013}. Additionally, \citet{Komacek2019} estimated that the strength of vertical mixing will increase for hotter planets. Particularly relevant to this work is the recent advances made in the field of brown dwarf atmospheres: \citet{Miles2020} study the strength of vertical mixing in cool brown dwarf atmospheres with temperatures of 250-750~K, and find that the cooler objects support mixing close to the theoretical maximum yet the warmer objects show weaker than predicted mixing. 

Additionally, the atmospheres of warm giant close-in exoplanets seem to be deficient in methane. According to equilibrium chemistry, methane is predicted to be abundant in the atmospheres of exoplanets with equilibrium temperatures cooler than 1100~K \citep{Madhusudhan2012}. In this context, \citep{Stevenson2010b} showed that the atmosphere of GJ436b is substantially methane deficient relative to chemical equilibrium models, suggesting the presence of nonequilibrium processes such as those induced by vertical mixing, which has been tested by follow-up studies \citep{Knutson2011, Lanotte2014}. Several other studies have attempted to model the methane depletion of GJ 436b: using nonequilibrium photochemical models \citep{Line2011}, high-metallicity (230-1000x solar) models \citep{Moses2013}, models with hydrogen depletion \citep{Hu2015}, and invoking tidal heating due to high eccentricity \citep{Agundez2014}. \citet{Morley2017} provide new data along with a reanalysis and new modeling, and confirm the methane depletion and find the best-fitting models have high metallicity, disequilibrium chemistry, and tidal heating resulting in an intrinsic temperature ($T_{\rm int}$) of 300-350~K. $T_{\rm int}$ characterizes the heat flux escaping from the planetary interior, which is written as $\sigma T_{\rm int}^4$. Recently, \citet{Fortney2020} suggested that the ongoing eccentricity damping of three warm Neptunes, including GJ 436b, heats their atmospheres and drives strong convective mixing resulting in a decreased \ce{CH4}/CO ratio.

Furthermore, methane depletion has been observed in a slew of other warm giant planets. HST/WFC3 observations of the transmission spectra of both WASP-107b and WASP-117 b reveal no detection of methane expected from chemical equilibrium, but only upper limits, suggesting a methane depletion in these atmospheres \citep{Kreidberg2018b, Spake2018, Carone2020}. Additionally, combined HST/WFC3 and Spitzer/IRAC transmission spectra observations of GJ3470 b \citep{Benneke2019}, HAT-P-11 b \citep{Chachan2019}, HAT-P-26 b \citep{Wakeford2017}, and WASP-39 b \citep{Wakeford2018} all have lower-than-expected abundances of methane given their temperatures. All in all, methane has only been sparsely detected in the atmospheres of a few exoplanets \citep{Swain2008,Tinetti2010,Guilluy2019}.

In this paper, we aim to statistically characterize a large sample of hot Jupiters using the two remaining active detectors on Spitzer/IRAC at 3.6~$\mu$m. and 4.5~$\mu$m. \citep{Fazio2004, Werner2004}. At these two wavelengths, we expect to see the absorption of methane (\ce{CH4}) and carbon monoxide or carbon dioxide (CO or \ce{CO2}) respectively. We uniformly analyze Spitzer/IRAC photometric transit light curves of a survey of 34 gas giant planets. This survey represents the largest analysis of Spitzer/IRAC observations of gas giants in transmission to date, and spans equilibrium temperatures from 500~K to 2700~K.

This paper is organized as follows: In Section \ref{P1:sec:observations} we describe the observations and the survey of planets. In Section \ref{P1:sec:Analysis} we describe the data reduction, photometric extraction, light-curve fitting, and the creation of our grid of 1D atmospheric models. Section \ref{P1:sec:Results} describes the results for the transit survey and the statistical survey comparison to the grid of models. In Section \ref{P1:sec:Discussion} we discuss the context and implications for the different trends and statistics that we observe. Additionally, in Section \ref{P1:sec:Discussion} we describe the collection and combination of the secondary eclipse data with GAIA distances and discuss our comparison between transits and eclipses.

\section{Observations}

\label{P1:sec:observations}

As part of the survey programs 90092 (PI Desert) and 13044 (PI Deming), we present the transit depth analysis using 70 transit light curves of 33 planets in the Post Cryogenic Warm \textit{Spitzer}/IRAC bandpasses of 3.6~$\mu$m and 4.5~$\mu$m. With the goal of gaining a stronger understanding of the origins and nature of the exoplanets already discovered, we designed the survey to probe a wide range of masses, radii, and equilibrium temperatures: ranging from cooler long-period gas giants ($\sim200$K) from the \textit{Kepler} mission to close-in hot Jupiters (up to 2300~K). Table \ref{P1:tab:obs} presents the observational information for the 33 planets in the survey. These exoplanets were selected due to their high expected signal-to-noise ratio and, in the case of the Kepler planets, their multiplicity. Additionally, we augment this sample with two extra planets to probe the coolest and the hottest regions of parameter space; these are WASP-121b from program 13044 (PI Deming) and WASP-107b from program 13052 (PI Werner). A full list of the observations is displayed in Table \ref{P1:tab:obs}.

All observations from our survey were taken in "peak-up" mode, meaning the main observation was preceded by a 30-minute peak-up observation allowing for accurate pointing, allowing us to obtain precise positioning of the target to within 0.1 pixels throughout the observations. This significantly reduces the ramp effect caused by the intrapixel sensitivity (discussed in Section \ref{P1:subsec:systematics}).

\begin{figure}
    \centering
    \includegraphics[width = \linewidth]{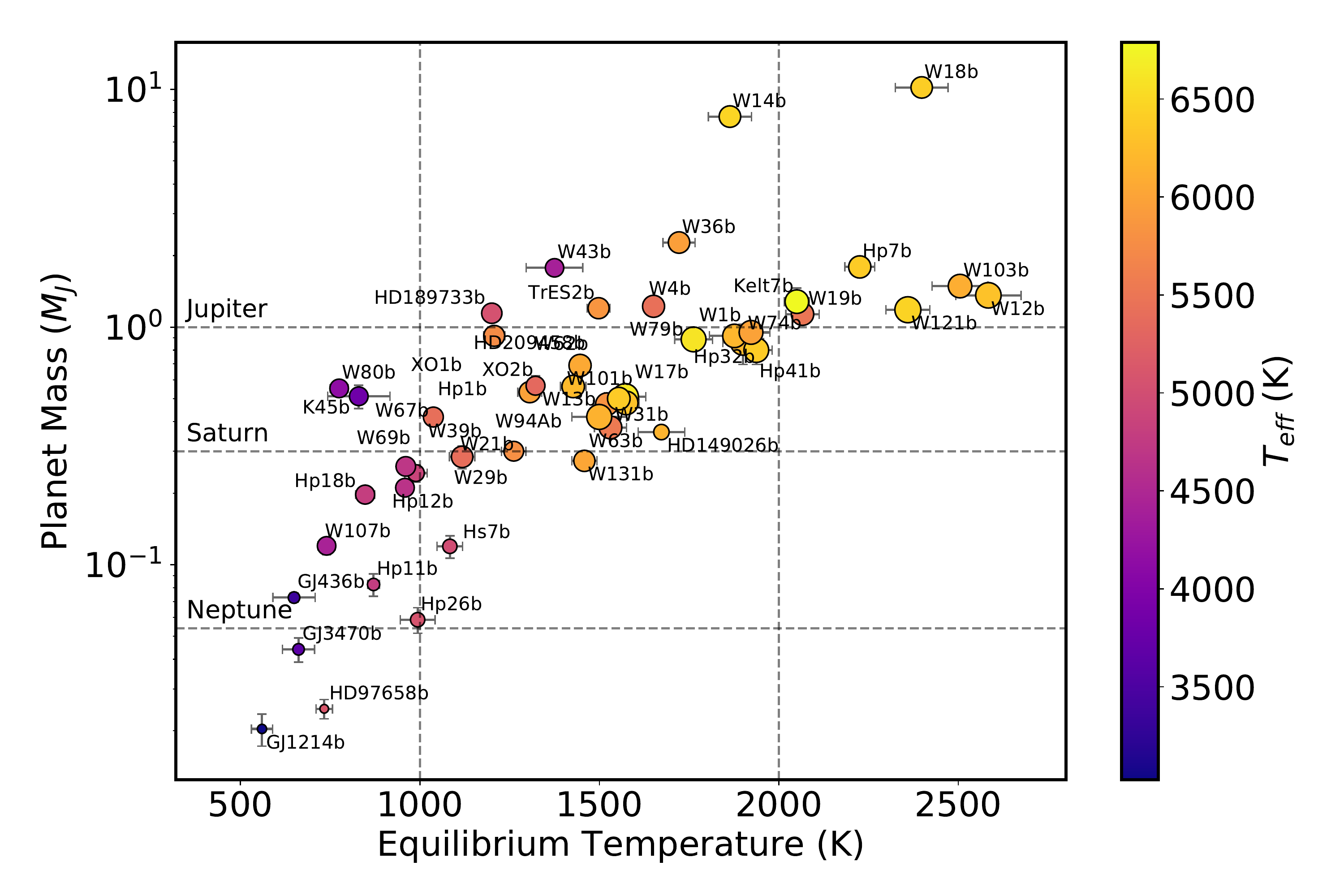}
    \caption{Planet mass ($M_{jup}$) versus equilibrium temperature in Kelvin (assuming zero albedo and full redistribution) for all the planets presented in the current survey. The color of the points shows the stellar temperature ($T_{\rm eff}$) in Kelvin and the size of the points is proportionate to the planetary radius. The gray dashed horizontal lines mark the masses of Jupiter, Saturn, and Neptune. The gray dashed vertical lines mark the temperature regions discussed in Section \ref{P1:subsec:TDvsTeq}.}
    \label{P1:fig:planets}
\end{figure}

We expand our survey to other transiting planets for which the transit depths in the Spitzer bandpasses are taken from the literature. First, we performed a search on exoplanets.org \citep{Wright2011} which yielded 3.6~$\mu$m and 4.5~$\mu$m transits for 16 additional planets. Combining these with our survey allows us to gain insights into the current state of infrared exoplanet transmission spectra in a statistical manner. These additional planets and their transit depths are listed in Table \ref{P1:tab:littransits}. Figure \ref{P1:fig:planets} presents a visualization of the parameter space covered by all the planets in our survey (analyzed and literature).

WASP-6b and WASP-34b are part of the original survey program 90092, but we exclude them from our analysis because the transits were missed. In the case of WASP-6b, the predicted mid-transit times had a large degree of uncertainty on the ephemeris, and the observed transits in both channels did not have sufficient baseline to gain accurate constraints on the atmosphere. In the case of WASP-34b, both transits were missed due to an error in the ephemeris.

\begin{table*}

\caption{Details of the Spitzer observations used in our survey analysis, showing the UT date of observation, the duration of observation in hours, and the program ID of each transit. \label{P1:tab:obs}}

\begin{tabular}{lllll|}
\hline\hline
Target & $\lambda$ & UT Start Date &  Duration &  Program ID\\
& $\mu$ m & & Hours &  \\
\hline
GJ3470 b    &               4.5 &   2013 Jan 01 &         4.4 &       90092 \\
GJ3470 b    &               3.6 &   2012 Dec 22 &         4.4 &       90092 \\
HAT-P-12 b  &               4.5 &   2013 Mar 11 &         4.5 &       90092 \\
HAT-P-12 b  &               3.6 &   2013 Mar 08 &         4.5 &       90092 \\
HAT-P-18 b  &               3.6 &   2013 Jun 17 &         5.0 &       90092 \\
HAT-P-18 b  &               4.5 &   2013 Jul 09 &         5.0 &       90092 \\
HAT-P-1 b   &               4.5 &   2013 Sep 20 &         5.2 &       90092 \\
HAT-P-1 b   &               3.6 &   2013 Sep 11 &         5.2 &       90092 \\
HAT-P-26 b  &               3.6 &   2013 Sep 09 &         4.5 &       90092 \\
HAT-P-26 b  &               4.5 &   2013 Apr 23 &         4.5 &       90092 \\
HAT-P-32 b   &               3.6 &   2012 Nov 18 &         5.4 &       90092 \\
HAT-P-32 b  &               4.5 &   2013 Mar 18 &         5.4 &       90092 \\
HAT-P-41 b   &               3.6 &   2017 Jan 18 &        12.1 &       13044 \\
HAT-P-41 b   &               4.5 &   2017 Feb 03 &        12.1 &       13044 \\
HATS-7 b     &               4.5 &   2016 Nov 04 &         5.2 &       13044 \\
HATS-7 b     &               3.6 &   2016 Nov 01 &         5.2 &       13044 \\
KELT-7 b     &               4.5 &   2017 Jan 04 &        10.3 &       13044 \\
KELT-7 b     &               3.6 &   2016 Dec 27 &        10.3 &       13044 \\
Kepler-45 b &               4.5 &   2013 Sep 29 &         4.5 &       90092 \\
Kepler-45 b &               3.6 &   2013 Sep 22 &         4.5 &       90092 \\
Kepler-45 b &               4.5 &   2013 Sep 12 &         4.5 &       90092 \\
Kepler-45 b &               3.6 &   2013 Sep 07 &         4.5 &       90092 \\
Kepler-45 b &               3.6 &   2013 Oct 16 &         4.5 &       90092 \\
Kepler-45 b &               4.5 &   2013 Nov 15 &         4.5 &       90092 \\
Kepler-45 b &               4.5 &   2013 Aug 21 &         4.5 &       90092 \\
Kepler-45 b &               3.6 &   2013 Aug 06 &         4.5 &       90092 \\
TrES-2 b     &               4.5 &   2012 Nov 26 &         4.3 &       90092 \\
TrES-2 b     &               3.6 &   2012 Nov 21 &         4.3 &       90092 \\
WASP-101 b   &               4.5 &   2017 Jan 17 &         8.0 &       13044 \\
WASP-101 b   &               3.6 &   2017 Jan 06 &         8.0 &       13044 \\
WASP-107 b  &               3.6 &   2017 May 02 &         8.7 &       13052 \\
WASP-107 b  &               4.5 &   2017 Apr 26 &         8.7 &       13052 \\
WASP-121 b   &               4.5 &   2017 Jun 05 &         8.5 &       13044 \\
WASP-121 b   &               3.6 &   2017 Jun 02 &         8.5 &       13044 \\
WASP-131 b   &               4.5 &   2017 Jun 04 &        11.3 &       13044 \\
WASP-131 b   &               3.6 &   2016 Nov 04 &        11.3 &       13044 \\
WASP-13 b   &               3.6 &   2013 Jul 07 &         7.5 &       90092 \\
WASP-13 b   &               4.5 &   2013 Jan 22 &         7.5 &       90092 \\
\hline
\end{tabular}
\begin{tabular}{lllll}
\hline\hline

Target & $\lambda$ & UT Start Date &  Duration &  Program ID\\
& $\mu$ m & & Hours &  \\
\hline
WASP-17 b   &               4.5 &   2013 May 14 &         8.2 &       90092 \\
WASP-17 b   &               3.6 &   2013 May 10 &         8.2 &       90092 \\
WASP-1 b    &               4.5 &   2013 Mar 20 &         6.8 &       90092 \\
WASP-1 b    &               3.6 &   2013 Mar 10 &         6.8 &       90092 \\
WASP-21 b   &               4.5 &   2013 Sep 01 &         6.1 &       90092 \\
WASP-21 b   &               3.6 &   2013 Aug 27 &         6.1 &       90092 \\
WASP-29 b    &               4.5 &   2017 Mar 14 &         7.8 &       13044 \\
WASP-29 b    &               3.6 &   2017 Feb 22 &         7.8 &       13044 \\
WASP-31 b   &               4.5 &   2013 Mar 19 &         4.6 &       90092 \\
WASP-31 b   &               3.6 &   2013 Mar 09 &         4.6 &       90092 \\
WASP-34 b   &               4.5 &   2013 Mar 25 &         4.5 &       90092 \\
WASP-34 b   &               3.6 &   2013 Mar 17 &         4.5 &       90092 \\
WASP-36 b    &               3.6 &   2017 Feb 20 &         7.3 &       13044 \\
WASP-36 b    &               4.5 &   2017 Aug 10 &         7.3 &       13044 \\
WASP-39 b   &               4.5 &   2013 Oct 10 &         5.0 &       90092 \\
WASP-39 b   &               3.6 &   2013 Apr 18 &         5.0 &       90092 \\
WASP-4 b    &               4.5 &   2012 Dec 31 &         4.3 &       90092 \\
WASP-4 b    &               3.6 &   2012 Dec 27 &         4.3 &       90092 \\
WASP-62 b    &               3.6 &   2016 Nov 24 &        11.3 &       13044 \\
WASP-62 b    &               4.5 &   2016 Dec 07 &        11.3 &       13044 \\
WASP-63 b    &               4.5 &   2017 Jun 17 &        15.8 &       13044 \\
WASP-63 b    &               3.6 &   2017 Apr 21 &        15.8 &       13044 \\
WASP-67 b    &               3.6 &   2017 Jan 22 &         5.6 &       13044 \\
WASP-67 b    &               4.5 &   2017 Aug 13 &         5.6 &       13044 \\
WASP-69 b    &               4.5 &   2017 Aug 30 &         6.5 &       13044 \\
WASP-69 b    &               3.6 &   2017 Aug 26 &         6.5 &       13044 \\
WASP-6 b    &               3.6 &   2013 Jan 21 &         4.6 &       90092 \\
WASP-6 b    &               4.5 &   2013 Jan 14 &         4.6 &       90092 \\
WASP-74 b    &               4.5 &   2017 Jan 16 &         6.7 &       13044 \\
WASP-74 b    &               3.6 &   2017 Jan 14 &         6.7 &       13044 \\
WASP-79 b    &               4.5 &   2016 Nov 27 &        11.1 &       13044 \\
WASP-79 b    &               3.6 &   2016 Nov 20 &        11.1 &       13044 \\
WASP-94 Ab   &               3.6 &   2017 Feb 10 &        13.3 &       13044 \\
WASP-94 Ab   &               4.5 &   2017 Aug 06 &        13.3 &       13044 \\
XO-1 b      &               4.5 &   2013 May 25 &         5.4 &       90092 \\
XO-1 b      &               3.6 &   2013 May 13 &         5.4 &       90092 \\
XO-2 b      &               3.6 &   2013 Jan 02 &         4.9 &       90092 \\
XO-2 b      &               4.5 &   2012 Dec 31 &         4.9 &       90092 \\
\hline
\end{tabular}

\end{table*}

\begin{table*}
\caption{Spitzer measurements at 3.6~$\mu$m and 4.5~$\mu$m for planets that have already been published. We include these measurements in our survey.}
\label{P1:tab:littransits}
\centering
\begin{tabular}{llllc}
\hline\hline
Planet &    $T_{eq}$ (K) & $\delta_{3.6}$ (\%) & $\delta_{4.5}$ (\%) & Reference \\
\hline
GJ 1214 b   &   560 $\pm$ 30 &     1.354 $\pm$ 0.009 &     1.367 $\pm$ 0.004 & 1 \\
GJ 436 b    &   649 $\pm$ 59 &     0.695 $\pm$ 0.011 &     0.705 $\pm$ 0.012 & 2, 16 \\
HAT-P-11 b  &   871 $\pm$ 16 &     0.338 $\pm$ 0.002 &     0.336 $\pm$ 0.003 & 3 \\
HAT-P-7 b   &  2225 $\pm$ 41 &     0.629 $\pm$ 0.024 &     0.604 $\pm$ 0.012 & 4 \\
HD 149026 b &  1673 $\pm$ 65 &     0.269 $\pm$ 0.004 &     0.253 $\pm$ 0.004 & 5 \\
HD 189733 b &  1200 $\pm$ 22 &     2.405 $\pm$ 0.008 &     2.416 $\pm$ 0.011 & 6 \\
HD 209458 b &  1446 $\pm$ 19 &     1.481 $\pm$ 0.012 &     1.466 $\pm$ 0.007 & 7 \\
WASP-103 b  &  2505 $\pm$ 78 &     1.401 $\pm$ 0.033 &     1.433 $\pm$ 0.026 & 8 \\
WASP-12 b   &  2584 $\pm$ 91 &      1.341 $\pm$ 0.02 &     1.306 $\pm$ 0.031 & 9 \\
WASP-14 b   &  1864 $\pm$ 60 &     0.887 $\pm$ 0.013 &     0.888 $\pm$ 0.013 & 10 \\
WASP-18 b   &  2398 $\pm$ 73 &     0.959 $\pm$ 0.057 &     0.972 $\pm$ 0.049 & 11 \\
WASP-19 b   &  2066 $\pm$ 46 &      1.957 $\pm$ 0.05 &     2.036 $\pm$ 0.051 & 4 \\
WASP-33 b   &  2694 $\pm$ 53 &     1.166 $\pm$ 0.022 &     1.061 $\pm$ 0.023 & 5 \\
WASP-43 b   &  1375 $\pm$ 79 &     2.496 $\pm$ 0.009 &     2.525 $\pm$ 0.016 & 12 \\
WASP-80 b   &   775 $\pm$ 25 &     2.937 $\pm$ 0.013 &     2.969 $\pm$ 0.014 & 13 \\
K2-25b      &   482 $\pm$ 20 &     1.143 $\pm$ 016 &     1.158 $\pm$ 018 & 14   \\
HD97658b    &   733 $\pm$ 23 &     074 $\pm$ 002 &      08 $\pm$ 002 & 15 \\
HAT-P-2 b   &  1540 $\pm$ 30 &      0.465 $\pm$ 01 &     0.496 $\pm$ 008 & 17 \\
\hline
\end{tabular}
\tablebib{
(1)~\citet{Fraine2013};
(2) \citet{Knutson2011}; (3) \citet{Chachan2019}; (4) \citet{Wong2016};
(5) \citet{Zhang2018a}; (6) \citet{Pont2013}; (7) \citet{Sing2016};
(8) \citet{Kreidberg2018}; (9) \citet{Stevenson2014c}; (10) \citet{Wong2015};
(11) \citet{Maxted2013}; (12) \citet{Stevenson2016a}; (13) \citet{Triaud2015};
(14) \citet{Thao2020}; (15) \citet{Guo2020}; (16) \citet{Morley2017}; 
(17) \citet{Lewis2013}; }

\end{table*}

\section{Analysis}
\label{P1:sec:Analysis}

\subsection{Transit light-curve analysis}

\subsubsection{Extracting Spitzer photometric light curves}
\label{P1:subsec:photometry}

We designed a custom pipeline to produce a photometric light curve from the Basic Calibrated Data frames produced by the Spitzer level 1 pipeline. As is standard for these data, our pipeline corrects dark current, flat fields, corrects for pixel nonlinearity, and converts to flux units. 

We first calculate the mid-exposure timing of each data point in our transit light curves using the UTC-based MBJD values from the headers of each fits file. Our custom pipeline then corrects transient bad pixels in the image time series by comparing each pixel intensity to a median of the 30 preceding and 30 following frames. We replace the pixel intensity with the median value if it is $\geq4\sigma$ from this value. The fraction of transient bad pixels that are corrected is displayed in Table \ref{P1:tab:tests}; this varies around 0.5\% and 0.06\% for channel 1 and channel 2, respectively. Our pipeline also consists of several different functions for three important steps in the data reduction: background sky subtraction, finding the centroid of the object, and performing aperture photometry. Additionally, in between these steps, a sliding $\sigma$clipping on any outliers is performed on the centroiding and on the resulting photometry.

Previous studies have demonstrated that the data reduction method chosen to produce the light curves can have significant effects on the resulting measured transit depths \citep{Ingalls2016}. We therefore optimized the background subtraction, centroiding, and aperture photometry methods by running the pipeline over a 3D grid of different methods for each step; we call these methods the pipeline parameters. We tested three methods of background subtraction:
\begin{enumerate}
    \item The "Box" method: median value from a 2x2 or 4x4 pixel box in all four corners of the frame.
    \item The "Annulus" method: the mean of an annulus centred on the star of radii 6 or 8 pixels and size 2 or 4 pixels (using photutils). 
    \item The "Histogram" method: fit a Gaussian to a histogram of all the pixels in the frame, excluding the star.
\end{enumerate}

We also tested three methods of centroiding: 
\begin{enumerate}
    \item The "Barycenter" method: center of light of a 3x3, 5x5 or 7x7 pixel box centered on the approximate position of the star. 
    \item The "Gaussian" method: Fit a 2D Gaussian function to the entire image using Astropy \citep{AstropyCollaborationandPrice-Whelan2018}. All of the parameters of the 2D Gaussian were let free ($A, x_0, y_0, \sigma_x, \sigma_y,\theta$) for each frame. The centroid position was the $x_0, y_0$ from the Levenberg Marquardt least squares fit. \citep{Agol2010}
    \item The "Moffat" method: same as above but instead a 2D Moffat function was fit to the entire image. 
\end{enumerate}
Finally, we varied the aperture radius from 2.5 to 5.0 pixels in increments of 0.25 pixels. For each instance of the grid and therefore each iteration of the data-reduction pipeline, we performed a least-squares fit to our model (transit + systematic) and calculated the reduced $\chi^2$. The parameters yielding the lowest reduced $\chi^2$ were used to create the light curve used for further analysis. There are a few exceptions to this. For example, some of the cooler planets have a lower signal-to-noise ratio (S/N) meaning the systematic errors dominate and there is therefore a larger scatter in the measured parameters at each pipeline iteration. These planets were examined manually and pipeline parameters were chosen by looking for both repeatable measurements and those with close to the minimum reduced $\chi^2$ . The optimum pipeline parameters including centroiding method, aperture size, background subtraction method, and data reduction information for each planet are detailed in Table \ref{P1:tab:pipeline}. Although the observations were made in "peak-up" mode, it is common that there is still some persistence at the beginning of the light curves. To correct for this, we devised a similar $\chi^2$ test for cutting out the ramp at the beginning of the observations. We performed a series of cuts at the beginning of the light curve and refitted the model. Similarly, we chose the time at which we trim off the beginning of the light curve as the one that gave the lowest reduced $\chi^2$ and root mean square (RMS) of the residuals. 

Prior to any further data analysis, the light-curve intensities are converted to electron counts following the method described in the Spitzer handbook (multiply by EXPTIME*GAIN/FLUXCONV). This allows us to calculate the photometric errors using Poisson statistics. 

\subsubsection{Instrumental systematic modeling}
\label{P1:subsec:systematics}

Spitzer light curves exhibit significant amounts of correlated noise, which has been extensively studied and documented in the literature \citep{Charbonneau2005, Agol2010, Seager2010, Stevenson2010a}. The dominant source of this red noise at 3.6 and 4.5 $\mu$m is caused by an intrapixel sensitivity. Variations in the telescope pointing combined with undersampling of the stellar PSF results in variations in the centroiding with time of $\sim10\%$ of a pixel. When combined with the intrapixel sensitivity, this results in variations in the photometric light curve on the order of 1\%, which is problematic because the atmospheric signal we are trying to extract is on the order of 0.01\%. There have been many different methods developed for dealing with these systematic errors \citep[e.g.,][]{Reach2005, Charbonneau2008, Ballard2010, Stevenson2012, Gibson2012, Morello2015, Deming2015}. \citet{Ingalls2016} presented the results of a data challenge on synthetic and real eclipse data of XO-3b, in which several systematic correction methods were tested against each other. These latter authors found that BLISS \citep{Stevenson2012}, Pixel Level Decorrelation \citep[PLD;][]{Deming2015}, and ICA techniques \citep{Morello2015} were the most precise for correcting the systematic errors of data of similar quality to XO-3b. In particular, PLD achieved the highest accuracy to the synthetic input data \citep{Deming2015}. We therefore present the results of the PLD function for correcting our systematic errors and, for comparison, we also test the polynomial function presented in \citet{Knutson2008}.

\paragraph{Pixel Level Decorrelation}
 
Unlike most methods of systematic correction, PLD does not use the centroid position of the stellar PSF on the pixel as input \citep{Deming2015}. Alternatively, PLD relates the intensities of the individual pixels directly to the photometry in one numerical step, whereas the other methods used two numerical steps: first finding the centroid position of the star on the detector and then relating that to the measured photometry with a different numerical process. To bypass this secondary measurement, PLD assumes that the measured brightness of the star is a smooth function of position. One can perform a Taylor expansion of this continuous and differentiable function such that the flux of the star can be expressed as a linear sum of the individual pixel fluxes (described fully in \citet{Deming2015}): 

\begin{equation}
    \Delta S^t = \sum_{i=1}^{N}c_i \hat{P}_i^t + DT(t) + ft + h,
\end{equation}

where $S^t$ is the flux measured over time and $\Delta$ represents the total fluctuations from all sources. $\hat{P}_i^t = \frac{P_i^t}{\Sigma_{i=1}^{N}P_i^t}$ represents the normalized flux from pixel $i$ at time $t$. Here, $i$ is an integer pixel number, where a 2D grid of pixels centered on the PSF is chosen, each pixel being indexed with a single number. The number of pixels included can be selected depending on the size of the PSF and the brightness of the star. In our survey, we uniformly take a 2D grid of 3x3 pixels containing the PSF of the star on the middle pixel. $DT(t)$ is the transit shape, and $ft+h$ is a temporal ramp which is a typical behavior seen in warm Spitzer light curves caused by the residual telescope pointing. 

\paragraph{Polynomial}

We also corrected the intrapixel variations using the polynomial function of the position presented in \citet{Knutson2008}. $F_{corr} = F(K_0 + K_1(x-x_0) + K_2(x-x_0)^2 + K_3(y-y_0) + K_4(y-y_0)^2 )$, where $x_0$ and $y_0$ are the integer pixel numbers plus 0.5, such that the polynomial is a function of the distance from the center of the pixel, where it is understood to be the most sensitive \citep{Stevenson2012}. Similarly to the PLD, we opted to use a linear function of time to correct the ramp over the entire light curve.

\subsubsection{Fitting light curves to obtain transit parameters}

\paragraph{Transit model}

The transit shape ($DT(t)$) was calculated using Batman \citep{Kreidberg2015}. Batman produces a transit light curve with nine tunable parameters: 
time of inferior conjunction (days), 
orbital period (days), 
planet radius (in units of $R_s$), 
semi-major axis (in units of $R_s$), 
orbital inclination (deg), 
eccentricity, 
angle of periastron (deg), 
limb darkening model, and limb darkening coefficients.

We fixed the orbital period for all of our planets to the values from the literature (Table \ref{P1:tab:jumpParams}). Several planets in our sample have reported values of the eccentricity and angle of periastron passage. As a test for these planets we ran a fit of both a circular and an eccentric orbit and found that the eccentricity did not affect the measured transit depth. We therefore fixed the eccentricity and angle of periastron to zero for the remainder of the analysis.

\paragraph{Limb darkening}

\citet{Southworth2008} demonstrated that the choice of limb darkening can affect the measured planetary radius. This is particularly important in the optical wavelengths where the limb-darkening effects are stronger, but we investigated the effects for each of our planets as a standard output of our pipeline. 
We started by using linear coefficients for the limb-darkening law, which were calculated using the 1D Atlas code from \citet{Sing2010} for the 3.6~$\mu$m and 4.5~$\mu$m Spitzer channels. We translated the interpolation routine from IDL to Python and interpolated the linear limb-darkening values and their 1$\sigma$ errors using the effective temperature, surface gravity, and metallicity of every star in our sample (Table \ref{P1:tab:jumpParams}). We were then able to vary the limb-darkening coefficients within the uncertainties and confirm that the limb-darkening does not have significant impact on the resulting measured transit depth at these wavelengths. For this reason, we fixed the limb darkening to the linear coefficients for the remainder of the analysis. 

This leaves four tunable parameters: the time of inferior conjunction ($t_0$), planet radius ($R_p/R_s$), semi-major axis ($a/R_s$), and the orbital inclination ($i$). The fixing and varying of these parameters is discussed in Section \ref{P1:subsec:fitting}.

\paragraph{Estimating uncertainties using MCMC}
\label{P1:subsec:fitting}

After determining the optimum pipeline parameters and the cutting time at the beginning and end of the light curve, we performed a full statistical analysis of the photometric transit light curves to estimate the uncertainties and study the co-variances of the parameters. 

Before performing any fitting, we normalized the light curves, which allowed us to directly compare the PLD co-efficients and the photon flux time-series. An initial normalization was done by taking the median of the first 100 data points in the light curve. We then performed an initial Levenberg-Marquardt least-squares fit to get the preliminary transit parameters, which were then used to cut out the transit, and we recalculated the normalisation scale such that the median of the out-of-transit flux was 1.

Following a second least-squares fit, we  performed a 4 $\sigma$ clip of the residuals to remove any outlying photometric points not captured in the centroiding clipping. We performed a final least-squares fit on the normalized $\sigma$clipped data to determine the initial guess for the parameters as an input for our Markov Chain Monte Carlo analysis. We first calculated the errors on the photometric points using Poisson statistics assuming photon noise ($\sqrt{N}$). Then, after the first initial least-squares fit, we determined how close we were to the photon noise for each fit and scaled up the uncertainties. The results are shown in Table \ref{P1:tab:tests}. As is commonly found for Spitzer time-series transit observations, our uncertainties are around 20-50\% above the photon noise limit for the whole survey. Scaling up the uncertainties on the photometric points by this factor before running the final fit results in a reduced $\chi^2$ of $\sim$1, which prevents us from underestimating the uncertainties on the physical transit parameters.

We estimated the uncertainties on the best-fit parameters using emcee, the open-source Affine-Invariant Metropolis-Hastings algorithm for Markov Chain Monte Carlo analysis developed by \citet{Foreman-Mackey2013}. We initialized the MCMC chains with 100 walkers, 1000 burn-in steps, and 2000 production steps. We also performed a prayer-bead analysis of the uncertainties as a sanity check, but here we adopt the results from the MCMC analysis because the sampling can be much larger. For each MCMC run, we checked for convergence with the emcee recommended acceptance fraction (0.2-0.5) and the Gelmin-Rubin statistic ($\leq1.1$) \citep{Gelman1992}. If the S/N of the data was low, sometimes the MCMC had an extremely low acceptance fraction. When this happened, we doubled the number of walkers until proper convergence was achieved. We derived the 1$\sigma$ error bars asymmetrically as 34\% above and below the median.

Our combined astrophysical and instrumental (PLD) noise model has 14 free parameters in total for the first fit. We treated the two distinct groups of planets slightly differently in our data reduction. The planets were split into two groups, lower S/N planets (generally cooler with longer periods) and higher S/N planets (short-period hot Jupiters). For the higher S/N planets, we let $t_0$, $R_p/R_s$, $a/R_s$, and $i$ free in the initial fit with uniform priors on all parameters. We then performed a second MCMC fit where we used a 1$\sigma$ Gaussian prior on $a/R_s$ and $i$ based on the results from the first fit. For the lower S/N planets, where it is difficult to detect the transit in each individual light curve, we need to fix $a/R_s$ and $i$ to the literature values for the fitting. For both of these methods, the walkers are initialized in a tight cluster around the best-fit Levenberg-Marquardt minimization. These lower S/N planets also had multiple transits in each band-pass;  each of these was analyzed, and the average transit depth was calculated using the weighted sum, where the weight is the inverse variance multiplied by an "over dispersion" factor as done in \citet{Ingalls2016}. The over dispersion factor allows for underestimation of the individual uncertainties; see \citet{Lyons1992} for further information.

\paragraph{Special cases}

The systematic errors caused by the intra-pixel variability of the IRAC detectors on one of the WASP-13b light curves were not properly captured by our pipeline, such that a bump at the end of the transit remained in the reduced light curve at 3.6~$\mu$m. This had the consequence of making our fitted transit depth shallower than it should be. We therefore removed these data from our fit and ran the pipeline again to get the optimal parameters and transit depths. In total, we removed 36 minutes from the last quarter of the in-transit flux, but the egress remained intact allowing us to still characterize the system.

Similarly, the 3.6~$\mu$m transit of WASP-131b showed a bump in the baseline before transit, likely a starspot occultation. Therefore, we also removed 50 minutes of flux in our MCMC fit.

Furthermore, our approach was slightly modified for HAT-P-26b due to its low S/N, and so we set  Gaussian priors on the semi-major axis and inclination in the initial fits based on the literature values. 

\subsection{Interpreting transmission spectrophotometry with 1D atmospheric modeling}
\label{P1:sec:modelgrid}

To interpret the results from our survey of transiting hot Jupiters, we compare the IRAC transit depths with a grid of simulated atmospheric spectra. First, double-gray analytical formulae are applied to construct the cloud-free temperature--pressure (T-P) profiles for a wide range of stellar irradiation \citet{Heng2014}. Second, we use a photochemical kinetics model (VULCAN, \citet{Tsai2017} see Section \ref{P1:subsec:VULCAN}) to compute the composition under the effects of photo-dissociation and vertical mixing. The T--P profiles and the chemical composition are not self-consistently computed; here, we focus on how the stellar flux impacts the disequilibrium chemistry. Last, a radiative transfer code (PLATON \citet{Zhang2019}, see Section \ref{P1:subsec:PLATON}) is used to create transmission spectra to compare with the observational data. Our fiducial model grid spans a range of equilibrium temperatures ($T_{\rm eq}$) from around 400~K to around 2400~K in $\sim$100~K steps, planetary surface gravities ($g_p$) 500, 1500, 5000~\cmss, planet radius ($R_p$) 0.5, 1, 1.5, and 2~$R_{jup}$ and stellar radius ($R_s$) 0.5, 1, 1.5, and 2~$R_{\odot}$ with 1x solar composition and equilibrium chemistry (no vertical mixing, no photo-chemistry, no boundary fluxes). We then expand our modeling in two dimensions. First, we incorporate nonequilibrium processes and capture vertical mixing in the form of an eddy diffusion coefficient ($K_{zz}$). Second, we test the effects of higher metallicity by creating the full set of grids with 30x solar metallicity. We describe the creation of these grids in full detail below. 

\subsubsection{Stellar irradiation and T-P profiles}
\label{P1:subsec:TPcreation}

\begin{figure}
    \centering
    \includegraphics[width=\linewidth]{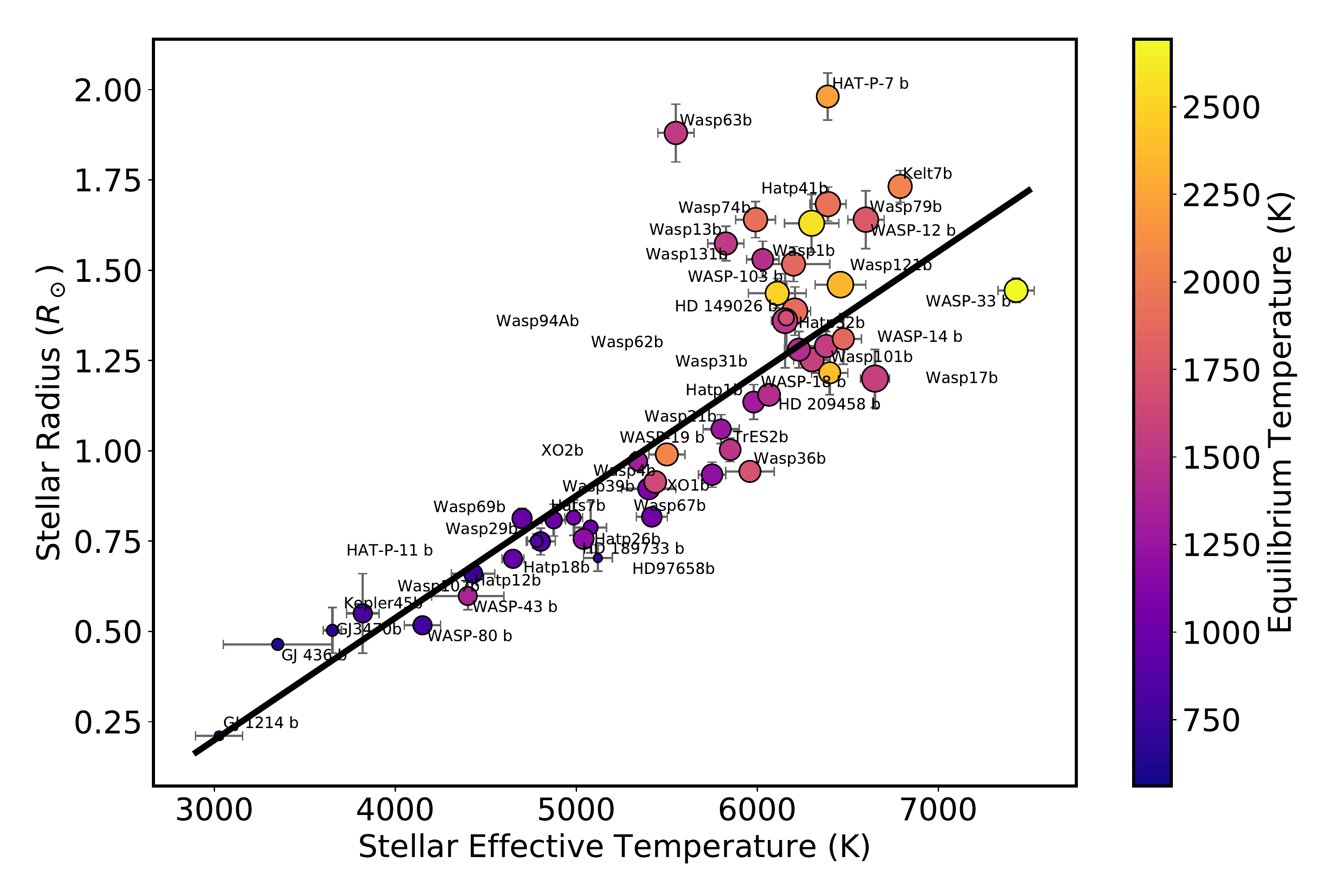}
    \caption{Stellar radius as a function of stellar effective temperature. The black line shows the first-order polynomial best fit used for creation of the T-P grid (see Section \ref{P1:subsec:TPcreation}). The color of the points shows the equilibrium temperature of the planets, and the size of the points is scaled to the planet radius.}
    \label{P1:fig:RsTs}
\end{figure}

In our survey, most hot Jupiters populate a small range of close-in orbits, $a_{mean}$ = 0.037 $\pm$ 0.013 AU, while the parent stars span several spectral types, from early M stars to F stars. In reality, the stellar temperature varies from about 3000~K to 7000~K, while the radius also increases with earlier stellar types. For a planet at orbital distance ($a$) around a star with radius ($R_s$) and effective temperature ($T_{\rm eff}$), the equilibrium temperature of the planet ($T_{\rm eq}$) is proportional to $T_{\rm eff} (R_s/a)^{1/2}$. Therefore, changes in the stellar flux play a bigger role than changes in the orbital distance or the stellar radius when determining the irradiation the planet received and its effective temperature. In this regard, we assume a fixed orbital distance for our model grid to isolate the effects of stellar irradiation. The mean value of the samples, 0.035 AU, is taken across the grid. In this setting, the effective temperature of the model planet is entirely determined by the stellar luminosity, and not by the orbital distance. We also note that the intrinsic temperature ($T_{\rm int}$) is negligible and thus the effective temperature ($T_{\rm eff}^4 = T_{\rm eq}^4 + T_{\rm int}^4$) is the same as the equilibrium temperature. Additionally, our assumed value of $T_{\rm int}$=150~K has little effect on our model temperature pressure profiles of the hot Jupiters. 
 
The next step is to determine the size and therefore the energy flux of the stars. For main-sequence stars, which follow the mass--luminosity and mass--radius relations, we fit a power-law relation between the radius and effective temperature; see Figure \ref{P1:fig:RsTs}. The power-law fitting to our sample yields the following expression for the stellar radius: $R_S = m*T_{\rm eff} + b $ where $m$ and $b$ are 0.0003381 and -0.81495, respectively. Once the effective temperature and the radius of the star are known, there is enough information to specify the incident irradiation of the model atmospheres. 

We then compute the T-P profiles for the given stellar irradiation using the analytical double-gray radiative equilibrium solutions in \citep[][; Equation 126]{Heng2014}.

The parameters used in this calculation are chosen to match the numerical radiative transfer results listed in Table \ref{P1:tab:TPparams}. 
Similar to the prescriptions in  \citet{Guillot2010} and \citet{Parmentier2014}, the opacities do not have a pressure dependence. 
We reiterate that this relation allows us to uniquely express the equilibrium temperature of a planet at a given orbital distance as a function of stellar temperature. Our stellar grid, with effective temperatures from 3250 to 7000~K, produces irradiated atmospheres of temperature from 446 to 2248 K at 0.035 AU. To reach the temperatures of the ultra-hot Jupiters we also run additional models with an orbital distance at 0.02 AU. The resulting temperature pressure profiles are shown in Figure \ref{P1:fig:TPgrid}.

The simple prescription allows us to explore the parameter space in a basic way and to focus on the correlation with stellar irradiation. Although the intrinsic temperature is held constant in our T-P profiles, the realistic interior can be potentially hotter. \citet{Tremblin2017} and \citet{Sainsbury-Martinez2019}  showed that circulation can transport entropy downward and leads to a hotter deep interior over time. \citet{Thorngren2019} suggested much higher $T_{\rm int}$ for observed hot Jupiters (with $T_{\rm eq}$ $\gtrsim$ 1300 K) than the 100K commonly assumed in GCMs. \citet{Fortney2020} also investigated the effects of heating from tidal dissipation for warm Jupiters (with $T_{\rm eq}$ $\gtrsim$ 1300 K) with simplified chemical timescale analysis. The upshot of the hotter interior is that it lowers the quenched [\ce{CH4}]/[\ce{CO}] ratio. In short, a hot deep interior changes the expectation for equilibrium chemistry in deep layers, hence the expectation for disequilibrium chemistry higher up. As the equilibrium abundance of \ce{CH4} generally increases with depth (at least in our solar and 30x solar models), lowering vertical mixing also results in a lower [\ce{CH4}]/[\ce{CO}] ratio and can effectively be degenerate with a hotter interior. However, we find high vertical mixing matches the hot Jupiters better, even with the lower $T_{\rm int}$ of 150 K. Increasing the interior temperature will reduce \ce{CH4}, so even higher vertical mixing would be required to recover the same \ce{CH4} abundance for these planets. As for the cooler planets, the signature leading to our inference of low vertical mixing can also be explained by a hot interior if there is actually no \ce{CH4} in the deep hot atmosphere. However, the sources of internal heating and their exact interior temperature for these cool planets are rather uncertain (see \citet{Fortney2020} for a  detailed discussion). 

In addition to this, our prescription for PT profiles is simplified compared to 1D radiative and convective models. Nevertheless, in this study, we are interested in the relative difference between two broad bandpasses (3.6 and 4.5~$\mu$m), and therefore the prescription used for the TP profiles is less critical than for absolute measurements. The relative difference between 3.6 and 4.5~$\mu$m is globally similar for our prescription as compared to the 1D RC models. We acknowledge that the vertical mixing is likely to be affected by the choice of TPs, but testing this difference is beyond the scope of our current study.

Our simple model emphasizes the importance of the degeneracies between vertical mixing, interior temperature, and equilibrium chemistry, but  also limits the possible interpretations. A more detailed approach than the simple model we used is required to study the impact of the various processes on the observations with greater accuracy. However, such a detailed study will also be limited by the unknown interior temperature. Therefore, we limit ourselves to a simple approach as a sophisticated analysis with more advanced temperature pressure profiles is beyond the scope of this paper.

\begin{table}[]
   \caption{Fixed parameters used in the double-gray radiative
equilibrium solution for creating TP profiles. For each TP profile we show the fixed irradiation temperature ($T_{\rm irr}$), intrinsic temperature ($T_{\rm int}$), longwave opacity ($\kappa_L$), shortwave opacity ($\kappa_S$), longwave scattering parameter ($\beta_{L}$), and the shortwave scattering parameter ($\beta_{S}$).}
   \centering
   \begin{tabular}{lllllll}
   \hline\hline
    $T_{\rm irr}$ & $T_{\rm int}$  & $\kappa_L$ & $\kappa_S$ &
$\beta_{L}$ & $\beta_{S}$ \\
   \hline
 631 & 150 & 0.02 & 0.00035 & 1 & 1 \\
 775 & 150 & 0.02 & 0.00068 & 1 & 1 \\
 919 & 150 & 0.02 & 0.001 & 1 & 1 \\
1069 & 150 & 0.02 & 0.0014 & 1 & 1 \\
1222 & 150 & 0.02 & 0.0017 & 1 & 1 \\
1379 & 150 & 0.02 & 0.0019 & 1 & 1 \\
1540 & 150 & 0.02 & 0.0022 & 1 & 1 \\
1706 & 150 & 0.02 & 0.0035 & 1 & 1 \\
1875 & 150 & 0.02 & 0.0038 & 1 & 1 \\
2049 & 150 & 0.02 & 0.004 & 1 & 1 \\
2227 & 150 & 0.02 & 0.0043 & 1 & 1 \\
2410 & 150 & 0.02 & 0.006 & 1 & 1 \\
2595 & 150 & 0.02 & 0.006 & 1 & 1 \\
2786 & 150 & 0.02 & 0.006 & 1 & 1 \\
2980 & 150 & 0.02 & 0.0061 & 1 & 1 \\
3179 & 150 & 0.02 & 0.0062 & 1 & 1 \\

   \hline
   \end{tabular}
   \label{P1:tab:TPparams}
\end{table}

\subsubsection{Grid of stellar spectra}

As the effective temperature of the star rises, the spectral energy distribution shifts to shorter wavelengths. We therefore adopted the stellar spectral grid from \citet{Rugheimer2013}, which ranges from 4250 to 7000 K and covers F0 to K7 spectral types. The models start with the synthetic ATLAS spectra \citep{Kurucz1979} and then we co-add the observed spectra from International Ultraviolet Explorer for UV (<= 300 nm); see \citep{Rugheimer2013} for the detailed stellar grid setup. Additionally, for late K and M stars ($T_{\rm eff}$ < 4250 K), we picked GJ 436 ($T_{\rm eff}$ = 3350 K) as our fiducial star. The high-resolution spectrum of GJ436 is taken from the MUSCLES survey \citep{France2016}\footnote{http://cos.colorado.edu/~kevinf/muscles.html} and scaled for the stellar fluxes with effective temperatures of 3250~K, 3500~K, 3750~K, and 4000~K. 

\subsubsection{Modeling the photo-chemical kinetics with VULCAN} 
\label{P1:subsec:VULCAN}

We explore the effects of photolysis, atmospheric mixing, and metallicity by using a photochemical kinetics model, VULCAN \citep{Tsai2017}\footnote{https://github.com/exoclime/VULCAN}. The code solves the steady-state chemical compositions for a given temperature-pressure profile and has been benchmarked for hot Jupiters. In this work, we use the updated version that includes nitrogen chemistry and photochemistry (Tsai et al. in preparation). The chemical model with updated nitrogen chemistry and photochemistry has been tested on nitrogen-dominated atmospheres for super-Earths \citep{Zilinskas2020}. The N-C-H-O network consists of about 600 thermal reactions (including forward and reverse) and 40 photodissociation reactions. We validate our updated model against the one-dimensional photochemical and thermochemical kinetics and diffusion model presented by \citet{Moses2011} for HD209458b, see Figure \ref{P1:fig:HD209}. 

Vertical mixing is simulated through means of an eddy diffusion co-efficient ($K_{zz}$), which assumes that atmospheric motion resembles diffusion when convection and turbulence occur on much smaller scales than the magnitude of the pressure scale height. We vary the eddy diffusion coefficient to explore various strengths of vertical mixing, with constant values of 10$^8$, 10$^{10}$, and 10$^{12}$~\cmcms. The choice of the values is consistent with those extracted from GCM simulations \citep{Moses2011, Parmentier2013, Zhang2018b, Komacek2019}. Furthermore, the elemental abundance of the atmosphere is assigned to two different metallicities: 1x solar and 30x solar \citep{Lodders2009}.

\subsubsection{Creating the transmission spectra with PLATON}
\label{P1:subsec:PLATON}

\begin{figure}
    \centering
    \includegraphics[width=\linewidth]{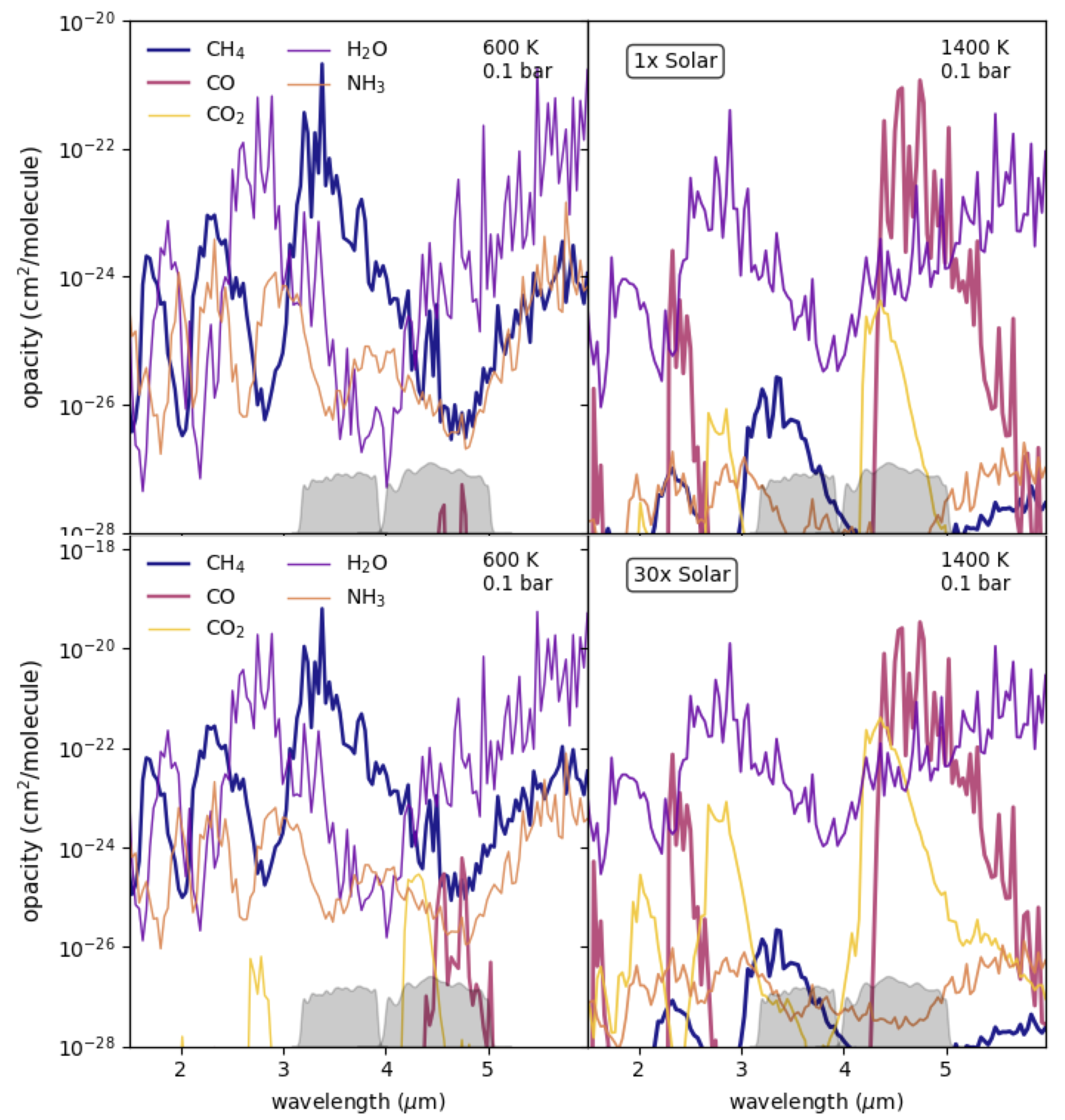}
    \caption{Opacities for a chemical equilibrium atmosphere at 600~K (left) and 1400~K (right) at 0.1 bar. Top panels show the abundance-weighted opacities for a solar composition atmosphere and the bottom panels show the abundance-weighted opacities for a 30x solar composition atmosphere. Carbon monoxide, water, methane, and carbon dioxide (for 30x solar) are the dominant absorbing species at the two IRAC channels (3.6 and 4.5~$\mu$m).}
    \label{P1:fig:opacities}
\end{figure}

Finally, transmission spectra are then simulated using the open-source, transit-depth calculator and retrieval tool, PLATON \citep{Zhang2019}\footnote{https://platon.readthedocs.io/en/latest/intro.html}. The code has been modified to take nonequilibrium compositions from our calculation, including \ce{CH4}, \ce{CO}, \ce{CO2}, \ce{C2H2}, \ce{H2O}, \ce{O2}, \ce{OH}, \ce{C2H4}, \ce{C2H6}, \ce{H2CO}, \ce{HCN}, \ce{NH3}, and \ce{NO}. The main opacities relevant for the wavelengths of Spitzer/IRAC are displayed in Figure \ref{P1:fig:opacities}. We assume chemical equilibrium for the rest of the species in PLATON. The details of the forward model can be found in \citet{Zhang2019}. We neglect stellar limb darkening in these models and the synthetic transit depth is expressed as $(R_p/R_s)^2$. 

\subsubsection{Calculating the model Spitzer/IRAC transit depths}

We integrate the simulated transmission spectra with Spitzer/IRAC spectral response functions and weight with the stellar flux using the following equation: 
\begin{equation}
\overline{\delta}_{\lambda} = \frac{\int_0^\infty  \delta(\lambda) \lambda R(\lambda) F_{s}(\lambda) d\lambda}{\int_0^\infty  \lambda R(\lambda) F_{s}(\lambda) d\lambda}
,\end{equation}
where $R(\lambda)$ is the spectral response function at either 3.6~$\mu$m or 4.5~$\mu$m [e-/photon] \citep{Quijada2004} and $\delta(\lambda)$ is the transmission spectrum from PLATON and $F_{s}(\lambda)$ is the stellar flux. The output, $\overline{\delta}_{\lambda}$, is the weighted average transit depth that would be observed with Spitzer/IRAC in either of the two bandpasses. 

Figure \ref{P1:fig:ultimateplot} shows the interpolated grid of fiducial models (solar composition, cloud-free with equilibrium chemistry). Here, we plot the normalized IRAC transit depth difference against the equilibrium temperature, and overplot the results from our transit survey. Figure \ref{P1:fig:tracks} shows the different tracks of the model grid that make up the shaded regions and Figure \ref{P1:fig:Kzzmodels} shows the different vertical mixing and metallicity interpolated grids with the data. For the cloudy grid, we simply assume a gray cloud opacity such that the spectra are flat and thus the transit depth difference would be zero, which is shown as a vertical line on Figures \ref{P1:fig:ultimateplot} and \ref{P1:fig:Kzzmodels}.

\begin{figure}[ht]
    \centering
    \includegraphics[width=\linewidth]{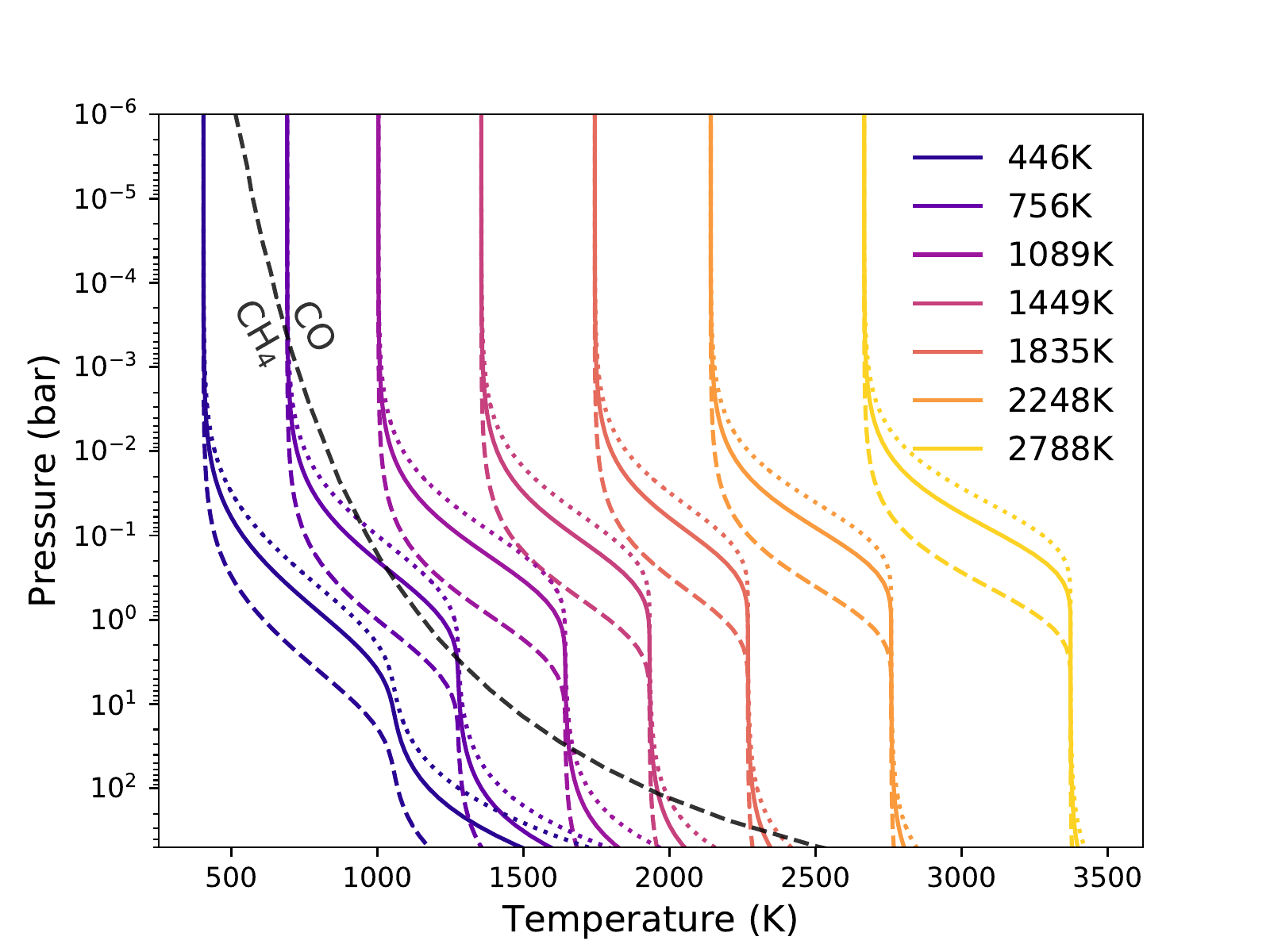}
    \caption{Analytical TP profiles for a grid of models spanning equilibrium temperatures $\sim$400-2800~K, showing every $\sim$400~K. For each temperature, we show three profiles where the surface gravity is varied: dotted, solid, and dashed lines represent 500, 1000, and 5000~\cmss~respectively. Grey dashed line represents the gas transition between \ce{CH4}- and \ce{CO}-dominated atmospheres when chemical equilibrium is assumed.}
    \label{P1:fig:TPgrid}
\end{figure}

\section{Results}
\label{P1:sec:Results}

\subsection{Measured transit depths and their ratios}

\subsubsection{Results of measured transit depths}

\begin{figure}
    \centering
    \includegraphics[width = \linewidth]{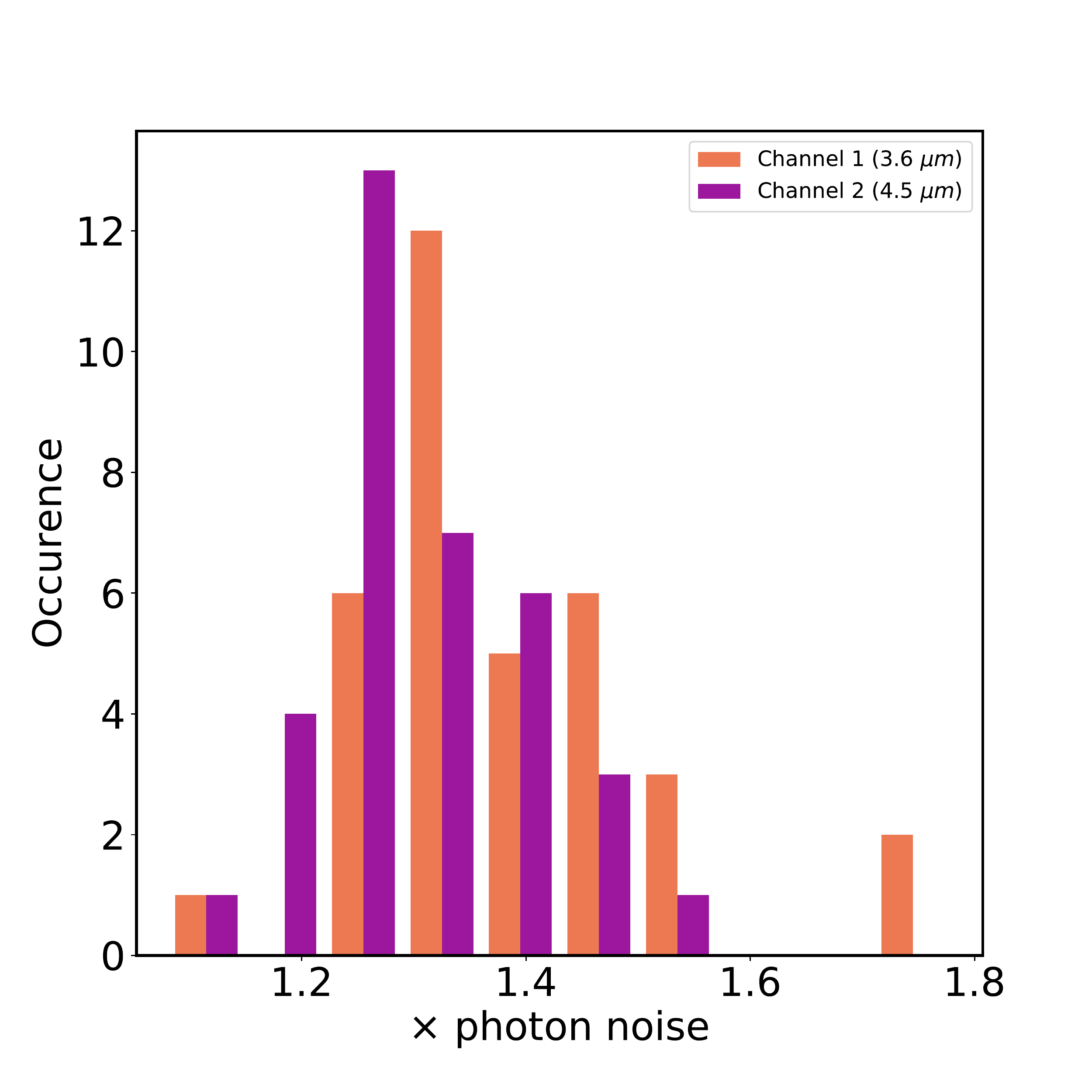}
    \caption{Histogram showing the percentage above photon noise for each of the individual light curves. Channel 1 (3.6~$\mu$m) is displayed in orange and channel 2 (4.5~$\mu$m) in purple.}
    \label{P1:fig:photonnoise}
\end{figure}

Table \ref{P1:tab:results} summarizes the results of the MCMC analysis of the light curves, and lists the final values and uncertainties for the transit depths, mid transit times, and impact parameters from the final fits as well as the inclination and semi-major axis obtained from the first fits. We checked that the initial fits of the semi-major axis and the inclination are in agreement with the literature values before fixing them with Gaussian priors for the second fit. The survey as a whole was in statistical agreement with the literature values within $<1\sigma$.

We also show the raw photometry with the best-fit model for each visit in Appendix \ref{P1:app:piperes}, and we show the corresponding plots of RMS versus bin size  in Appendix \ref{P1:app:piperes}. Figure \ref{P1:fig:normlc1} shows the reduced, normalized, and systematic corrected transit light curves for all planets in our sample for both channel 1 and channel 2 with the best-fit model resulting from the MCMC. We calculate the residuals, the $\chi^2$, and the RMS of the residuals as sanity checks for each light curve (Table \ref{P1:tab:tests}).

As mentioned in Section \ref{P1:subsec:fitting}, before performing a complete MCMC analysis, we first check the fraction above photon noise and scale up the errors accordingly. Figure \ref{P1:fig:photonnoise} displays a histogram of the fraction above photon noise for all analyzed light curves. The histograms have a median of 1.36 and 1.27 times photon noise for 3.6~$\mu$m and 4.5~$\mu$m respectively, which is typical for what has been achieved with Spitzer in the past \citep{Ingalls2016}.

\subsubsection{Comparison to literature}

Several of the planets from our survey have had their Spitzer light curves previously analyzed  \citep[e.g.,][]{Sing2016, Garhart2020}. We compare our results with those from \citet{Sing2016} and \citet{Garhart2020}. Our measured transits are consistent within 3-$\sigma$ with those from the literature apart from a couple of outliers described below. Two of the largest outliers are the channel 2 transit depth of KELT-7b and the channel 1 transit depth of WASP-62b, both analyzed in \citet{Garhart2020} with PLD. We interpret the differences as due to the brightness of the host stars, and more specifically as due to the number of pixels selected for the pixel level decorrelation. These stars are bright and therefore 12 pixels are selected to model the systematic errors in \citet{Garhart2020} whereas we use 9 pixels uniformly for the entire survey (e.g., see Figure \ref{P1:fig:rmsvsbin}). We emphasize that these differences do not affect the general conclusion of the paper. 

\subsubsection{Transit depth ratio}
\label{P1:subsec:transit}

We combine our results with transit measurements from the literature, which results in a survey of transit depths at 3.6 and 4.5~$\mu$m for 49 planets spanning a large range of equilibrium temperatures. We now compare all targets in our survey in a statistical manner. To do this, we opt to use a metric that is as free as possible from any assumptions: the normalized difference of the transit depths:

\begin{equation}
     \bar{\Delta}_{tr} = \frac{(\delta_{ch2} - \delta_{ch1})}{\delta_{ch1}}
.\end{equation}

With this calculation, we tested for correlations with a number of other parameters:  stellar parameters (Teff, logg, Fe/H, $R_s$), orbital parameters (semi-major axis (AU), eccentricity, inclination), and planetary parameters ($T_{\rm eq}$, logg, $R_p$, $M_p$, scale height). We looked for correlations between these parameters using two statistical methods. First, we calculated the Pearson correlation coefficient (r) and its associated chance probability (p). We then fit a straight line using an orthogonal distance regression (ODR) to account for the errors on both the abscissa and ordinate values as in \citet{Boggs1989}; we note the resulting residual variance of the fits.

\subsubsection{Searching for trends in the difference of transit depths}

We analyze our Spitzer survey by looking at the normalized difference in the transit depths. Our normalized transit depth difference metric confers the advantage that it does not include any additional assumptions on the composition of the atmosphere.  Several studies look at the number of scale heights crossed at different wavelengths, including the strength of the water feature in the HST/ WFC3 bandpass \citep[e.g.,][]{Sing2016}. Including the scale height requires an assumption on the mean molecular weight, which includes errors from the surface gravity and equilibrium temperature. Furthermore, our metric is also independent of the stellar radius,  unlike the difference in transit depths ($\delta_{ch2} - \delta_{ch1}$). Ultimately, this metric is a proxy for the ratio of the optical depths at these two wavelengths. We expect that the strength and magnitude of this metric can be used to test how the dominant expected atmospheric opacities change with the equilibrium temperature of the planets; see Section \ref{P1:subsec:Chemistry}.

\setcounter{table}{4}
\begin{table}
    \caption{Correlations between parameters and the transit depth ratio. We show the Pearson correlation coefficient ($r$), the associated chance probability ($p$), and the residual variance from an ODR linear fit to the data. }
    \label{P1:tab:pearson}
\begin{tabular}{lccc}
\hline\hline
Parameter &  $r$ &  $p$ &  Res Var \\
\hline
$T_{\rm eq}$ (a=0)      &                            -0.35 &            0.01 &               7.23 \\
$T_{\rm eff}$           &                            -0.34 &            0.02 &               7.14 \\
Stellar log(g)          &                             0.13 &            0.36 &               6.98 \\
$[$Fe/H$]$  &                            -0.21 &            0.15 &               4.48 \\
$R_p$              &                            -0.26 &            0.07 &               8.11 \\
Inclination            &                             0.20 &            0.18 &               7.03 \\
a (AU)         &                             0.07 &            0.63 &               8.46 \\
Planetary log(g)          &                             0.01 &            0.92 &               8.47 \\
$M_p$ ($M_J$)      &                             0.09 &            0.56 &               8.78 \\
H (km)             &                            -0.17 &            0.25 &               8.50 \\
$R_s$ ($R_{\odot}$)      &                            -0.40 &            0.00 &               7.06 \\
Radius Anomaly &                            -0.25 &            0.14 &               6.86 \\
\hline
\end{tabular}
\end{table}

We search for any correlations that could be present between the calculated normalized transit depth difference and the physical parameters of the planetary systems that we are exploring. Table \ref{P1:tab:pearson} summarizes the correlations for each of the parameters. The three parameters with the strongest Pearson correlation coefficients and the lowest chance probabilities are $T_{\rm eq}$, $T_{\rm eff}$, and $R_s$. Both $T_{\rm eff}$ and $R_s$ are incidentally included in the calculation of the equilibrium temperature, $T_{\rm eq}$ (in our case with zero albedo and full redistribution). We also observe that the weakest correlations are with the planetary mass, planetary radius, and semi-major axis. This is not surprising because our sample is highly biased towards hot Jupiters with a relatively small range of radii and masses, and with similarly close-in orbits. This means that the span of these parameters is small and therefore the uncertainties will be large and the correlations will not be obvious. 

\subsubsection{Transit depth versus equilibrium temperature}
\label{P1:subsec:TDvsTeq}

\begin{figure*}
    \centering
    \includegraphics[trim={0 4cm 0 0},clip, width=\textwidth]{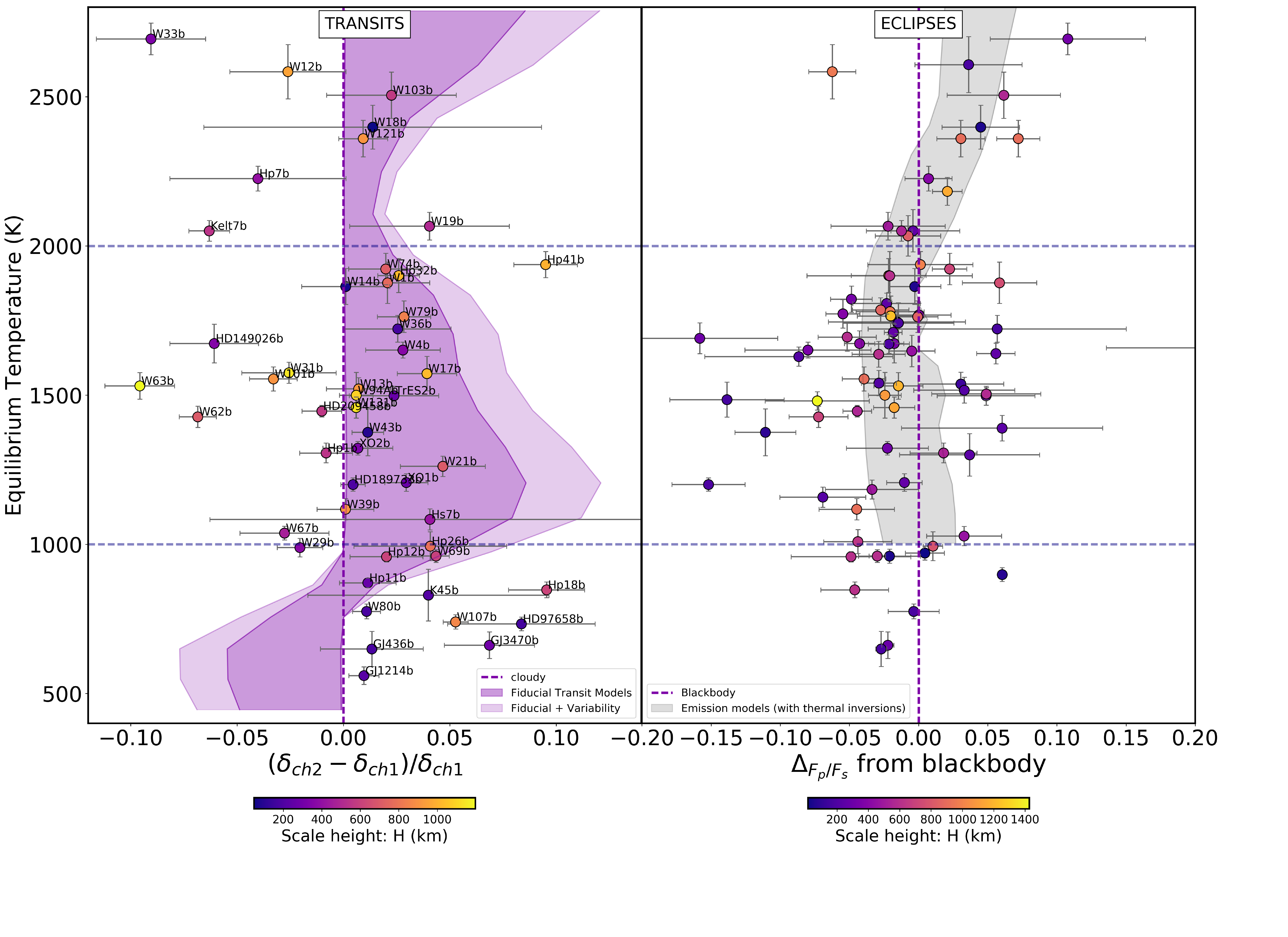}
    \caption{Left panel: Difference in transit depths between 4.5~$\mu$m and 3.6~$\mu$m normalized to the transit depth at 3.6~$\mu$m plotted against equilibrium temperature. The 1-$\sigma$uncertainties are shown in gray, and the color bar depicts the scale height in km. The shaded area on the left panel shows our grid of solar composition cloud-free equilibrium chemistry models; the extended lighter region is corrected for stellar variability. The dashed purple vertical line at zero is where planets with low-pressure gray clouds would lie. Right panel: Deviation from the black body in emission against the equilibrium temperature, presented in \citet{Baxter2020}. The color scale shows the scale height in km and the shaded region shows the grid of models containing temperature inversions.}
    \label{P1:fig:ultimateplot}
\end{figure*}

In Figure \ref{P1:fig:ultimateplot} (left panel), we plot the normalized transit depth difference ($(\delta_{ch2} - \delta_{ch1}) / \delta_{ch1}$) against the equilibrium temperature for all planets in our sample. This plot contains 49 planets with masses 0.02 - 10.2 $M_{jup}$, radii 0.24 - 1.9 $R_{jup}$, and equilibrium temperatures 550 - 2690 K. The color scale on the data points shows the scale height ($H$) of each planet, ($H = kT_{\rm eq}/\mu g$) calculated assuming a hydrogen-dominated atmosphere with mean molecular weight ($\mu$) of 2.3, equilibrium temperature ($T_{\rm eq}$) calculated with zero albedo and zero redistribution, and planetary surface gravity ($g$) from the literature. 

In Table \ref{P1:tab:significance} we show the weighted mean of the normalized transit depth difference and the corresponding number of scale heights for each temperature bin in Figure \ref{P1:fig:ultimateplot}. We also calculate the weighted mean of the absolute value of the normalized transit depth difference and the number of scale heights. 

We find that the weighted mean of the absolute value normalized transit depth difference and the number of scale heights to be significant to 8.0$\sigma$ and 7.5$\sigma$ respectively. This means that we are statistically detecting the atmosphere with a very high significance. 

All nine of the cool (<1000~K) planets lie on the positive side of the transit depth metric with a weighted mean transit depth of $0.029 \pm 0.007$, 4.0~$\sigma$ from zero (gray assumption). We also find that the weighted mean transit depth difference and the number of scale heights of the 1000-2000~K planets and the >2000~K planets are not significant ($<3\sigma$). We therefore treat all planets >1000~K as one sample. These 36 hot planets have an absolute value weighted mean that is 0.3 $\sigma$ from zero (cloudy) assumption. In total, 14 of these planets are consistent with the cloudy models (zero) within 1$\sigma$. However, as these hot planets span both positive and negative values of the transit depth difference, it is unsurprising that their weighted mean transit depth is only marginally deviating from zero. The weighted mean of the absolute value of the difference in the transit depths for the hot planets is $0.025 \pm 0.004$ (5.9 $\sigma$) and is more scattered than the cooler planets. 

\begin{table*}
    \centering
    \setlength{\tabcolsep}{3.5pt}
    \caption{Weighted means of the normalized transit depth difference ($(\delta_{ch2} - \delta_{ch1}) / \delta_{ch1}$), the absolute value of the normalized transit depth difference, the corresponding number of scale heights (NH), and its absolute value. This is shown for the different temperature ranges (<1000~K, 1000-2000~K, and >2000~K) presented in Figure \ref{P1:fig:ultimateplot}. The intermediate columns labeled N$\sigma$ indicate the significance of the previous weighted mean and weighted error.   }
    \label{P1:tab:significance}
\begin{tabular}{lllllllll}
\hline\hline
Planet Selection & $(\delta_{ch2} - \delta_{ch1}) / \delta_{ch1}$ & N$\sigma$ & $|(\delta_{ch2} - \delta_{ch1}) / \delta_{ch1}|$ & N$\sigma$ & NH & N$\sigma$ & |NH| & N$\sigma$ \\ 
\hline 
All planets &   0.010 $\pm$ 0.005 & 1.9 $\sigma$ &        0.028 $\pm$ 0.003 & 8.0 $\sigma$ &    0.2201 $\pm$ 0.0935 & 2.4 $\sigma$ &      0.5032 $\pm$ 0.0669 & 7.5 $\sigma$ \\
<1000~K &        0.029 $\pm$ 0.007 & 4.0 $\sigma$ &        0.032 $\pm$ 0.006 & 5.1 $\sigma$ &    0.4515 $\pm$ 0.1179 & 3.8 $\sigma$ &      0.4900 $\pm$ 0.1043 & 4.7 $\sigma$ \\
1000-2000~K &    0.002 $\pm$ 0.006 & 0.3 $\sigma$ &       0.023 $\pm$ 0.005 & 5.0 $\sigma$ &    0.0130 $\pm$ 0.1343 & 0.1 $\sigma$ &     0.4840 $\pm$ 0.0968 & 5.0 $\sigma$ \\
>2000~K &        -0.032 $\pm$ 0.015 & 2.2 $\sigma$ &       0.042 $\pm$ 0.010 & 4.1 $\sigma$ &    -0.5907 $\pm$ 0.4271 & 1.4 $\sigma$ &     0.9239 $\pm$ 0.3322  & 2.8 $\sigma$ \\
\hline
\end{tabular}
\end{table*}

\subsection{Results from the 1D grid of model transmission spectra}

\subsubsection{General trends observed in the grids of models}

\begin{figure}
    \centering
    \includegraphics[width=\linewidth]{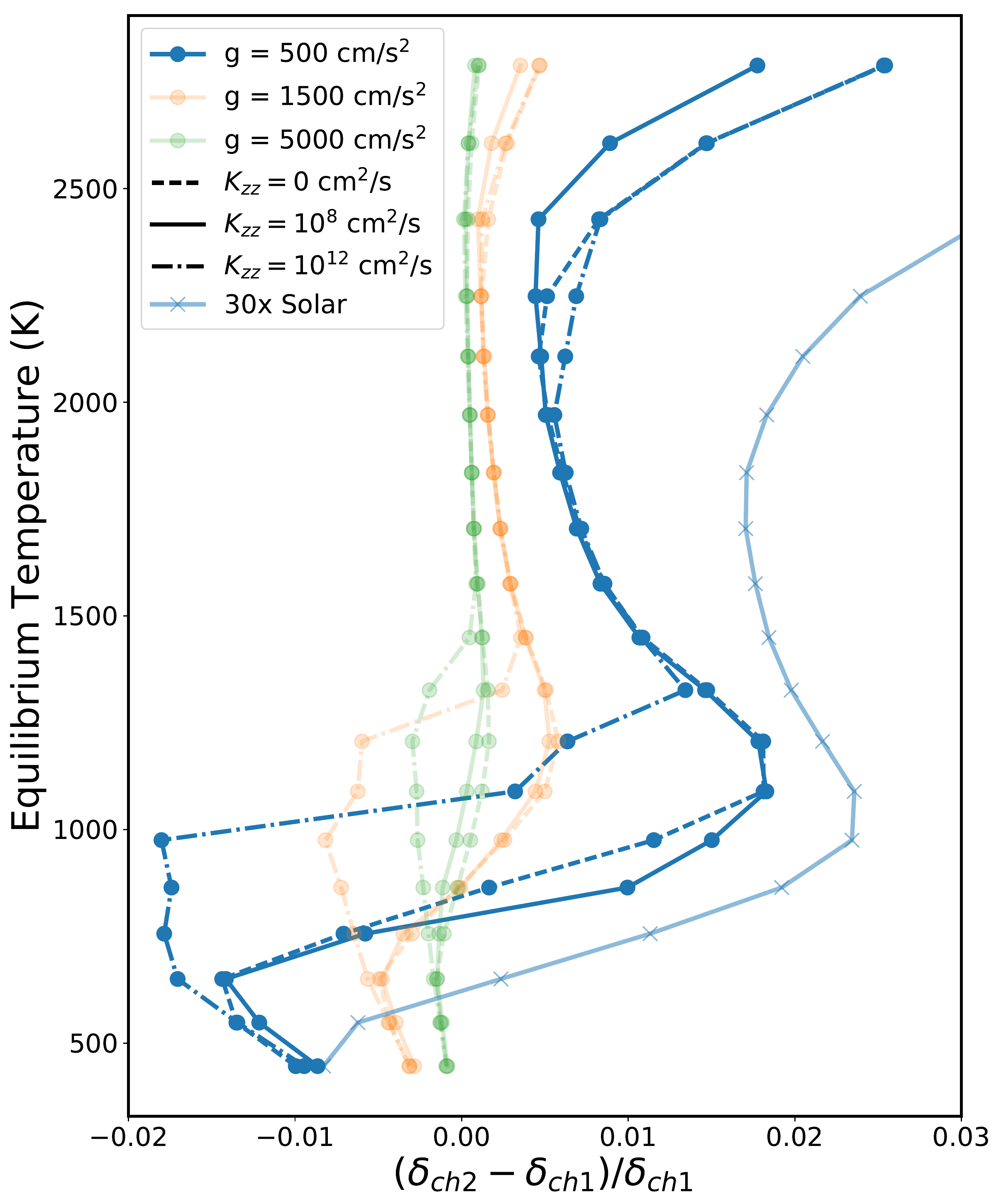}
    \caption{Normalized Spitzer transit depth difference as a function of the equilibrium temperature for a selection of the grid tracks created with our atmospheric model framework described in Sections \ref{P1:subsec:VULCAN} and \ref{P1:subsec:PLATON}. We show a selection of grids with 1x solar composition and $R_p = 2R_J$. Different colors show different surface gravities: blue is g = 500~\cmss, orange is g = 1500~\cmss~, and green is g = 5000~\cmss. Different line styles show the effect of vertical mixing: solid line shows equilibrium chemistry, dashed is $K_{zz}=10^8$~\cmcms,~and dot-dashed is $K_{zz}=10^{12}$~\cmcms. The lighter blue line with `x' markers shows a 30x solar track with $R_p = 2R_J$, g = 500~\cmss, and $K_{zz}=0$~\cmcms.}
    \label{P1:fig:tracks}
\end{figure}

\begin{figure*}
    \centering
    \includegraphics[width=\linewidth]{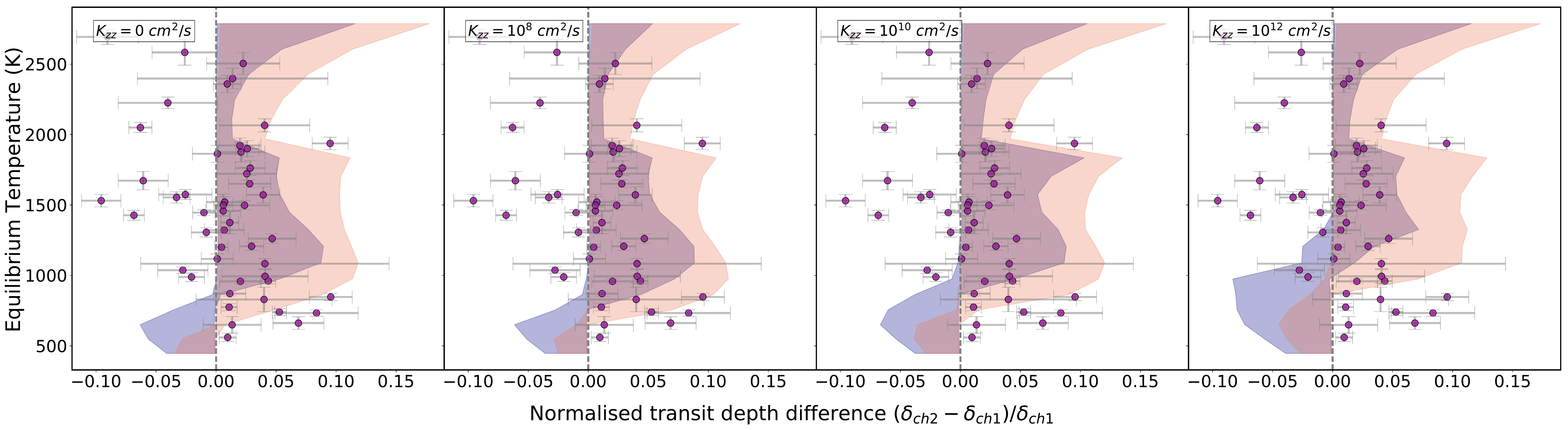}
    \caption{Normalized Spitzer transit depth difference as a function of the equilibrium temperature for the complete grids of transmission models created with our atmospheric model framework described in Sections \ref{P1:subsec:VULCAN} and \ref{P1:subsec:PLATON}. At each temperature, we show the models where the surface gravity is representative of our survey; i.e., at $T_{\rm eq}>1800$~K we only plot g = 1500 and 5000~\cmss. Panels from left to right show equilibrium chemistry (no vertical mixing), $K_{zz} = 10^8$, $K_{zz} = 10^{10}$ and $K_{zz} = 10^{12}$~\cmcms. The blue translucent shaded region shows the 1x solar composition and orange translucent shaded region shows the 30x solar composition (overlap in purple). Gray dashed line represents a gray opacity source showing no spectral features. Planets from our sample are overplotted in purple circles with their $1\sigma$ error bars.}
    \label{P1:fig:Kzzmodels}
\end{figure*}

In Figure \ref{P1:fig:tracks} we show a selection of tracks from the complete grid of models and in Figure \ref{P1:fig:Kzzmodels} we show each interpolated grid as a shaded region in comparison with the survey data. The fiducial model grid (1x solar and equilibrium chemistry, no vertical mixing $K_{zz}=0$) plotted in Figure \ref{P1:fig:tracks} shows the effect of increasing equilibrium temperatures on the transit depths. At $\sim$900~K, the model grid switches from a negative transit depth difference to a positive transit depth difference. 

The interpolated grid shows a spread in the expected difference in the two transit depths. An important aspect of the model grid that largely influences the spread is the surface gravity. Lower surface gravities result in larger scale heights and lead to a larger signal in the difference of the two Spitzer/IRAC transit depths. The surface gravity also changes the shape of the TP profile as seen in Figure \ref{P1:fig:TPgrid}. Figure \ref{P1:fig:tracks} shows the effect of different surface gravities. We designed the model grid to span the parameters of the survey, notably with surface gravities of g = 500, 1000, 1500, and 5000~\cmss. However, for the ultra-hot model planets with low surface gravity of g = 500~\cmss~, the upper atmosphere exceeds the Hill radius. These models do not represent any planets in our survey because the hottest planets in our survey tend to have larger surface gravity (g$\sim1000$~\cmss. We therefore discard these model planets from Figure \ref{P1:fig:Kzzmodels}.

The effect of vertical mixing can be seen in Figure \ref{P1:fig:tracks}. A large amount of mixing results in the transition between \ce{CH4} and CO occurring at higher temperatures. Increasing the metallicity to 30x solar has the effect of lowering the temperature of the transition between negative and positive transit depth difference. Increased metallicity also results in a stronger positive signal for the hotter planets >1000~K.  

\subsubsection{Statistical comparison of planet atmospheres with the model grid}
\label{P1:sec:gridstats}

We compare the data with the grids of models quantitatively by calculating the average number of standard deviations (based on the 1$\sigma$ uncertainties) between each of the planets and their corresponding model grid point with the closest input parameters ($T_{\rm eq}$, log($g_p$), $R_s$ and $R_p$). We then compute a weighted average for the whole grid, such that we can express the statistical significance of each grid with one number. We split this comparison into different temperature regimes based on the expected carbon chemistry. We compare the data to a transit depth difference of zero, representing a gray cloud opacity. Additionally, we also compare the data with the grids of models qualitatively by interpolating a shaded region between grid points, allowing us to visually compare the models with the Spitzer/IRAC transit depth difference; for example see Figure \ref{P1:fig:ultimateplot}.

In Section \ref{P1:subsec:TPcreation}, we fix the orbital distance to 0.035 AU in our model grid creation. We do this because in our sample of planets the equilibrium temperature has a much larger correlation with the stellar effective temperature than with the semi-major axis. The range of semi-major axes in our sample spans $\sim$0.017 to $\sim$0.06 AU. We explore how much our choice of model parameterization (fixing the orbital distance to 0.035 AU) affects our results with the following two tests. We start by creating models with the minimum and maximum orbital distance of our sample, 0.017 and 0.06 AU. 

\begin{figure}
    \centering
    \includegraphics[width=\linewidth]{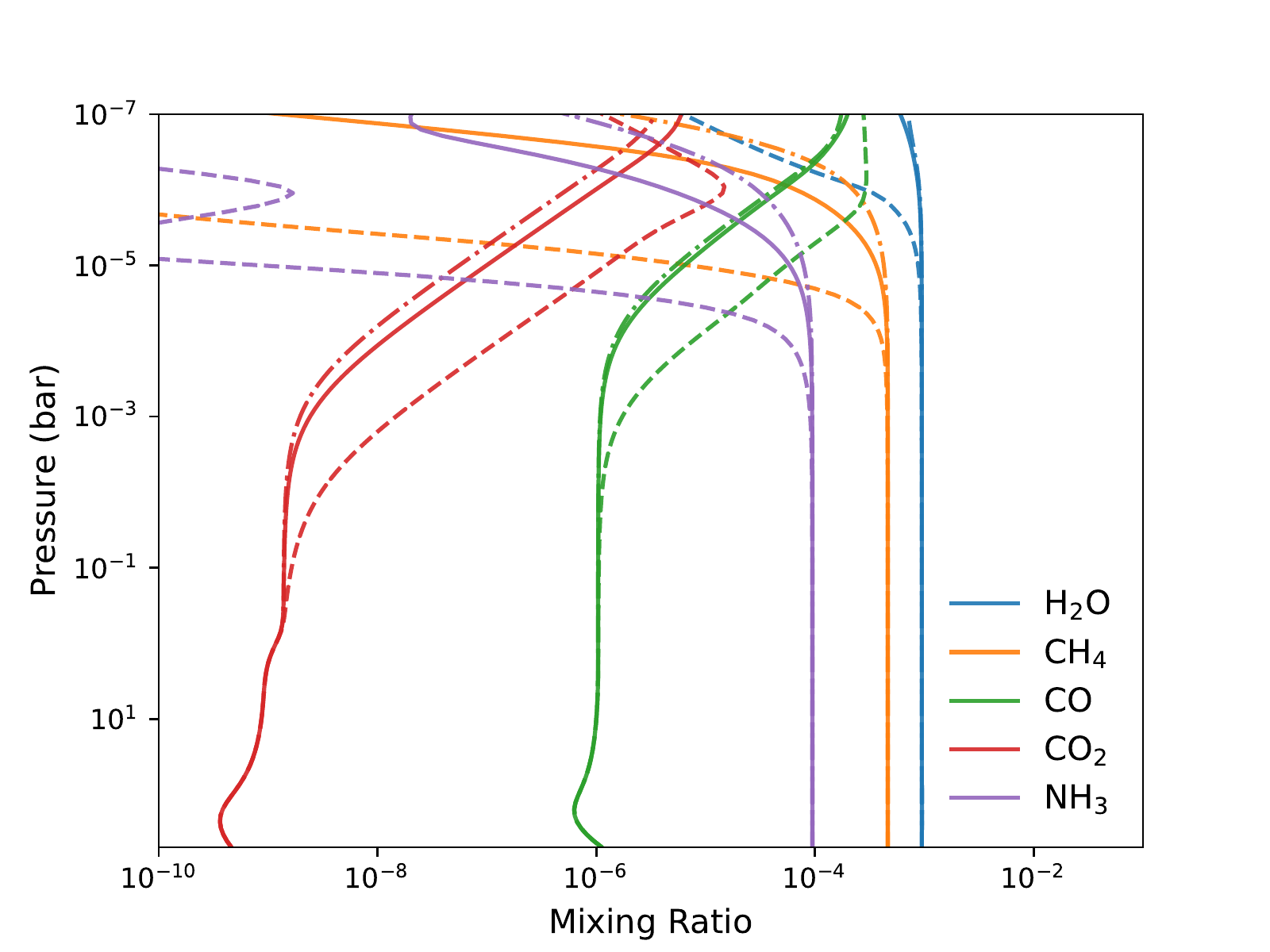}
    \caption{Abundance mixing ratios at different pressures for the main species in the Spitzer bandpasses. The solid line shows the nominal situation (a = 0.035 AU, $T_{\rm eff}$ = 3750K), the dashed line shows a = 0.06 AU, $T_{\rm eff}$ = 4250K, and the dashed-dotted line shows a = 0.017 AU, $T_{\rm eff}$ = 4250K.}
    \label{P1:fig:Tefftest}
\end{figure}

In the first test, we match the equilibrium temperature by changing the effective temperature of the star. For a 650 K planet, an orbital distance of 0.017 AU corresponds to a stellar effective temperature of 3250 K and 0.06 AU corresponds to 4250 K. Figure \ref{P1:fig:Tefftest} shows the effects on chemistry, where the star with higher $T_{\rm eff}$ provides greater flux even at larger orbit and leads to more photolysis. Nevertheless, it mainly impacts the main species at the lower pressures (P < 1 mbar). We find that the resulting difference in our transit depth metric for a planet placed at the minimum and maximum orbital distance is 0.0025. This is a factor of ten smaller than the mean error bar in our sample, so we do not expect this to change our results. 

In the second test, we match the equilibrium temperature by changing the stellar radius. This time the resulting difference in our transit depth metric is 3.2e-6, which is three orders of magnitude smaller than the mean error bar of our sample. As the changes in the models are so small compared to the size of the uncertainties, we do not expect that the different orbital distances are the reason behind the scatter seen in Figure \ref{P1:fig:ultimateplot}.

\begin{figure}
    \centering
    \includegraphics[width=\linewidth]{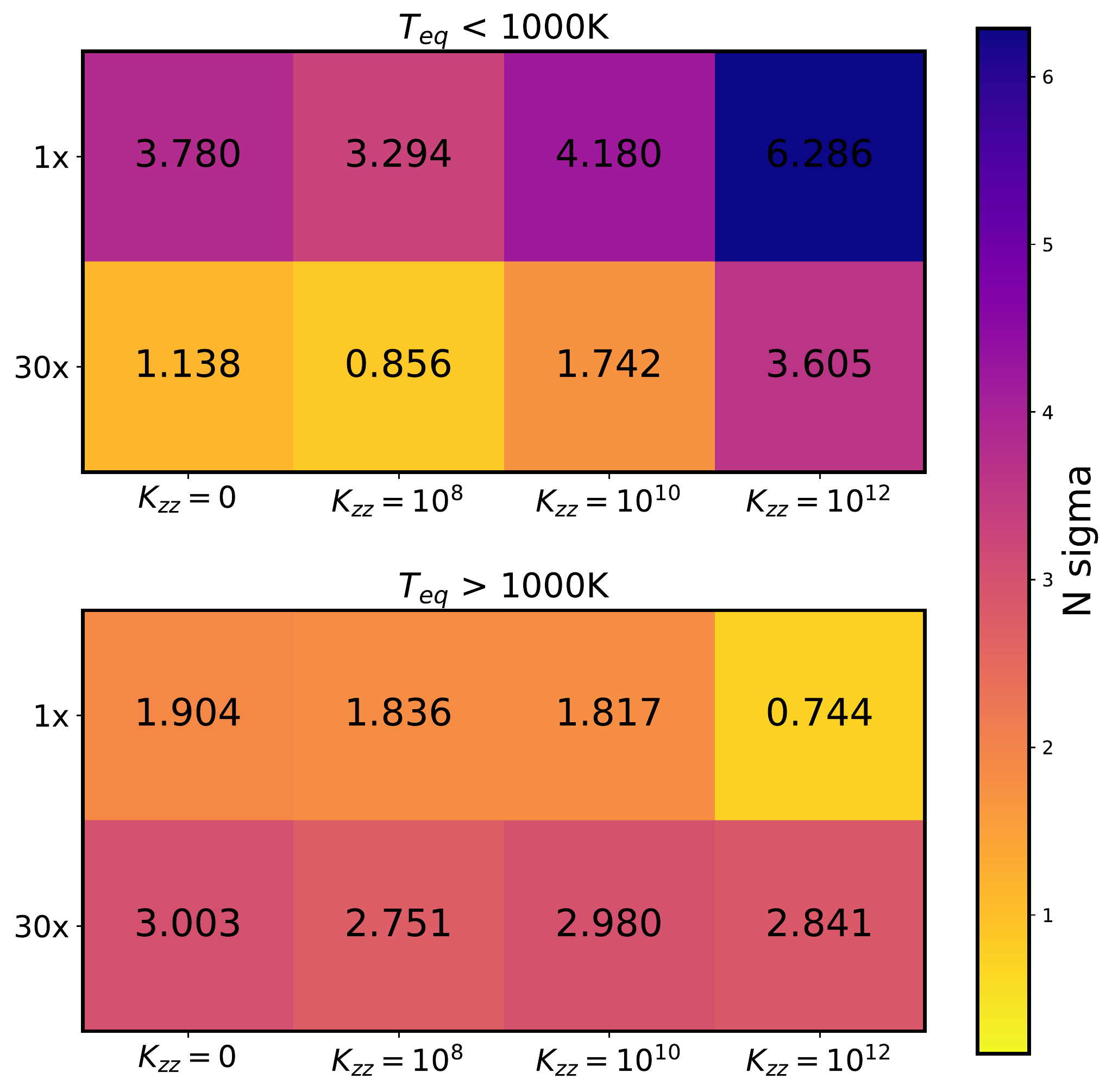}
    \caption{Number of sigmas between the data and each model grid. We do this for eight grids and three sets or subsets of planets. The eight model grids are composed of two different metalicities (1x and 30x Solar) and four different vertical mixing scenarios ($K_{zz} = 0, 10^{8}, 10^{10}$ and $10^{12}$~\cmcms). The color bar represents the average number of sigmas  between each model grid and the set of data; a lower N sigma (yellow) means a better fit. The top panel shows the results for the cool planets ($T_{\rm eq}$<1000~K) and the bottom panel shows the  hot planets ($T_{\rm eq}$>1000~K). The number of sigmas is written on each cell.}
    \label{P1:fig:gridstats}
\end{figure}

Figure \ref{P1:fig:gridstats} displays the results of the statistical comparison of each model grid with the planets in our survey. Each planet transit depth measurement is compared to the corresponding transmission model with the closest parameters ($T_{\rm eq}$, log($g_p$), $R_s$ and $R_p$). We calculate the statistical significance for a set of planets, which is quantified by the average number of sigmas, for all eight grids of models. In the two panels of Figure \ref{P1:fig:gridstats} we show the results of the cool planets ($T_{\rm eq}$<1000~K), followed by the hot planets ($T_{\rm eq}$>1000~K). We find that the hot planets are best fit by 1x solar and high vertical mixing, $K_{zz} = 10^{12}$~\cmcms. We rule out high-metallicity models for these planets to $\sim3\sigma$ confidence. 

On the other hand, we find that the cool planets are best fit by 30x solar and a low amount of vertical mixing ($K_{zz} = 10^{8}$ or $K_{zz} = 0$~\cmcms). We find that the 1x solar composition and high amounts of vertical mixing ($K_{zz} = 10^{12}$~\cmcms) are ruled out with $>3\sigma$ confidence for these cool planets. 

Comparing the model grids to the full sample, we find that the full sample mimics the cool sample. This is because the different grids of models are divergent at the cool temperatures, and so the results from the cool temperatures drive the statistical results for the full grid. 

\section{Discussion}
\label{P1:sec:Discussion}

\subsection{Expected opacities at 3.6 and 4.5~$\mu$m}
\label{P1:subsec:Chemistry}

The features we see in the transmission spectra are a result of the underlying chemistry at the pressures probed by our observations. Figure \ref{P1:fig:opacities} shows the abundance-weighted opacities for the dominant opacity sources in the grid of models at the wavelengths of the Spitzer bandpasses. The dominating absorbing molecules in the Spitzer bandpasses are \ce{CH4} and \ce{H2O} at 3.6~$\mu$m, and \ce{CO}, \ce{H2O,} and \ce{CO2} (for high metallicities) at 4.5~$\mu$m. As \ce{H2O} opacity is about equally present in both IRAC bandpasses, the two Spitzer transit depths can be used to understand the relative abundance of CO and \ce{CH4}. The following summary chemical reaction plays an important role in determining the dominating carbon-bearing species in an atmosphere \citep[e.g.,][]{Visscher2010, Moses2011, Visscher2011, Ebbing2016}:
\\
\ce{ \centering CH4 + H2O <=> CO + 3H2}. 
\\

At temperatures higher than $\sim1100$~K the forward reaction is favored (CO creation) for nominal pressures of $\sim$ 1 bar, whereas at temperatures lower than $\sim1100$~K the reverse reaction is favored \citep[\ce{CH4} creation; e.g.,][]{Madhusudhan2012, Molliere2015, Molaverdikhani2019}. The gas transition between \ce{CH4} and CO is plotted as a function of temperature in Figure \ref{P1:fig:TPgrid}. This shows where the abundance of \ce{CH4} and CO are the same \citep{Visscher2012}. A temperature pressure profile crossing this line results in CO or \ce{CH4} becoming the dominant absorber.

We therefore expect that the atmospheres of planets in thermochemical equilibrium with temperatures above $\sim1100$~K have \ce{CO} as the dominating carbon-bearing species and the cooler atmospheres have \ce{CH4}. The result of this on the normalized difference of the transit depths (Figure \ref{P1:fig:ultimateplot}) is that the \ce{CH4} planets would have a negative difference whereas CO planets have a positive difference. The transition from negative to positive transit depth differences seen in Figures \ref{P1:fig:ultimateplot}, \ref{P1:fig:tracks}, and \ref{P1:fig:Kzzmodels} shows the changing carbon chemistry (\ce{CH4} to \ce{CO}) with increasing equilibrium temperature. We find that the equilibrium temperature of the transition in the fiducial model grid (thermochemical equilibrium, 1x solar) is slightly lower than the 1100~K presented in a previous study \citep[e.g.,][]{Madhusudhan2012}. We emphasize that the transition from \ce{CH4} to \ce{CO} depends on the temperature and pressure of the layer being probed with Spitzer/IRAC transmission photometry, and that this temperature is not necessarily at the  equilibrium temperature of the planet.

\subsection{The  transit survey}
\subsubsection{Comparing transit depths to fiducial model grid}

Figure \ref{P1:fig:ultimateplot} shows the normalized difference of the two Spitzer transit depths with the fiducial grid of models. The fiducial models are calculated with opacities from thermochemical equilibrium and 1x solar composition. The sample of planets with temperatures hotter than 1000~K follow the fiducial models, but we see that the cool planets appear to deviate from this model grid. As we see that different chemical and physical processes are likely occurring at these different equilibrium temperatures, we proceed by splitting Figure \ref{P1:fig:ultimateplot} into three temperature regimes based on the expected chemistry from our model grid: the cooler, methane planets (<1000~K), the hotter, carbon monoxide hot planets (1000~K - 2000~K), and the few ultra-hot planets where molecular dissociation can occur ($T_{\rm eq}$>2000~K).

There are 13 planets in our survey with $T_{\rm eq}<$1000~K. Our fiducial (1x solar and no vertical mixing) models demonstrate that the predicted carbon-bearing species for planets in this temperature regime is methane, which results in the models occupying the negative side of Figure \ref{P1:fig:ultimateplot}. However, we find that the data show the opposite trend: all planets lie on the right side of Figure \ref{P1:fig:ultimateplot}. We find that this equilibrium chemistry grid is ruled out at 3.8$\sigma$, which is statistically capturing the dearth of methane in the sample of coolest planets; see Section \ref{P1:sec:gridstats}. This supports previous individual studies of cool gas giants with HST/WFC3 and indicates that there are more complex physical processes happening that are not included in the fiducial models.

There are 28 planets in the mid-temperate/hot range (1000-2000~K) and 8 planets in the hot/ultra-hot range (>2000~K) of Figure \ref{P1:fig:ultimateplot}. Of these 36 hot/ultra-hot planets, 14 are consistent to less than 1 $\sigma$ with the cloud-free solar composition model grid. In Section \ref{P1:sec:gridstats} we show that these planets are consistent with the fiducial model grid to 2$\sigma$. Additionally, we find that there is only 1 of these 36 hot/ultra-hot planets with a stronger positive signal than the fiducial model grid, meaning that a model grid with a higher CO abundance (e.g., 30x solar) is not required to explain our sample of observations. We find that 30x solar is ruled out with 3 $\sigma$ confidence for the hotter planets.

There are several effects not included in the fiducial grid of models that contribute to the statistical deviation. For example, we assume solar metallicity, no vertical mixing, and cloud-free atmospheres. We compare the survey of planets to the model grids in a statistical manner and discuss the effects of each of these in detail below.

\subsubsection{Effect of metallicity in hot Jupiter atmospheric spectra}
\label{P1:sec:metallicitydisc}

The metallicity of a planet contributes to the atmospheric molecular abundances. Our fiducial model grid assumes 1x solar composition and solar metallicity. Increasing the metallicity would increase the amount of CO in the atmosphere \citep[e.g.,][]{Venot2014}. Figure \ref{P1:fig:tracks} shows a 30x solar track and Figure \ref{P1:fig:Kzzmodels} shows the whole interpolated grid (with no vertical mixing; see the first panel). Increasing the metallicity to 30x solar results in a lower temperature at which the model atmospheres transition between \ce{CH4} and CO. This transition occurs at a temperature of around 600~K, much lower than the transition of 900~K for the fiducial grid. 

In Section \ref{P1:sec:gridstats} we show that the cool planets lack the methane signature and are better fit with 30x solar composition models, with a significance of >2.5$\sigma$. This is the case for the lower values of vertical mixing ($K_{zz} = 0$,  $10^8$ and $10^{10}$~\cmcms) discussed in more detail in Section \ref{P1:sec:Kzzdisc}. These cool planets are also generally lower mass planets because of the detection biases for these systems; see Figure \ref{P1:fig:planets}. Lower mass planets typically have higher metallicities \citep{Fortney2013, Welbanks2019}. Therefore, a higher average metallicity in the 13 planets with temperatures <1000~K likely explains the lack of methane. Our findings support the predicted high metal enrichment in cool gas giants presented by \citet{Espinoza2017}. These latter authors predict C/O ratios for a sample of 50 gas giants with $T_{\rm eq}<1000$~K; 6 of our 13 planets in this temperature range are also in their sample. Furthermore, our finding of high metallicity for these coolest warm giant planets supports the individual high-metallicity measurements of several planets in the literature: HAT-P-12b \citet{Line2013}, HAT-P-26b \citep{Wakeford2017}, GJ 436b \citep{Morley2017} and HAT-P-11b \citep{Mansfield2018}.  All of these exoplanet atmospheres are found to have super-solar metallicities, except for GJ 3470b which is suggested to have a relatively low atmospheric metallicity for its planet mass \citep{Benneke2019}.

On the other hand, the planets with equilibrium temperatures >1000~K are consistent with the 1x solar composition models to less than 2$\sigma$ for all values of $K_{zz}$. The higher metallicity grid is less favored for these planets (2.6$\sigma$ deviation). Similar to the high abundance of CO at cooler temperatures, the high-metallicity model grid shows stronger CO features throughout the entire temperature range, which is not favored by the planets in our survey. We do not find it necessary to statistically invoke high metallicity to explain the near-infrared spectral features of hot Jupiters. 

Figure \ref{P1:fig:opacities} shows the opacities for the 1x and 30x metallicity used in the creation of our model grids. In practice, differences in the opacities for the two cases would also affect the temperature pressure profile. However, in our analysis we do not compute the temperature pressure profiles self consistently. Nevertheless, we can predict what effect this might have. Higher metallicities would result in hotter temperatures in our TP profiles, which would in turn result in a larger \ce{CO}/\ce{CH4} ratio. This means that we could explain the dearth of methane with less extreme enhancements in the metallicity of the models. 

\subsubsection{Vertical mixing and nonequilibrium effects}
\label{P1:sec:Kzzdisc}

Another aspect not included in our fiducial model grid is the presence of nonequilibrium effects such as photochemistry, advection, convection, and turbulence in the atmosphere. To capture some of these nonequilibrium atmospheric processes, we introduce an eddy diffusion coefficient, $K_{zz}$, into our modeling (see Section \ref{P1:subsec:VULCAN}). Theory suggests that, for hot Jupiters, $K_{zz}$ can range from $10^8$ to $10^{12}$~\cmcms~based on the estimation from the mean vertical wind in GCMs \citep{Moses2011}. We create four different grids of models spanning the range of eddy diffusion coefficients: equilibrium chemistry, $K_{zz} = 10^8$, $K_{zz} = 10^{10}$, and $K_{zz} = 10^{12}$~\cmcms. 

The models incorporating different $K_{zz}$ show that the transition between \ce{CH4} and CO being the dominating carbon bearer in these atmospheres occurs at higher temperatures for larger values of $K_{zz}$. This is because with larger values of $K_{zz}$, the mixing penetrates deeper into the atmosphere and can therefore dredge up methane to the observable pressures of hotter planets where methane is not expected. The models on Figure \ref{P1:fig:Kzzmodels} (right panel) demonstrate that $K_{zz} = 10^{12}$~\cmcms~can dredge up \ce{CH4} for planets up to 1300~K.

For the cool planet data (T<1000~K), we find that the models containing low amounts of vertical mixing are significantly favored over high vertical mixing for both metallicities. For 30x solar metallicity, the low mixing $K_{zz}$ = $10^8$~\cmcms~fits marginally better than equilibrium chemistry ($K_{zz}$ = 0) and is a 3$\sigma$ better fit than the high vertical mixing ($K_{zz}$ = $10^{12}$~\cmcms). On the other hand, for the hot planets we find that $K_{zz}$ = $10^{12}$~\cmcms~is favored over the lower mixing or no mixing for both the 1x and 30x solar metallicities. 

\citet{Komacek2019} showed that for tidally locked hot Jupiters, vertical mixing increases with increasing equilibrium temperature and rotation rates: starting at $K_{zz} = 10^{7}-10^{8}$~\cmcms~for the coolest (500~K) planets and going to $K_{zz} = 10^{11}-10^{12}$~\cmcms~for the hottest (1500-3000~K). We find that the cool planets support these results, with a vertical mixing of $K_{zz} = 10^{8}$~\cmcms~favored by the data. However, the hotter planets seem to suggest a lower level of mixing than theory predicts, our models with $K_{zz} = 10^{10}$~\cmcms~are marginally supported over the equilibrium and $K_{zz} = 10^{8}$~\cmcms~grids, which is lower than the theoretical maximum of $K_{zz} = 10^{12}$~\cmcms. These findings are in line with the findings of \citet{Miles2020} for nonequilibrium processes in brown dwarfs. These latter authors found warmer brown dwarfs showed lower mixing than theory predicts, yet the cooler objects were close to the theoretical maximum.

Additionally, our nonequilibrium chemistry models include the effects of photochemical reactions. For hot planets ($T_{\rm eq}$ > 1000 K), CO is only dissociated in the upper atmosphere due to its strong bond, which has negligible influence on the Spitzer bandpasses. For cooler planets ($T_{\rm eq} \leq$ 1000 K), \ce{CH4} is dissociated by atomic hydrogen produced by photolysis. This destruction of \ce{CH4} can penetrate down to around 0.1 mbar with lower mixing ($K_{zz}=10^{8}$~\cmcms). Nevertheless, the competing effects of mixing can overtake and efficiently transport methane to the upper atmosphere. HCN is also produced by photochemistry and can reach abundances close to CH4 in some cases. Nevertheless, HCN absorbs similarly at the two IRAC wavelengths, and so we do not expect it to have significant effects on the normalized transit depth difference. 

As vertical mixing is responsible for dredging up \ce{CH4} to observable pressures in the hotter planets, and not for dredging CO to observable pressures in the cooler planets, we conclude that the dearth of methane is not due to strong atmospheric mixing, but is likely due to the higher metallicity of these atmospheres. Another possible factor affecting the lack of methane signatures in the cool planets could be the amount of interior heating; see \citet[e.g.,][]{Fortney2020}. We find that several of the coolest planets are eccentric (see Table \ref{P1:tab:jumpParams}), which could cause some tidal heating. Our temperature pressure profile calculation assumes an interior heating of $T_{\rm int}= 150$~K. However, substantial interior heating, $T_{\rm int}$>300~K, could result in pushing the deeper layers of these atmospheric TP profiles towards the CO regime \citep{Morley2017, Benneke2019, Thorngren2019, Thorngren2020}. If the interior is more CO dominated, then vertical mixing could dredge up CO in the cooler planets, resulting in a dearth of methane \citep[e.g.,][]{Moses2013}. We did not test this as it is beyond the scope of our paper. 

\subsubsection{Effects of clouds on the cool and hot Jupiter atmospheric spectra}

Clouds are ubiquitous in transiting exoplanet atmospheres \citep{Sing2016}.
There are several mechanisms responsible for producing homogeneous and inhomogeneous clouds on tidally locked planets \citep{Parmentier2013, Parmentier2021, Helling2016, Helling2019a, Helling2019b}.
An example can be found in \citet{Line2016} in which HD 189733b and HAT-P-11b can be explained by patchy clouds without the need to invoke global clouds or high mean molecular weight atmospheres. 

Hazes are expected to be prominent in the cooler atmospheres. \citet{Morley2015} predicted that a transition between haze-free and hazy atmospheres will occur at 800-1100~K, implying that any planet below this temperature might show no molecular features. \citet{Gao2020} showed that the amplitude of the HST/WFC3 water feature on planets with temperatures <900~K is such that these atmospheres become dominated by haze formation. However, \citet{Kawashima2019} predict that molecular features such as CO and \ce{CH4} are still detectable in the infrared for their sample of warm Jupiters (<1000~K) with hazy atmospheres. 

Furthermore, if all planets with temperatures <1000~K in our survey were characterized by a gray cloud opacity, then we would expect the transit depth difference to be evenly distributed around zero in Figure \ref{P1:fig:ultimateplot}. However, these 13 planets have a mean transit depth of $0.026 \pm 0.008$. This rules out a gray cloud (flat spectrum) at 4.0~$\sigma$ confidence for all planets, suggesting that these planets cannot be characterized by a gray cloud opacity, and that there is a molecular feature. 

\citet{Molaverdikhani2020} suggested that clouds could play a role in the heating of the atmosphere, resulting in a lack of \ce{CH4}. However, such clouds would also dampen the \ce{CO} feature significantly. This effect could be the reason for the few planets consistent with zero, but we do not expect that this effect explains the 4.0$\sigma$ detection for the sample of cool planets (<1000~K). 

There are 14 planets with equilibrium temperature >1000~K that have transit depth differences consistent with zero (flat spectrum). The weighted mean transit depth difference of all these planets is -0.002 $\pm$ 0.006, only 0.3$\sigma$. However, the weighted mean of the absolute value of the transit depth difference is 0.025 $\pm$ 0.004 (5.9$\sigma$). 

Based on the prediction by \citet{Morley2015} we would not expect hazes at these temperatures. However, \citet{Gao2020} show that the HST water feature is dampened when compared to a cloud-free atmosphere, and they find the data is better fit by their models containing silicate clouds. Furthermore, \citet{Line2016} suggested that patchy cloud cover can mimic the spectral features of a high-mean-molecular-weight atmosphere, resulting in a flatter transmission spectrum. Additionally, due to the varying temperature across the day and night sides of tidally locked highly irradiated hot Jupiters, clouds and hazes may behave differently at the east and west terminators of the planet \citep{Kempton2017b} such that photochemically generated hazes formed on the day side can be blown over to the nightside and dampen the transmission features. We therefore expect clouds to indeed play a role in dampening the spectral features in some of our planets, namely those in the temperature region predicted to be cloudy by \citet{Gao2020} (>1000~K).

However, as there is still a strong signal in the absolute value of the transit depths of these planets (5.9$\sigma$), there is indication that the population cannot be captured by a completely featureless model. Mie scattering theory results in a drop off in cloud opacity at ~2-3~$\mu$m \citep[e.g.,][]{Benneke2019}. As we are detecting molecular features between 3 and 5~$\mu$m, it may be that any possible cloud particles exhibit Mie scattering in this regime. Including Mie scattering as a cloud prescription in our transmission-spectrum forward modeling is beyond the scope of this paper. However,  cloud opacity that is lower at 4.5~$\mu$m than it is at 3.6~$\mu$m would result in a negative transit depth metric, similar to the expected methane signature. However, we do not find planets with a negative transit depth metric, and hence find no evidence for Mie scattering clouds.

\subsubsection{Outliers and the effect of nightsides}
\label{P1:subsec:additionaleffects}

According to our grids of models, we do not expect any of the planets above 1400~K to have \ce{CH4} as the dominating carbon-bearing species in any of the metallicity or mixing scenarios. However, there are two hot planets that are significantly on the left: HD149026b and WASP-33b (2.9$\sigma$ and 3.5$\sigma$ from zero respectively). HD 149026b has previously been discrepant from models. For example, \citet{Zhang2018a} found that they needed 30x solar metallicity to reproduce the Spitzer 3.6 and 4.5~$\mu$m phase curves. Furthermore, the biggest outlier, WASP-33 b, is a planet that is orbiting a $\delta$ Scuti star, with pulsating periods close to the transit duration \citep{Herrero2011}. Both of these planets indicate that there may be additional factors that could significantly affect the transit light curves, but statistically it is not unexpected to have a couple of outliers. In Section \ref{P1:sec:stellarVariabilitiy} we discuss how we treat stellar variability for the whole survey. 

Additionally, three of the seven hottest planets above 2000~K (WASP-33b, WASP-121b, and WASP-18b) show evidence of a temperature inversion \citep{vonEssen2015, Haynes2015, Evans2017, Arcangeli2018}. However, transmission spectroscopy is not as sensitive to the temperature profile at low resolution \citep[e.g.,][]{Brown2001} and so we do not expect to see the effect of temperature inversions in the transit depth difference of the two Spitzer/IRAC bandpasses. Additionally, the \ce{H-} opacity seen at the WFC3 bandpass \citep[e.g.,][]{Arcangeli2018} does not become important at the Spitzer/IRAC bandpasses until equilibrium temperatures as high as 3500~K. 

\subsubsection{Radius anomaly}

Our sample subsequently spans a very large range of scale heights ranging from HAT-P-2b with a scale height of 26~km to WASP-31b with a scale height of 1150~km. Figure \ref{P1:fig:ultimateplot} demonstrates that there is no trend with the atmospheric scale height and the strength of the spectral features indicated by the magnitude of the transit depth metric. Furthermore, the radius anomaly is thought to correlate with incident flux, with hotter planets having a more inflated radius \citep{Thorngren2018}. However, we do not find a trend with the radius anomaly and the strength of the spectral features seen with Spitzer (see Figure \ref{P1:fig:RadiusAnomaly}).

\subsubsection{Stellar variability}
\label{P1:sec:stellarVariabilitiy}

Contamination of the transmission spectrum from starspots, faculae, and flares generates brightness temperature differences between the disk-integrated spectra of the star and the region occulted by a transiting planet \citep{Desert2011d, Pont2008, Sing2011}. If a planet occults a star spot at a different temperature to the photospheric one, it can change the shape of the light curve by appearing as a change in the flux during transit. On the other hand, if the star spot is not occulted, then the disk-integrated spectrum of the star is  fainter or brighter, depending on the spot properties, which can cause the measured transit depth to be different from the nominal one. Stellar variability can occur when star spots rotate in and out of view of the integrated stellar disk, which depends on the rotation period of the star. 

To estimate the possible effect of stellar variability on our results, we aim to provide a quantitative estimate of how this would affect the sample as a whole by expanding the interpolated model grid. We do this by looking at a worst-case-scenario variable star, HD 189733, which has a peak-to-peak variability of $\sim 3$\% in the visible \citep{Henry2008}. We follow the method in \citet{Desert2011d, Sing2011} and \citet{Berta2012} to calculate the effect of this variability on the transit depth metric. We first translate this 3\% in V-band to 0.8\% at 3.6~$\mu$m using the ratio of black bodies; we set 2.8\% spot coverage with spots at 1000~K less than the stellar photosphere. We can propagate this to a relative error on the transit depth and assuming it will affect 4.5~$\mu$m as much as 3.6~$\mu$m (in reality it will be a smaller effect), we can then propagate this to an error on the transit depth metric ($\delta_{ch2} - \delta_{ch1} / \delta_{ch1}$). This leads to a 42\% maximum error on the transit depth metric, and we therefore extend the models positive and negative by this percentage, which is plotted in Figure \ref{P1:fig:ultimateplot}. 

Despite choosing the worst-case scenario to expand our model grid, the features arising from the changing chemistry with equilibrium temperature can still be clearly distinguished in the grid of models in Figure \ref{P1:fig:ultimateplot}, that is, the transition regions at $\sim1000$~K and $\sim 2200$~K remain clear. The relative size of the variability region is on average one-third of the size of the average uncertainty on the data points. Nevertheless, this is a conservative upper estimate and will not apply as strongly to all planets in our sample, because not all stars are as variable as HD 189733. If the temperature difference between spot and photosphere is less or if the spot covering fraction is smaller, then this would result in a lower variability amplitude and smaller effect on the transmission spectrum. Additionally, 3\% is the maximum peak-to-peak variability, and this would only apply if the observations were taken at the peak and trough of the variability period. As we designed the observations to have as few orbital periods as possible to be within one variability period of the star, it is unlikely that we reach the maximum variability between our two Spitzer observations.

\subsection{Comparing transmission and emission with warm Spitzer/IRAC}
\label{P1:sec:TvsE}

Several of the planets from our survey have published secondary eclipse measurements. We utilize the secondary eclipse literature survey from \citet{Baxter2020} (and references therein) which contains 3.6~$\mu$m and 4.5~$\mu$m eclipses for 78 planets in total. Several of the eclipse depths presented in Table 1 of \citet{Baxter2020} were taken from \citet{Garhart2020}, and some of these planets had dilution corrections due to companions in the field of view. The dilution corrections were applied before any analysis in \citet{Baxter2020}, but this was not reported in their Table 1. We therefore report the eclipse depths with dilution corrections in Table \ref{P1:tab:eclipses} of this work using the dilution correction factors presented in Table 4 of \citet{Garhart2020}. We use the eclipse depth, equilibrium temperature, brightness temperatures at 3.6~$\mu$m and 4.5~$\mu$m, and the deviation from the black body presented in \citet{Baxter2020}. The deviation from the black body probes the temperature pressure profile. A positive deviation is either methane in absorption at 3.6~$\mu$m with a nominal TP profile or CO in emission at 4.5~$\mu$m if the TP profile is inverted. Twenty-four of the planets in \citep{Baxter2020} are also in our transmission survey, which allows us to statistically compare the two samples. To better understand the dearth of methane planets presented in Section \ref{P1:sec:gridstats}, we compare the difference in brightness temperature from emission with the normalized difference in transit depths for planets with both emission and transmission observations. Additionally, we create a color--magnitude plot and compare the emission to the brown dwarf spectral sequence with a focus on the coolest planets.

\subsubsection{Probing different pressures with emission and transmission}
\begin{figure}
    \centering
    \includegraphics[width = \linewidth]{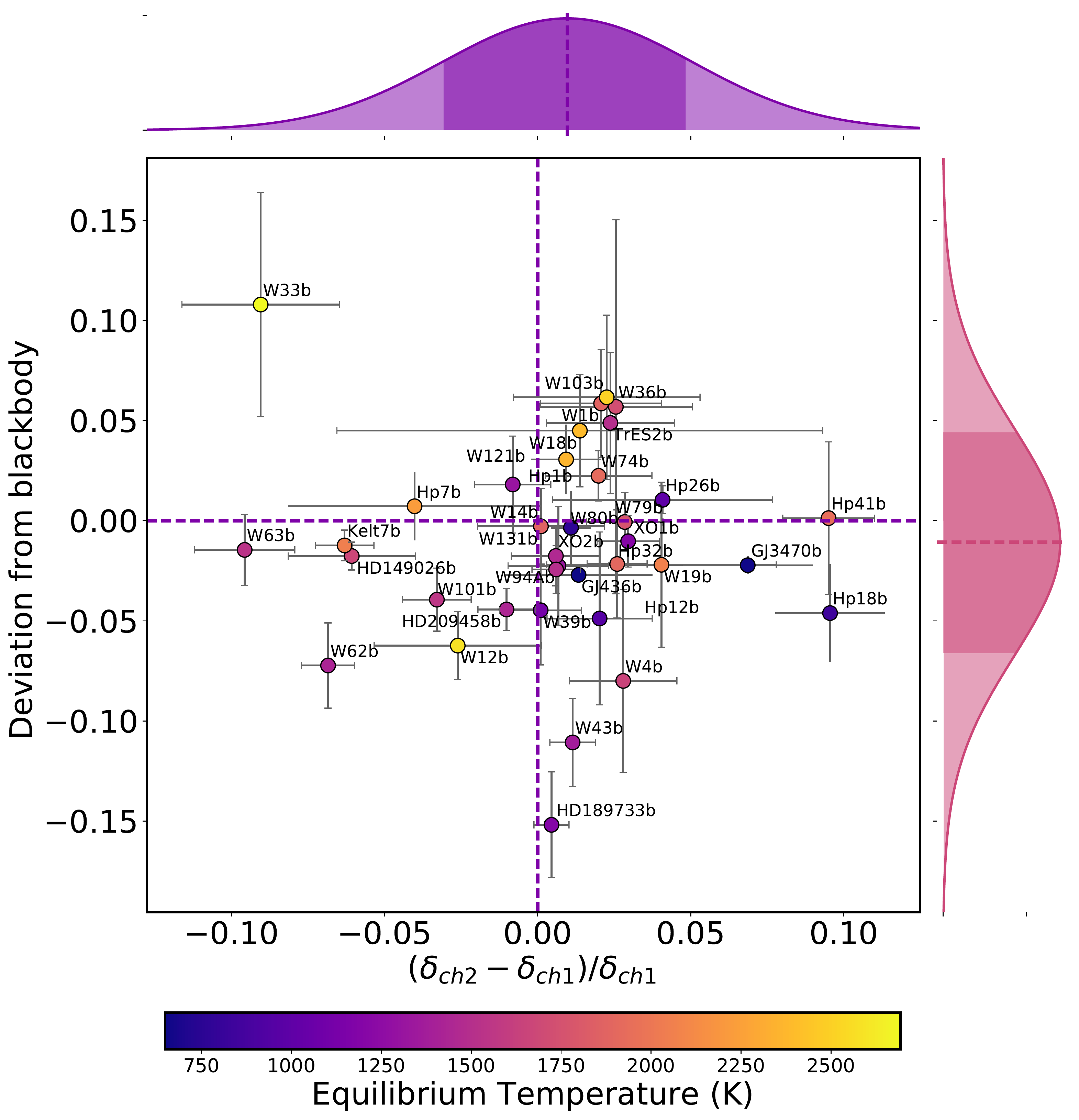}
    \caption{Deviation from a black body calculated from emission against the normalized difference in the transit depth presented in Figure \ref{P1:fig:ultimateplot}. Histograms on each axis show the mean and standard deviation of each axis. The equilibrium temperature of each planet is shown with the color scale.}
    \label{P1:fig:TvsE}
\end{figure}

In \citet{Baxter2020}, we demonstrate that the relative opacities in the two Spitzer/IRAC bandpasses can act as a probe of the atmospheric temperature structure when observing the dayside emission. The deviation from black body metric described in \citet{Baxter2020} plotted against the equilibrium temperature shows that ultra-hot Jupiters show statistical evidence for thermal inversion. In a noninverted atmosphere, a positive deviation indicates that the 3.6~$\mu$m brightness temperature is lower than that at 4.5~$\mu$m due to methane absorption at 3.6~$\mu$m and a negative deviation indicates that the 4.5~$\mu$m $T_b$ is lower due to CO absorption at 4.5~$\mu$m. On the other hand, if the atmosphere is inverted, CO being the dominating carbon-bearing species in the atmosphere would result in a positive deviation due to seeing CO in emission at 4.5~$\mu$m. In this work we are focusing on the \ce{CH4}-to-\ce{CO} transition temperature and thus we do not account for temperature inversions. 

In Figure \ref{P1:fig:TvsE} we compare the difference of the two IRAC brightness temperatures against the normalized difference in the transit depth. There appear to be no trends in the emission and transmission of the planets in our survey with both eclipses and transits, and we also find that the top left quadrant is almost empty, with the main outlier being WASP-33b. Given that the deviation from the black body can be positive or negative for a CO-dominated atmosphere depending on the TP profile, we test for trends in the planets with equilibrium temperatures below 1800~K, which are not expected to have thermal inversions in their atmospheres. We find that the top-left quadrant, which indicates \ce{CH4} in transmission and \ce{CH4} in emission, is empty, meaning that any planets that show signs of methane in their emission  do not show it in their transmission, and vice versa. For example, HD 149026b lies in the bottom-left quadrant, and has a negative deviation from a black body, which indicates a CO absorption feature in emission (assuming a noninverted TP profile). The expected corresponding transmission spectrum would predict a positive transit depth difference. However, this is not what we see. Possible reasons for these differences include: longitudinal abundance differences, more complex atmospheric processes such as atmospheric mixing, different cloud composition/abundances at different layers in the atmosphere, or changes in the thermal structure between emission and transmission \citep[e.g.,][]{Fortney2005}. Additionally, similar to the results for the planets in transmission, we do not find a correlation between the deviation of the black body and the radius anomaly.

\subsubsection{Comparing to brown dwarfs with a color-magnitude diagram} 

We create also a color-magnitude plot using these Spitzer secondary eclipses. Our work expands on that presented in \citet{Triaud2014b} by extending their survey from 37 planets to the 78 planets presented in \citet{Baxter2020} and by using the newly released GAIA dr2 for more accurate distances \citep{GaiaCollaborationandBrown2018}. We calculate the planetary apparent magnitudes using the apparent stellar magnitudes from the WISE spacecraft \citep{Cutri2012} in combination with the planet-to-star-flux ratio from Spitzer. The two WISE channels W1 and W2 are known to overlap the two remaining Spitzer channels \citep{Kirkpatrick2011}. We then use the GAIA dr2 distances which were calculated using a Bayesian prior from \citet{GaiaCollaborationandBrown2018} and \citet{Bailer-Jones2018} to calculate the planetary absolute magnitudes. The equilibrium temperature, GAIA distances, and WISE magnitudes used are provided in Table \ref{P1:tab:eclipses}. Errors are propagated fully throughout the calculation from the errors on all input properties.

\begin{figure}
    \centering
    \includegraphics[width = \linewidth]{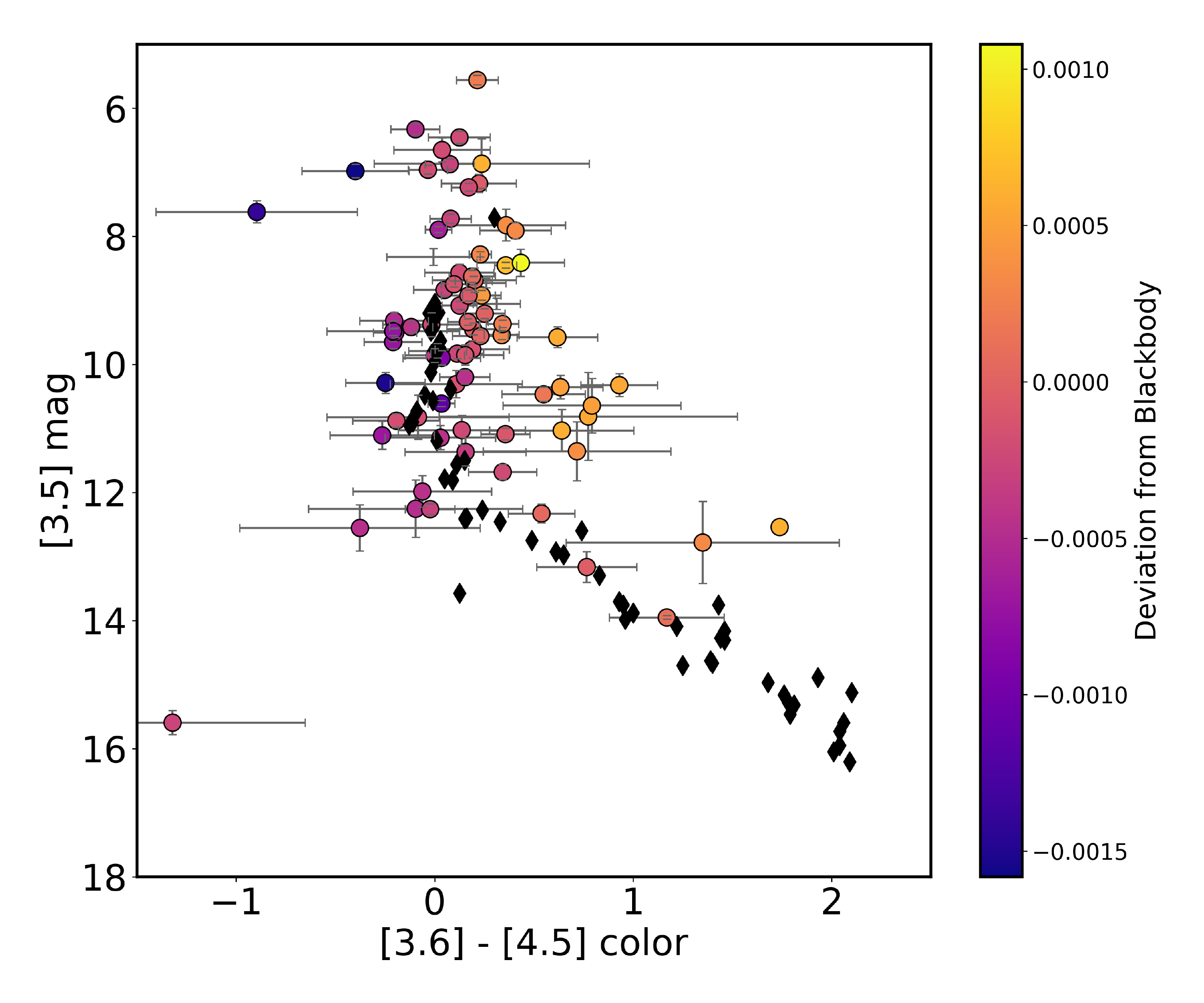}
    \caption{[3.6] - [4.5] color vs. [3.6] magnitude diagram of exoplanets and brown dwarf planets with available eclipse depth measurements in the two warm Spitzer band-passes. Brown dwarf colors (estimated from the WISE catalog) are shown as black diamonds from \citet{Dupuy2012}. The color scale is the deviation from a black body metric described \citet{Baxter2020}.}
    \label{P1:fig:colourmag}
\end{figure}

Figure \ref{P1:fig:colourmag} shows the [3.6] - [4.5] color versus [3.6] magnitude diagram. We have over-plotted the survey of brown dwarfs spanning M, L, and T spectral classes from \citet{Dupuy2012} for comparison. The planets plotted in Figure \ref{P1:fig:colourmag} show an increasing scatter with increasing 3.6~$\mu$m magnitude. This is unlike the brown dwarf spectral sequence which follows a very tight L/T transition. This increase in scatter confirms that seen in \citet{Triaud2014a, Beatty2014, Melville2020} and \citet{Dransfield2020}, which is suggested to be due to an increase in atmospheric diversity.

Figure \ref{P1:fig:colourmag} shows that the increase in scatter is driven by a small family of planets which lie redder than the brown dwarf spectral sequence. The color scale shows that this family of planets shows a positive deviation from a black body \citep{Baxter2020}, which indicates methane absorption (nominal TP profile) or CO emission (inverted TP profile). As these planets are around 1200-1500~K, we do not expect their atmospheres to be inverted, and we therefore think that these warmer planets could be displaying a signature of methane in their atmospheres. In Section \ref{P1:subsec:TDvsTeq} we show that the cooler planets (<1000~K) deviate from equilibrium chemistry models by not showing signatures of methane in their atmospheres. Similarly, these warmer planets are not expected to have a high methane abundance given their equilibrium chemistry, and again we have to invoke nonequilibrium processes such as vertical mixing. Warmer planets are expected to have greater vertical mixing than their cooler counterparts \citep{Komacek2019}, creating an ideal scenario for methane dredge-up. 

Furthermore, brown dwarfs are expected to undergo less mixing than gas giant planets, with $K_{zz}$ ranging from $10^4$ to $10^{8.5}$ for brown dwarfs and $10^7$ to $10^{12}$ for gas giants \citep{Zahnle2014, Leggett2017, Miles2020}. Although this prediction is based on nonirradiated, higher gravity objects with mostly convective atmospheres, GCMs of highly irradiated, radiative atmospheres of hot Jupiters do display stronger mixing \citep{Parmentier2013, Komacek2019}. We therefore propose that the increased atmospheric diversity of planets compared to their brown dwarf counterparts seen in Figure \ref{P1:fig:colourmag} could be due in part to the diversity of the processes involved, such as the presence of vertical mixing.

\section{Conclusion}
\label{P1:sec:conclusion}

We performed the data analysis of 70 light curves and presented a total of 49 planets with transit depths at 3.6 and 4.5~$\mu$m with Spitzer/IRAC. This survey represents the largest analysis of Spitzer/IRAC observations of gas giant transits to date, and spans equilibrium temperatures from 500~K to 2700~K. We implemented our custom Spitzer/IRAC data analysis pipeline which thoroughly searches over a grid of data-reduction parameters before employing pixel-level decorrelation \citep{Deming2015} to correct for the strong Spitzer systematic errors and extract the transit depths using an MCMC transit-fitting algorithm. 

We then statistically studied the sample of all planets with transmission in these two bandpasses. We create a fiducial cloud-free 1D atmospheric model grid with 1x solar composition and equilibrium chemistry spanning the parameters of the planets in our sample. We compare the survey of planets with this model grid and note a family of outliers with equilibrium temperature <1000~K; they do not show the expected methane abundance from these equilibrium chemistry models. 

Next, we expanded our grid in two dimensions by extending to 30x solar metallicity and incorporating nonequilibrium effects with different values of an eddy diffusion co-efficient ($K_{zz}$). We find that the best-fitting grid for the cool planets (T<1000~K) has high metallicity (30x solar) and low or no vertical mixing ($K_{zz}=0$ or $10^8$~\cmcms). On the other hand, we find that the hot planets (T>1000~K) are best explained with 1x solar composition with a marginally better fit with the high vertical mixing model ($K_{zz}=10^{12}$~\cmcms). We conclude that the cool planets are better fit by models with higher metallicity due to an observational bias resulting in lower masses. We find evidence supporting nonequilibrium chemistry in a survey of planets and find that our work agrees with the theory that hotter planets have higher vertical mixing. 

Furthermore, we combine our transits with our previous literature eclipse survey. We do not find any trend between eclipses and transits, and propose that this is due to several effects: clouds at different pressures or more complex atmospheric processes. We created a color--magnitude diagram using the emission observations and compared to L/T transition brown dwarfs. With a larger sample size than previous studies, we also see the increase in scatter with increasing magnitude first seen byn \citet{Triaud2014b}. We see that the increase in scatter is driven by a family of mid-temperate planets showing a methane signature that is not expected from equilibrium chemistry. We propose that this increase in scatter is due to methane being dredged up due to high levels of vertical mixing in the atmosphere, which supports the theory that brown dwarfs have approximately 100 times lower levels of vertical mixing than planets.

\begin{acknowledgements}
This work is based on observations made with the Spitzer Space Telescope, which is operated by the Jet Propulsion Laboratory, California Institute of Technology under a contract with NASA. Support for this work was provided by NASA through an award issued by JPL/Caltech. J.M.D acknowledges support from NASA grant NNX16AC64G, the Amsterdam Academic Alliance (AAA) Program, and the European Research Council (ERC) European Union’s Horizon 2020 research and innovation program (grant agreement no. 679633; Exo-Atmos). S-M. Tsai thanks Michael Zhang for supporting the free-chemistry version of PLATON. S-M. Tsai acknowledges support from the European community through the ERC advanced grant EXOCONDENSE (PI: R.T. Pierrehumbert). This work has made use of the MUSCLES Treasury Survey High-Level Science Products; doi:10.17909/T9DG6F. The analysis done in this project made use of the SciPy stack \citep{Jones2001}, including NumPy \citep{Oliphant2006}, Matplotlib \citep{Hunter2007}, Pandas \citep{reback2020pandas}. It also made use of Astropy \citep{AstropyCollaborationandPrice-Whelan2018}, pysynphot \citep{STScIDevelopmentTeam2013}, sklearn \citep{scikit-learn}, photutils \citep{larry_bradley_2020_4044744}, batman \citep{Kreidberg2015}, emcee \citep{Foreman-Mackey2013} and corner \citep{Foreman-Mackey2016}. Models in this paper made use of the VULCAN code which is available at \url{https://github.com/exoclime/VULCAN}. 
\end{acknowledgements}


\bibliographystyle{aa.bst}
\bibliography{bib_out}

\begin{appendix}

\section{Pipeline results Figures and Tables} 
\label{P1:app:piperes}

\begin{table*}
 \caption{\label{P1:tab:jumpParams} Jump Parameters used as starting points for the MCMC analysis. Eccentricity was fixed to zero for all planets because it did not affect the resulting transit depths. Stellar parameters (Teff, logg and [Fe/H]) are used to calculate linear limb-darkening parameters. }
 \setlength{\tabcolsep}{4pt}
\begin{tabular}{llllllllll}
\hline \hline
Planet & $a/R_s$ & inc & $R_p/R_s$ & Period &  Eccentricity &        $T_{\rm eff}$ & $log(g_*)$ & [Fe/H] & Ref  \\
 &  & $^{\circ}$ &  & days &   & Kelvin & $\log_{10}(cm/s^2)$ & dex &          \\
\hline
HAT-P-32 b    &   6.05(4) &     88.9(4) &       0.1508(4) &         2.150008(1) &       0.16(6) &   6207(88) &      4.33(1) &  -0.04(8) &                                13          \\
XO-1 b       &  11.24(9) &    88.8(2) &       0.1320(5) &         3.94150685(91) &    0 &         5750(75) &      4.50(1) &   0.02(8) &                                  6 \\
HAT-P-1 b     &   9.85(7) &    85.63(06) &     0.1180(2) &         4.46529976(55) &    0 &         5980(49) &      4.36(1) &   0.13(1) &                                19       \\
WASP-17 b    &   7.05(7) &    86.83(68) &     0.13 &              3.7354380(68) &     0.03(2) &   6650(80) &      4.16(3) &  -0.19(9) &                               1        \\     
WASP-39 b    &  11.37(24) &   87.75(27) &     0.1457(16) &        4.0552765(35) &     0 &         5400(150) &     4.4(2) &   -0.12(10) &                            18,10      \\ 
HAT-P-12 b    &  11.77(21) &   89.0(4) &       0.1406(13) &        3.2130598(21) &     0.03(3) &   4650(60) &      4.61(1) &  -0.29(5) &                               11      \\  
HAT-P-18 b    &  16.04(75) &   88.8(3) &       0.1365(15) &        5.508023(6) &       0.08(5) &   4803(80) &      4.57(4) &    0.10(08) &                                14    \\         
TrES2 b     &    7.90(2) &   83.87(02) &     0.1254(5) &         2.47061317(9) &     0.02(2) &   5850(50) &      4.43(2) &   -0.15(10) &                                9, 22   \\       
WASP-4 b     &   5.46(2) &    88.52(39) &     0.1544(2) &         1.33823204(16) &    0.003(7) &  5436(34) &      4.46(5) &  -0.05(4) &                                  15,20    \\      
XO-2 b       &   8.18(3) &    88.9(7) &       0.1 &               2.615857(50) &      0 &         5340(32) &      4.48(5) &   0.45(2) &                                         5  \\
GJ3470 b    &  13.94(49) &   88.88(72) &     0.0764(4) &         3.3366487(43) &     0.02(2) &   3652(50) &      4.78(12) &   0.17(6) &                                 3        \\  
WASP-21 b    &   9.62(17) &   87.12(24) &     0.1030(8) &         4.3225126(22) &     0 &         5800(100) &     4.2(1) &  -0.46(11) &                               21,4        \\  
WASP-31 b    &    8.00(19) &  84.41(22) &     0.127 &             3.4059096(50) &     0 &         6302(102) &     4.31(2) &   -0.20(09) &                               2       \\
WASP-1 b     &   5.69(6) &    90.0(1.3) &     0.1036(8) &         2.5199454(5) &      0.01(3) &   6200(200) &     4.3(3) &     0.1(2) &                            17,7       \\               
HAT-P-26 b    &  13.44(83) &   88.6(9) &       0.0737(12) &        4.234516(15) &      0.12(6) &   5079(88) &      4.56(6) &  -0.04(8) &                               12     \\   
WASP-107 b   &    18.2(1) &   89.56(08) &     0.1446(2) &         5.72149242(46) &    0 &         4430(120) &     4.5(1) &    0.02(10) &                                 23 \\
WASP-13 b    &   7.58(15) &   85.64(24) &     0.0922(8) &         4.353011(13) &      0 &         5826(100) &     4.04(20) &     0.0(2) &                                 24,25    \\         
WASP-121 b   &   3.75(3) &    87.6(6) &       0.1245(5) &         1.27492550(25) &    0 &         6459(140) &     4.24(1) &   0.13(9) &                                 26  \\        
WASP-69 b    &  11.96(17) &   86.71(20) &     0.1336(16) &        3.8681382(17) &     0 &         4700(50) &      4.54(2) &   0.15(8) &                                35    \\   
WASP-67 b    &  13.42(13) &   85.8(3) &       0.1345(48) &        4.61442(1) &        0 &         5417(85) &      4.53(2) &   0.18(6) &                                36       \\
HATS7 b     &  10.59(51) &   87.92(75) &     0.0711(19) &        3.185315(5) &       0 &         4985(50) &      4.54(5) &   0.25(8) &                               37  \\
WASP-29 b    &  12.15(44) &   88.8(7) &       0.101(2) &          3.922719(7) &       0.03(5) &   4875(65) &      4.54(4) &   0.11(14) &                                38   \\       
HAT-P-41 b    &   5.45(18) &   87.7(1.0) &     0.1028(16) &        2.694050(4) &       0 &         6390(100) &     4.14(2) &    0.21(10) &                                32\\
WASP-101 b   &    8.45(30) &  85.0(2) &       0.1140(9) &         3.585720(4) &       0 &         6380(120) &     4.31(8) &    0.20(12) &                                32\\
WASP-131 b   &   8.53(9) &    85.0(3) &       0.0815(7) &         5.322023(5) &       0 &         6030(90) &      4.09(3) &  -0.18(8) &                                39       \\
WASP-36 b    &   5.85(6) &    83.15(13) &     0.1368(6) &         1.53736596(24) &    0 &         5959(134) &     4.49(1) &   -0.26(10) &                                40     \\        
WASP-63 b    &    6.59(30) &  87.8(1.3) &     0.0781(11) &        4.378080(6) &       0 &         5550(100) &     4.01(3) &   0.08(7) &                                32\\
WASP-79 b    &   7.03(36) &   85.4(6) &       0.1049(24) &        3.662380(5) &       0 &         6600(100) &     4.20(15) &    0.03(1)&                                32\\
WASP-94 Ab   &    7.3(7) &    88.7(7) &       0.1094(8) &         3.9501907(44) &     0 &         6153(75) &      4.18(1) &   0.26(15) &                          41          \\               
WASP-74 b    &    4.86(20) &  79.81(24) &     0.0964(7) &         2.137750(1) &       0 &         5990(110) &     4.39(7) &    0.03(10) &                                33 \\
WASP-62 b    &   9.55(41) &   88.30(75) &     0.1095(9) &         4.411950(3) &       0 &         6230(80) &      4.45(10) &   0.04(6) &                                34 \\
Kepler-45 b  &    10.6(1.0) & 87.0(7) &       0.179(2) &          2.455239(4) &       0.11(10) &  3820(90) &      3.1(1) &   0.13(13) &                                16      \\ 
\hline
\end{tabular}
\tablebib{
(1)~\citet{Anderson2011};
(2) \citet{Anderson2011b}; (3) \citet{Biddle2014}; (4) \citet{Bouchy2010};
(5) \citet{Burke2007}; (6) \citet{Burke2010}; (7) \citet{CollierCameron2007};
(8) \citet{Doyle2011}; (9) \citet{Esteves2015}; (10) \citet{Faedi2011};
(11) \citet{Hartman2009}; (12) \citet{Hartman2011b}; (13) \citet{Hartman2011};
(14) \citet{Hartman2011c}; (15) \citet{Hoyer2013}; (16) \citet{Johnson2012};
(17) \citet{Maciejewski2014}; (18) \citet{Maciejewski2016}; (19) \citet{Nikolov2014};
(20) \citet{Petrucci2013}; (21) \citet{Seeliger2015}; (22) \citet{Sozzetti2007};
(23) \citet{Mocnik2017}; (24) \citet{Barros2012}; (25) \citet{Skillen2009};
(26) \citet{Delrez2016}; (27) \citet{Holman2010}; (28) \citet{Torres2011};
(32) \citet{Stassun2017}; (33) \citet{Stassun2018}; (34) \citet{Stassun2019};
(35) \citet{Anderson2014}; (36) \citet{Hellier2012}; (37) \citet{Bakos2015}; 
(38) \citet{Hellier2010}; (39) \citet{Hellier2017}; (40) \citet{Mancini2016};
(41) \citet{Neveu-VanMalle2014}
}
\end{table*}

\begin{figure*}[!b]
  \includegraphics[trim={0 2cm 0 0},clip,width=\textwidth]{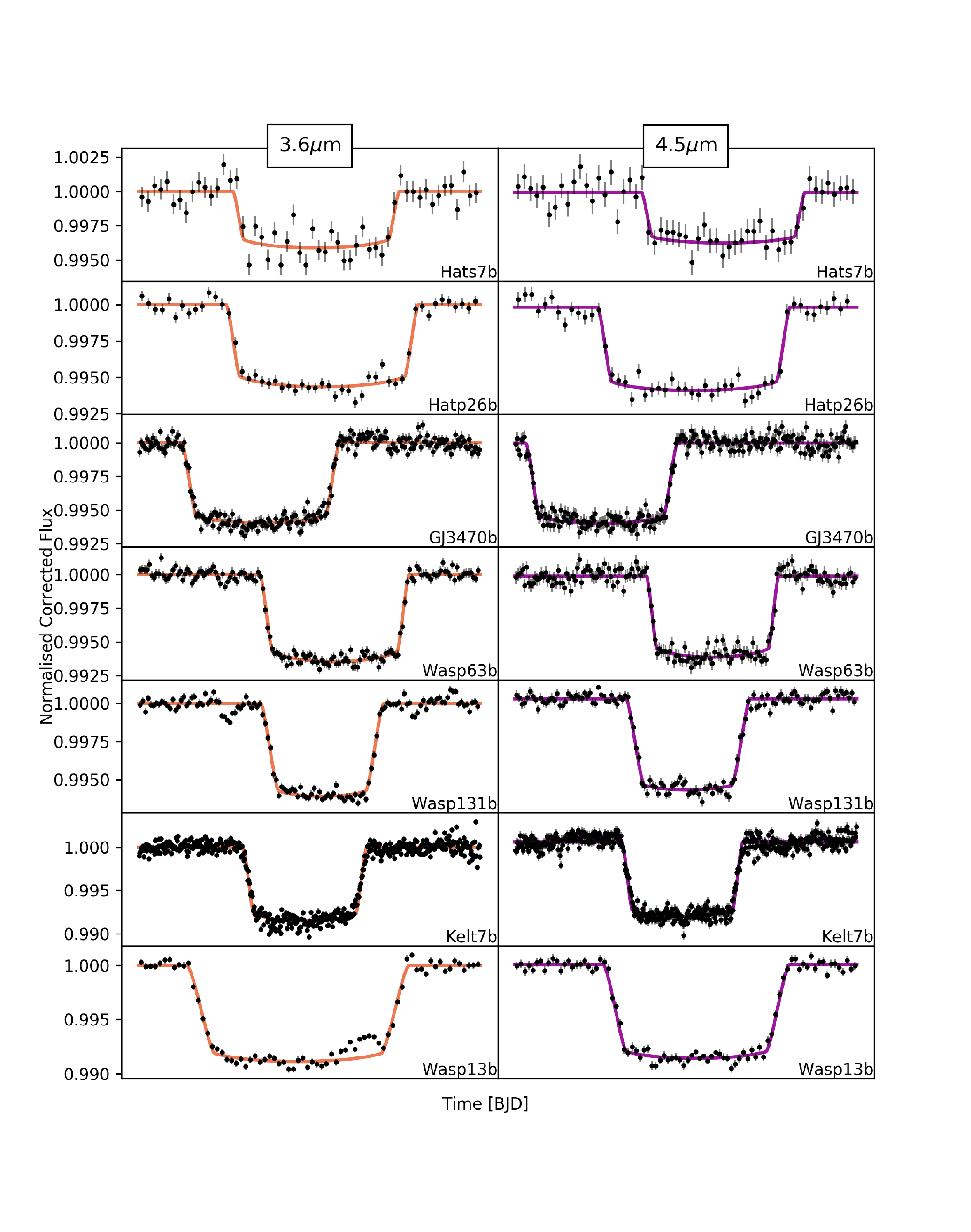}
  \caption{Normalized and systematic-error-corrected transit light curves for each planet  at 3.6 (left column, orange) and 4.5~$\mu$m (right column, purple). 1 $\sigma$error bars are those originally calculated from scaled photon noise. The data and the error bars are binned in five-minute intervals for display purposes. Continuous curves show the best-fit transit models in each band-pass for comparison. Kepler-45b displays the result of four phase folded light curves in each channel.}
  \label{P1:fig:normlc}
\end{figure*}

\addtocounter{figure}{-1}
\begin{figure*}
  \includegraphics[trim={0 2cm 0 0},clip,width=\textwidth]{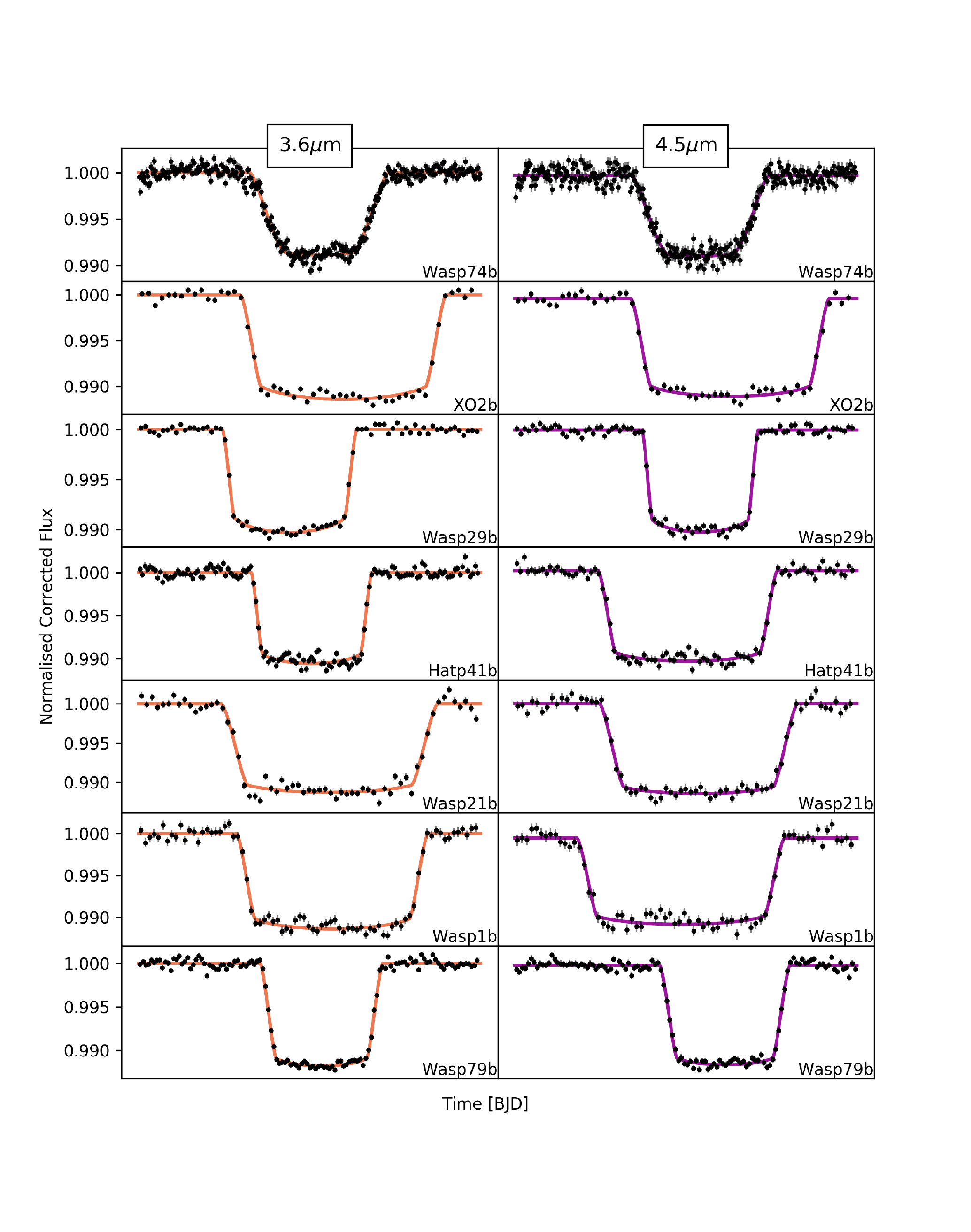}
  \caption{}
  \label{P1:fig:normlc1}
\end{figure*}

\addtocounter{figure}{-1}
\begin{figure*}
  \includegraphics[trim={0 1cm 0 0},clip,width=\textwidth]{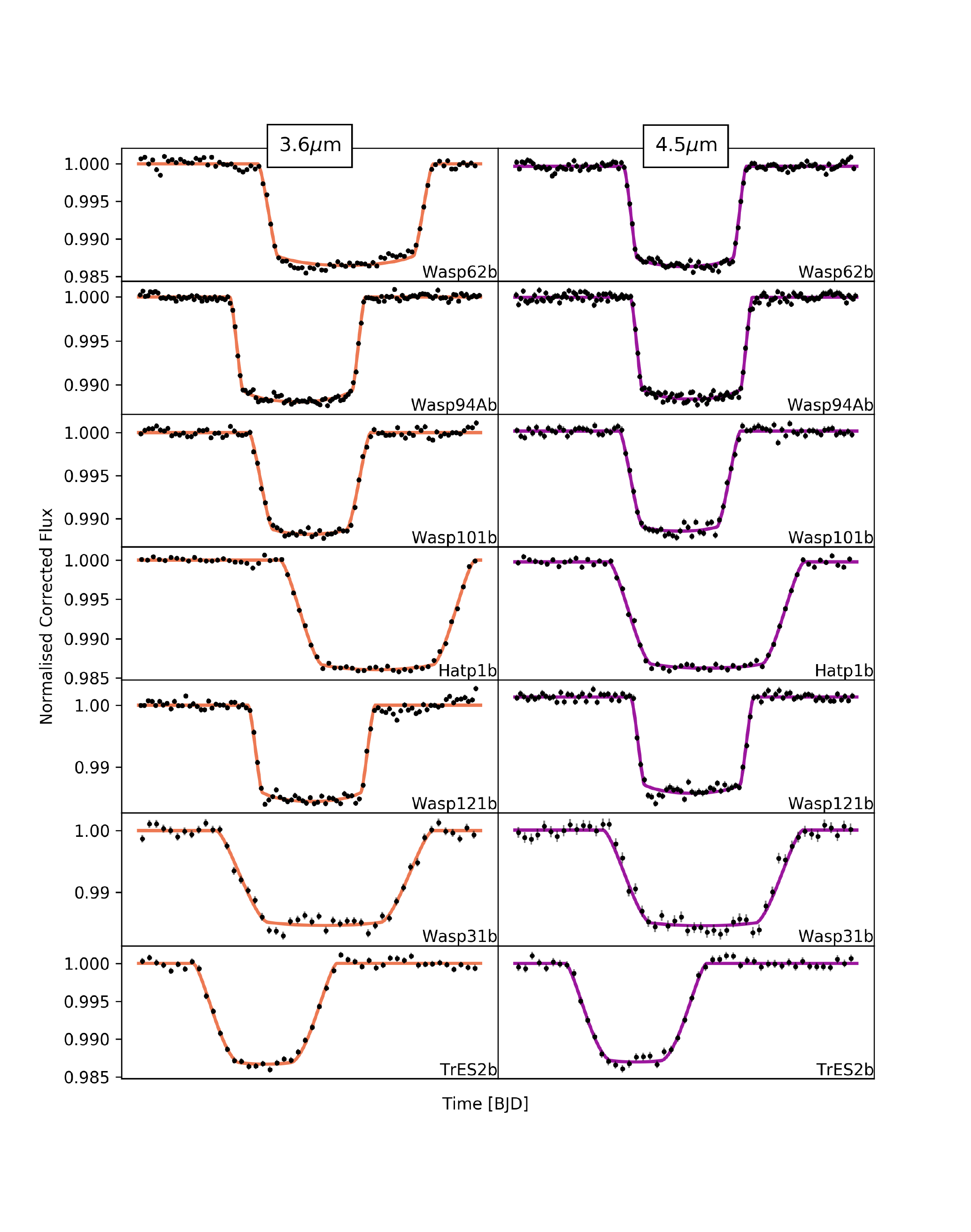}
  \caption{\textit{Continued.}}
  \label{P1:fig:normlc2}
\end{figure*}

\addtocounter{figure}{-1}
\begin{figure*}
  \includegraphics[trim={0 1cm 0 0},clip,width=\textwidth]{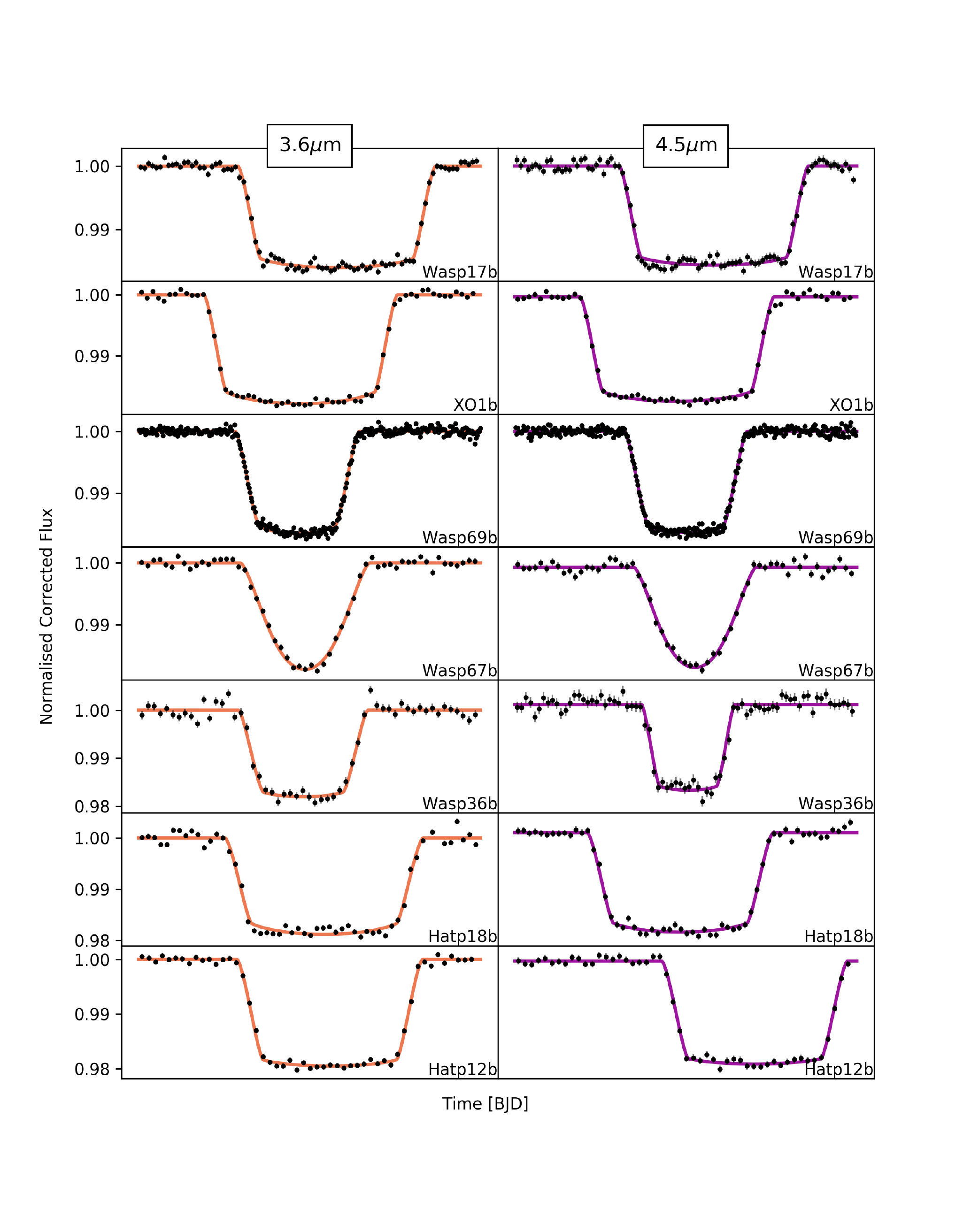}
  \caption{\textit{Continued.}}
  \label{P1:fig:normlc3}
\end{figure*}

\addtocounter{figure}{-1}
\begin{figure*}
  \includegraphics[trim={0 1cm 0 0},clip,width=\textwidth]{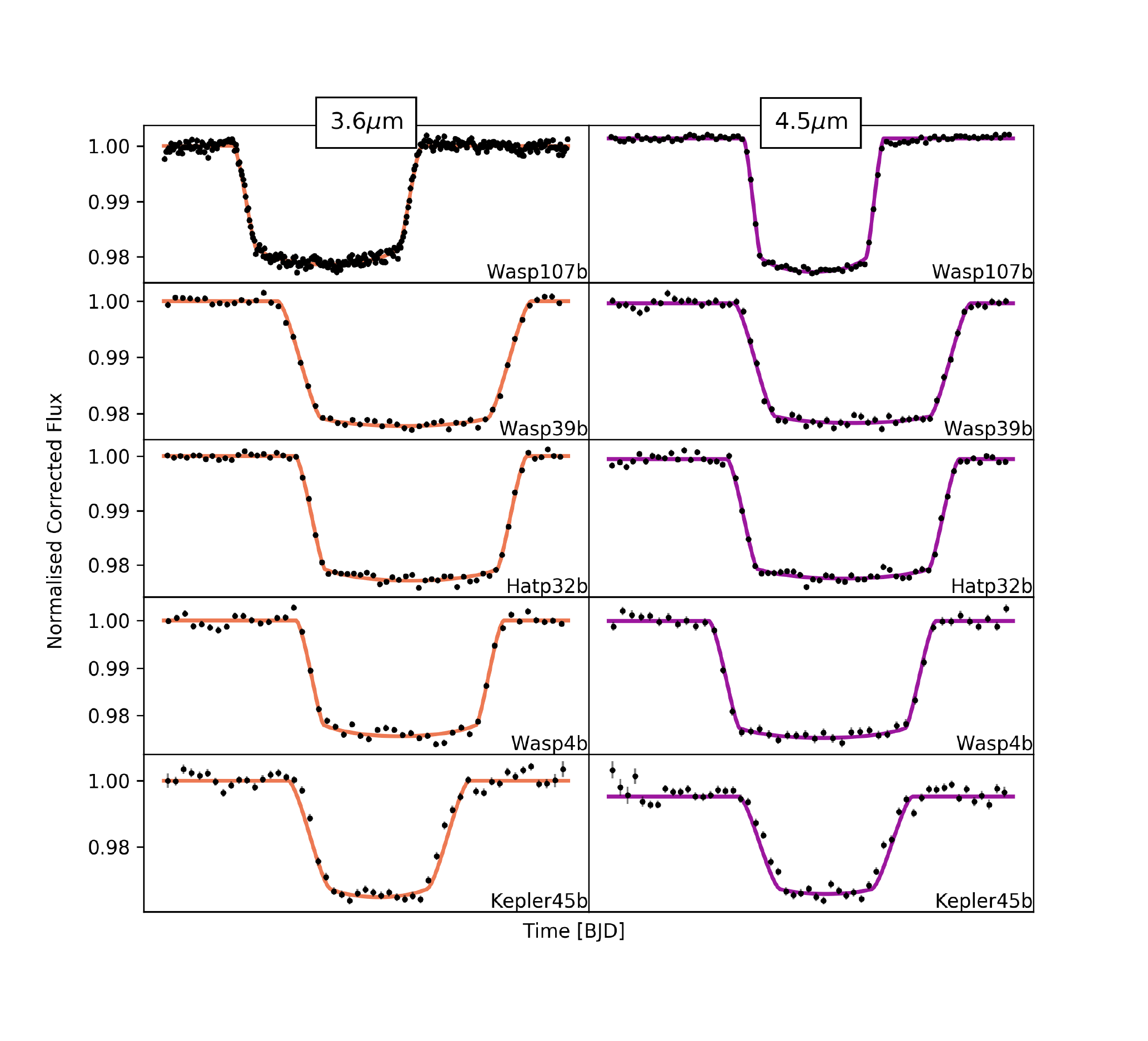}
  \caption{\textit{Continued.}}
  \label{P1:fig:normlc4}
  \end{figure*}


\begin{figure*}
  \label{P1:fig:rawlc}
  \includegraphics[width=\textwidth]{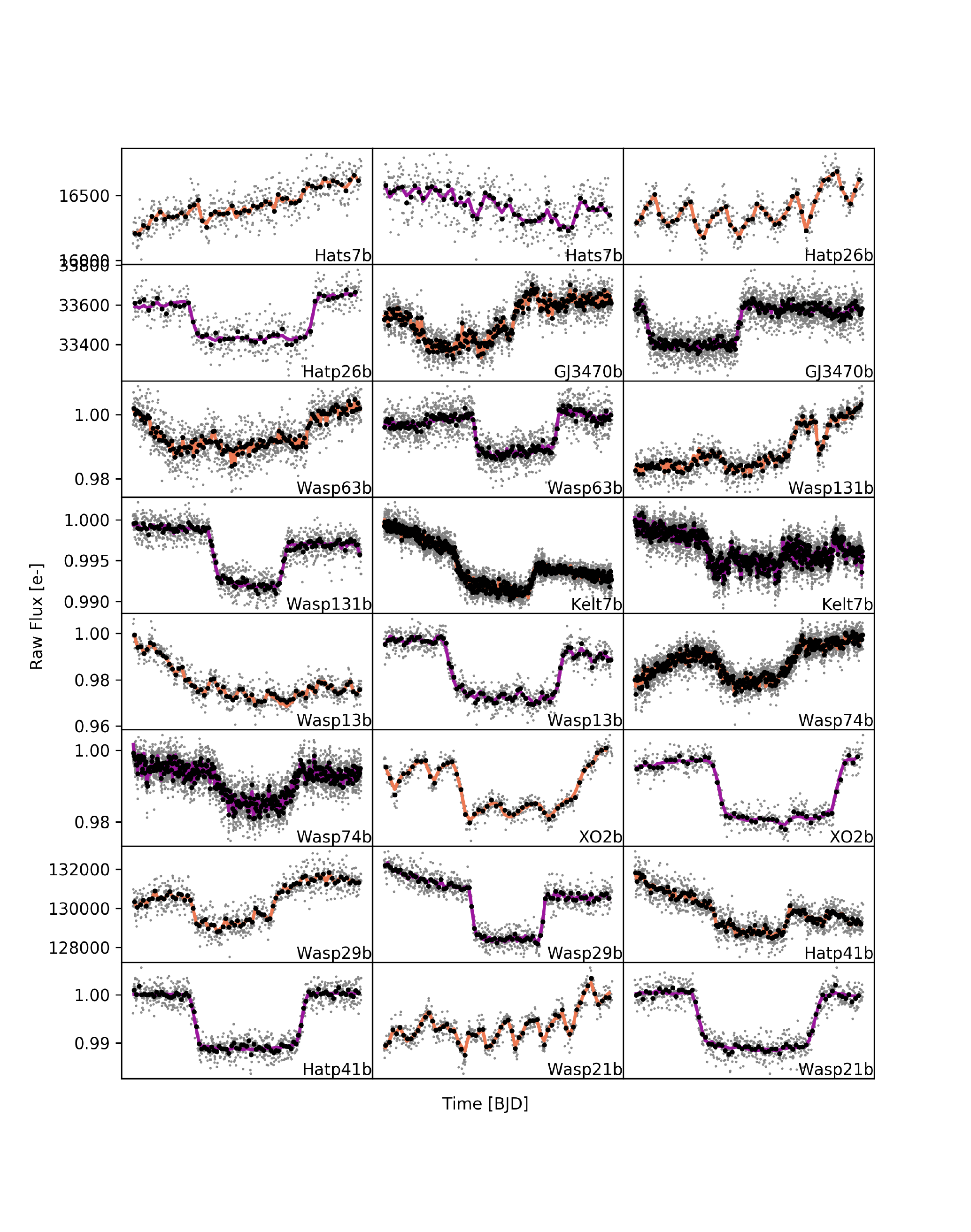}
  \caption{Raw light curves for each planet. Flux binned in 5 minutes is shown in black and 30 seconds is shown in gray. Colored lines indicate the best-fit instrumental and transit model from our MCMC analysis.}
\end{figure*}

\addtocounter{figure}{-1}
\begin{figure*}
    \label{P1:fig:rawlc1}
  \includegraphics[width=\textwidth]{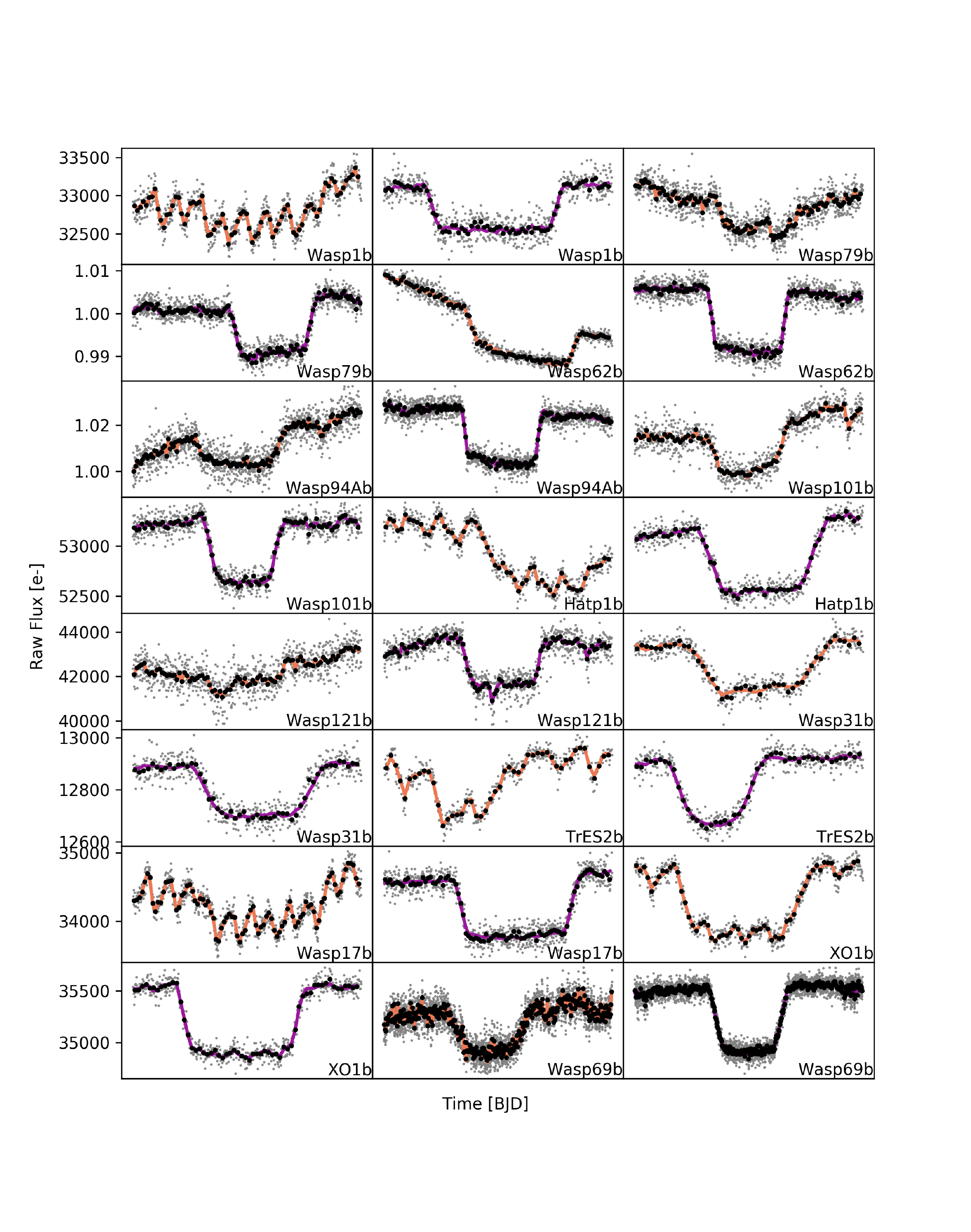}
  \caption{\textit{Continued.}}
\end{figure*}

\addtocounter{figure}{-1}
\begin{figure*}
    \label{P1:fig:rawlc2}
  \includegraphics[width=\textwidth]{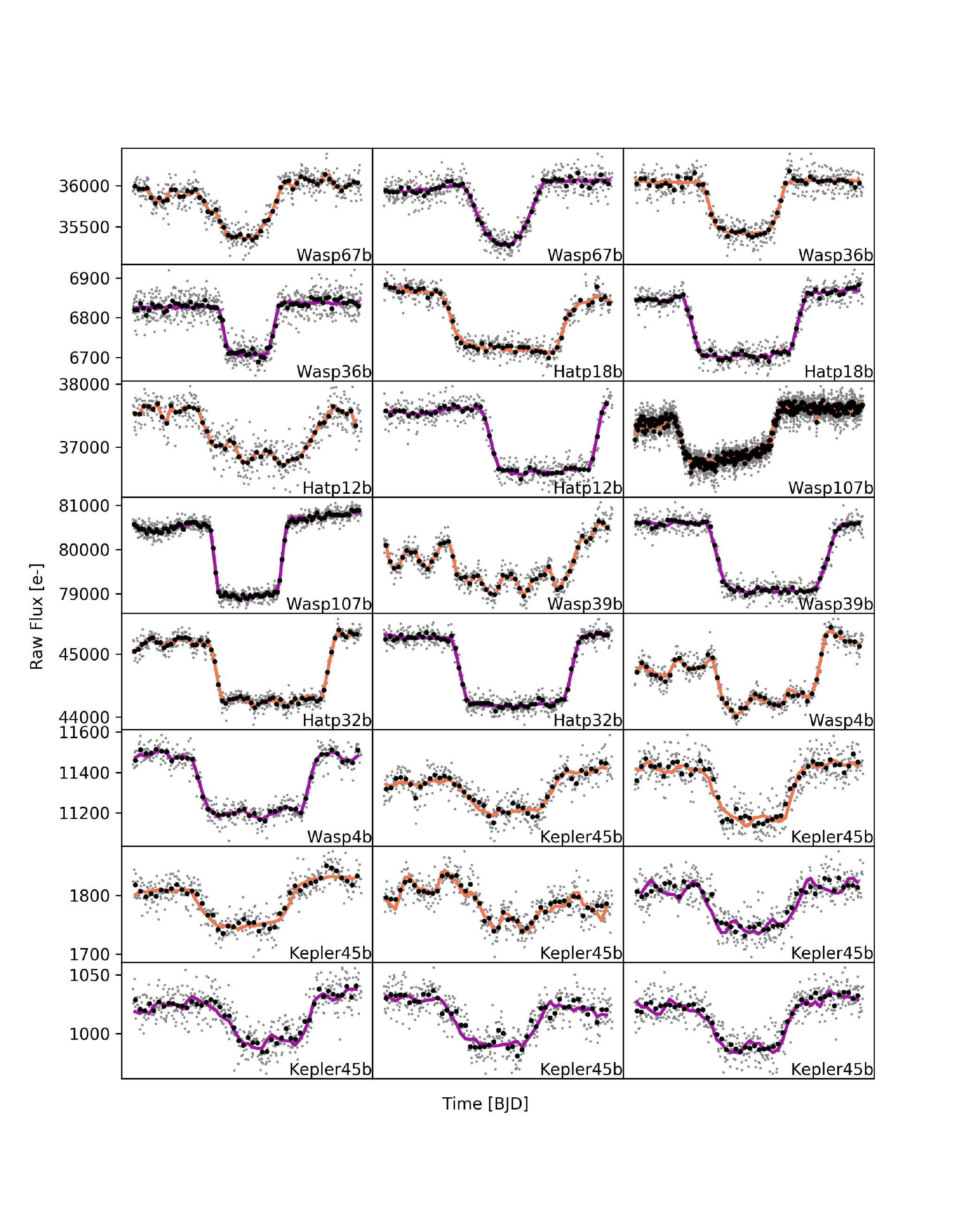}
  \caption{\textit{Continued.}}
\end{figure*}


\begin{figure*}
  \label{P1:fig:rmsvsbin}
  \includegraphics[width=\textwidth]{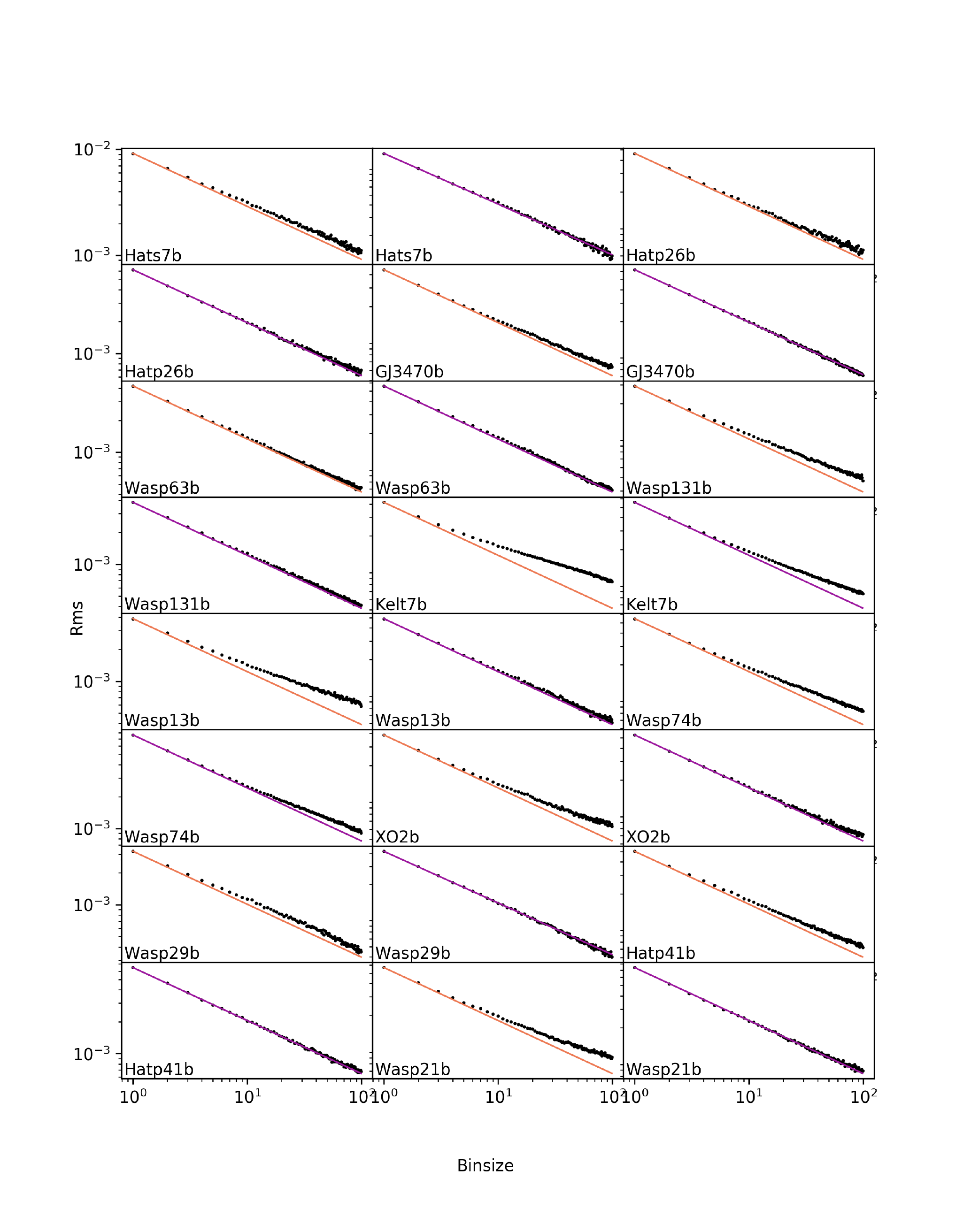}
  \caption{RMS vs. normalized bin size of each of the fitted light curves. Straight line is the sqrt(N) theoretical value.}
\end{figure*}

\addtocounter{figure}{-1}
\begin{figure*}
    \label{P1:fig:rmsvsbin1}
  \includegraphics[width=\textwidth]{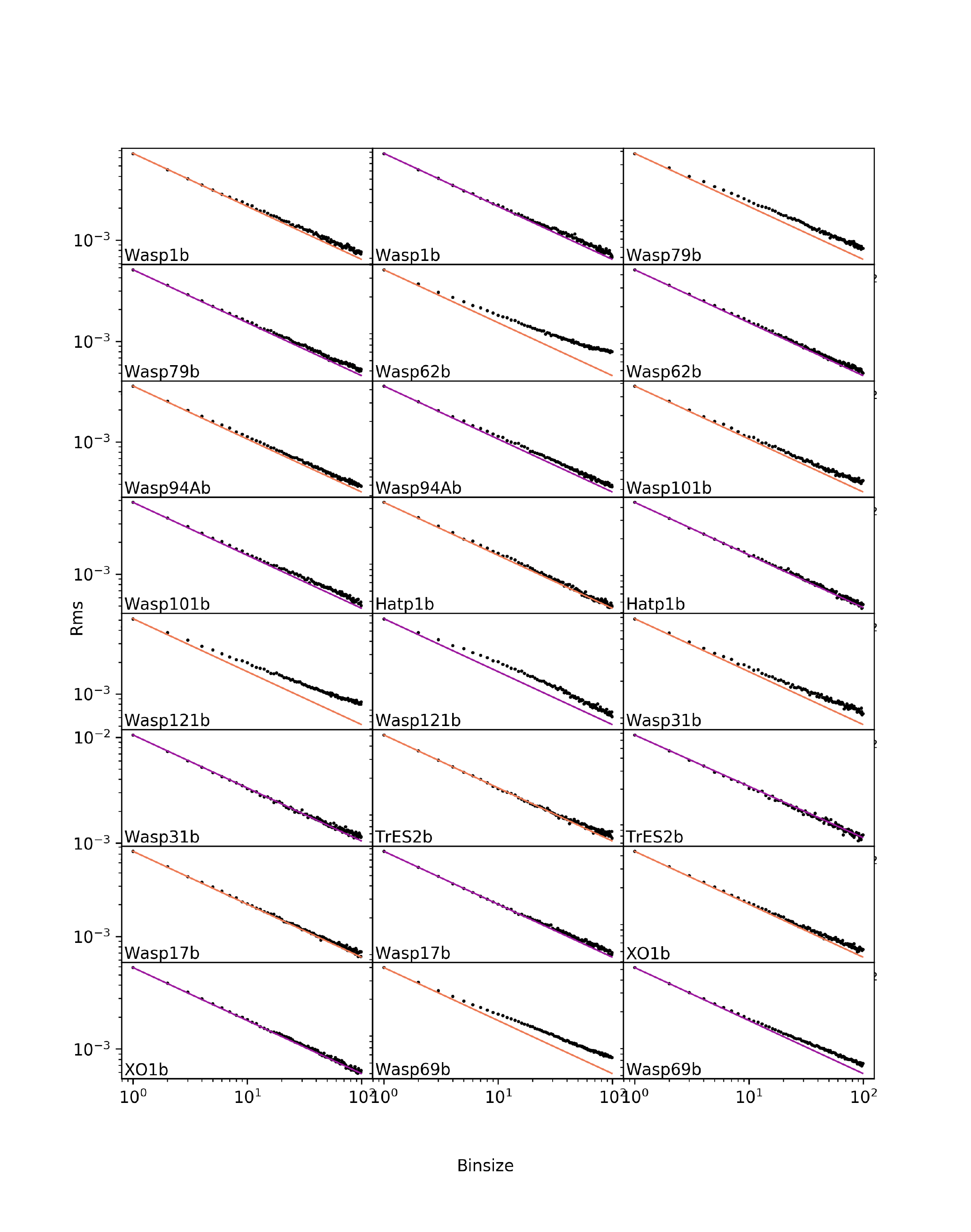}
  \caption{\textit{Continued.}}
\end{figure*}

\addtocounter{figure}{-1}
\begin{figure*}
    \label{P1:fig:rmsvsbin2}
  \includegraphics[width=\textwidth]{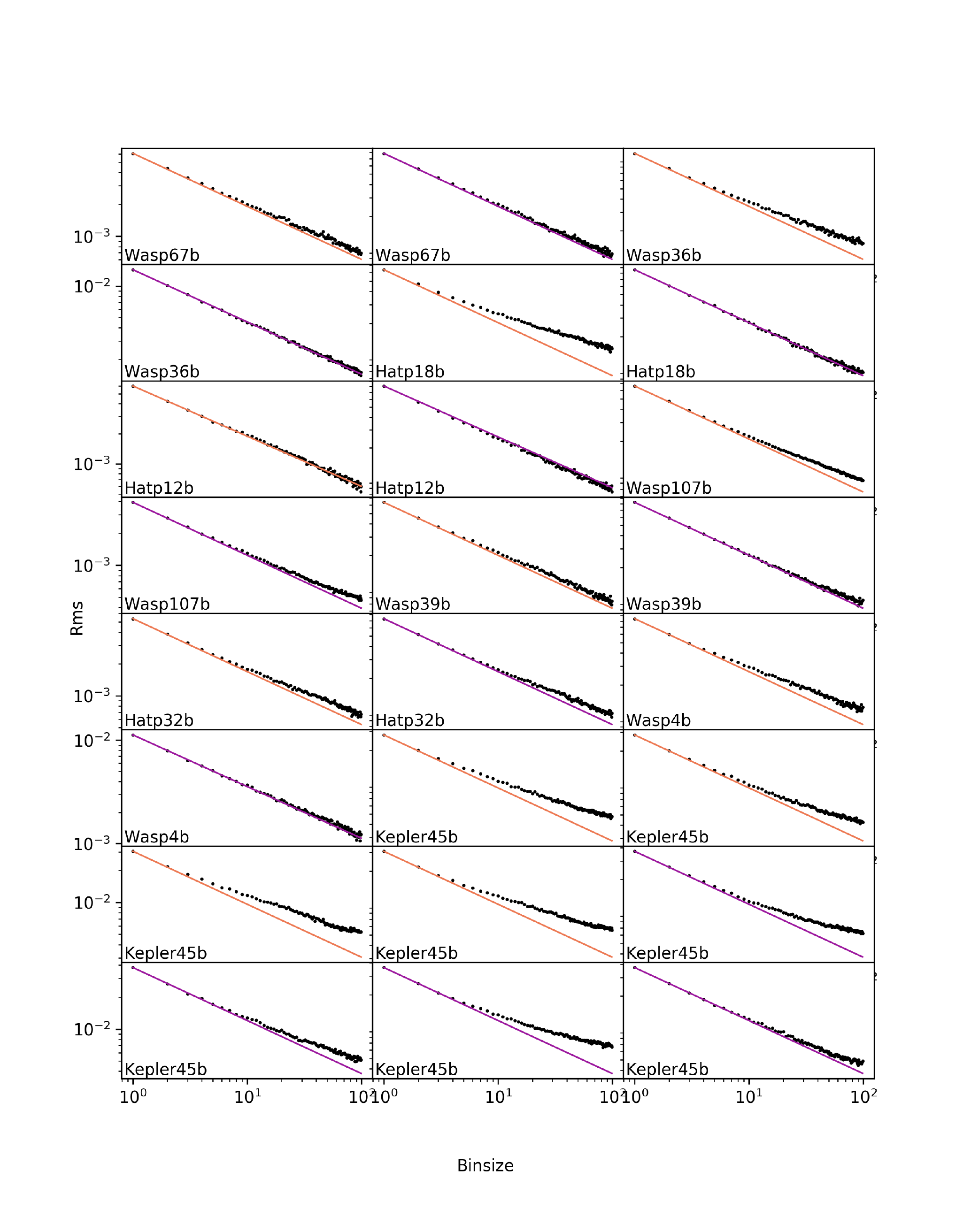}
  \caption{\textit{Continued.}}
\end{figure*}

\section{VULCAN validation on HD 209458b}

\begin{figure}
    \centering
    \includegraphics[width = \linewidth]{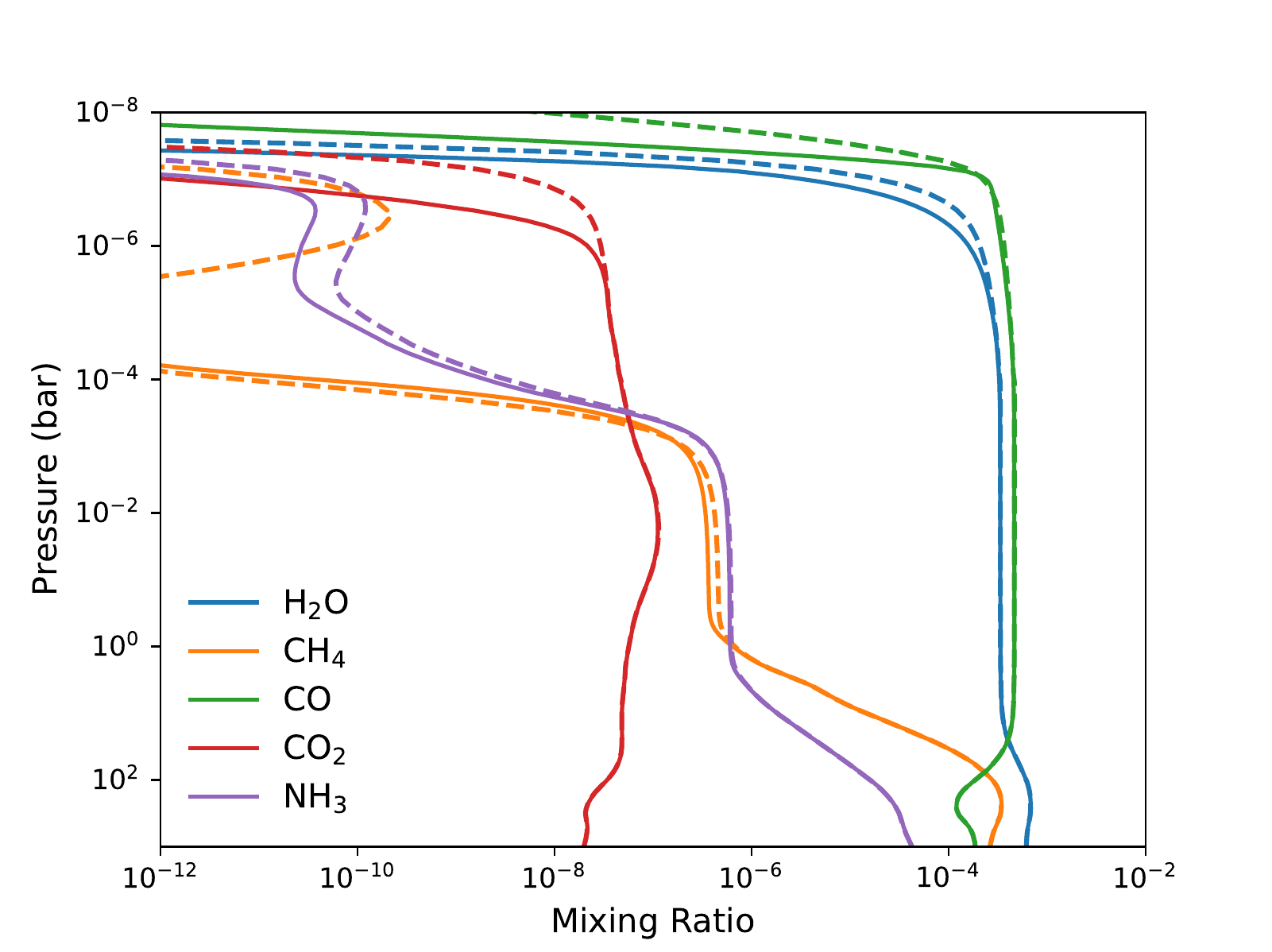}
    \caption{Abundance mixing ratios at different pressures for the main species in the Spitzer bandpasses in HD 209458 b. The solid line shows the results from our VULCAN calculation and the dashed line the results from \citet{Moses2011}. The temperature and eddy-diffusion structure are taken the same as the dayside-average P-T profile in \citet{Moses2011}. The solar flux is also used as an analog for HD 290458 at a distance of 0.04747 AU.}
    \label{P1:fig:HD209}
\end{figure}

\section{Radius anomaly}

\begin{figure*}
    \centering
    \includegraphics[width = \textwidth]{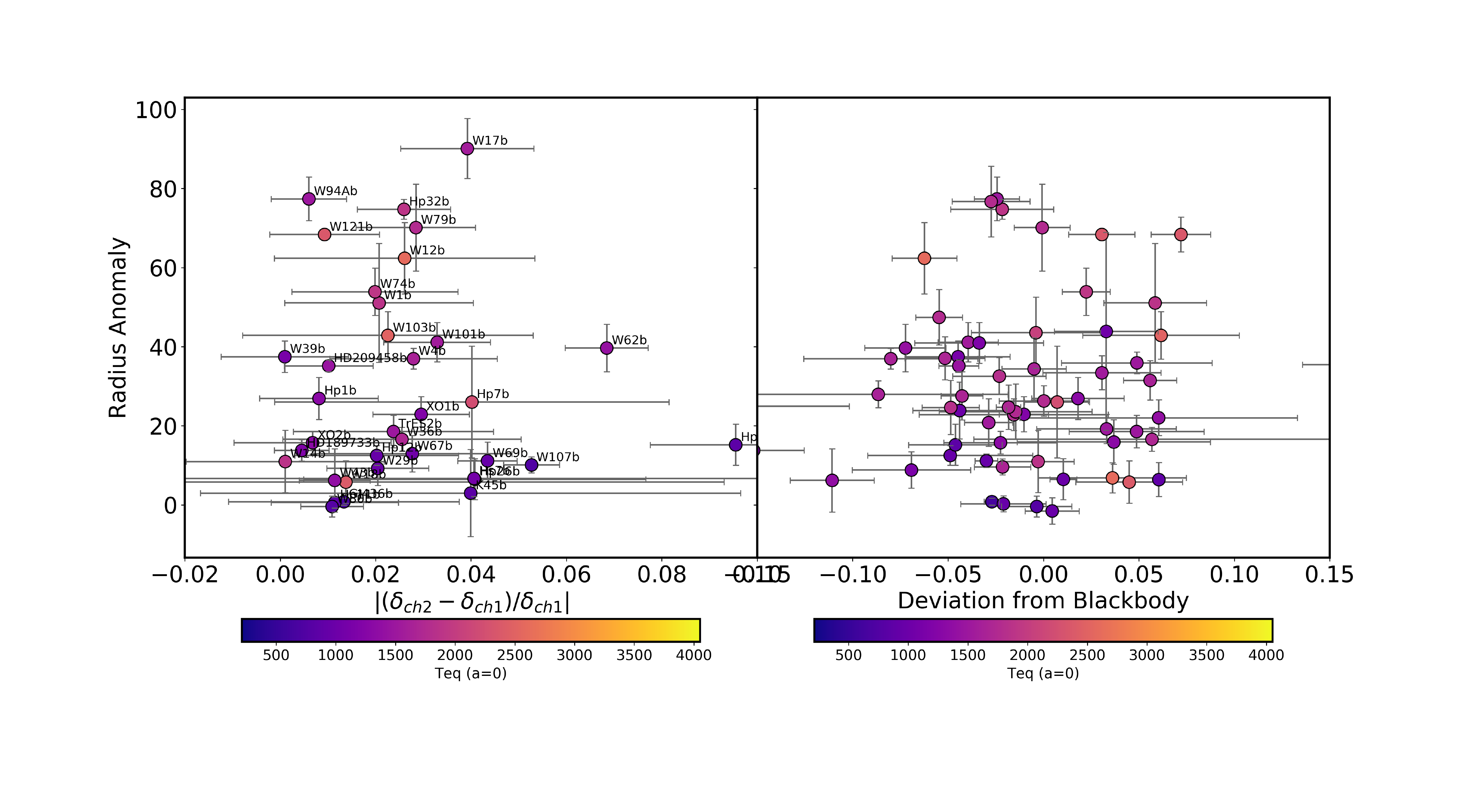}
    \caption{Radius anomaly (calculated using models from \citet{Thorngren2016} and \citet{Thorngren2018} against the absolute value of the normalized difference in transit depths for the available planets. The color scale is the equilibrium temperature of the planet.}
    \label{P1:fig:RadiusAnomaly}
\end{figure*}

\clearpage
\onecolumn
\setcounter{table}{1}
\def\thetable{A.\arabic{table}}

\begin{landscape}
\setlength{\tabcolsep}{3pt}
\begin{longtable}{llllllllllc}

\caption{\label{P1:tab:eclipses} Eclipse depths at 3.6 and 4.5$\mu$m collected from the literature ( $(F_p/F_s)_{3.6}$ and $(F_p/F_s)_{4.5}$ respectively). $T_{eq}$ is the equilibrium temperature assuming no redistribution and zero albedo. Distance is the estimated distance in parsecs taken from \citet{Bailer-Jones2018}. $m_{W1}^{s}$ and $m_{W2}^{s}$ are the stellar apparent  magnitudes. $M_{3.6}^{pl}$ and $M_{4.5}^{pl}$ and the planetary absolute magnitudes in the wise 1 and wise 2 bandpasses (which are equivalent to the spitzer channel 1 and channel 2 respectively) calculated using the planet-to-star-flux ratio. Color is the 4.5 minus 3.6 micron magnitude color of the planets. The final column details the reference for each of the eclipse depths in the literature, the majority of which were collected from exoplanets.org \citep{Wright2011} and have been individually verified.} \\

\hline\hline
Planet & $(F_p/F_s)_{3.6}$ & $(F_p/F_s)_{4.5}$ & $T_{eq}$ (a=0) &  Distance  & $M_{3.6}^{pl}$ &    $M_{4.5}^{pl}$ &  $m_{W1}^{s}$ &  $m_{W2}^{s}$ &         color  &  Eclipse Reference \\
& ppm & ppm & Kelvin & parsec & mag & mag & mag & mag & mag & \\
\hline
\endfirsthead
\caption{continued.} \\
\hline\hline
Planet & $(F_p/F_s)_{3.6}$ & $(F_p/F_s)_{4.5}$ & $T_{eq}$ (a=0) &  Distance  & $M_{3.6}^{pl}$ &    $M_{4.5}^{pl}$ &  $m_{W1}^{s}$ &  $m_{W2}^{s}$ &         color  &  Eclipse Reference \\
& ppm & ppm & Kelvin & parsec & mag & mag & mag & mag & mag & \\ 
\hline
\endhead
\hline
\endfoot
HAT-P-32 b     &    3640 $\pm$ 160 &    4380 $\pm$ 200 &   1901 $\pm$ 57 &    289.2 $\pm$ 5.3 &   8.7 $\pm$ 0.1 &    8.5 $\pm$ 0.07 &    9.9 $\pm$ 0.02 &   9.91 $\pm$ 0.02 &    0.2 $\pm$ 0.09 &        1 \\
XO-1 b        &      860 $\pm$ 70 &     1220 $\pm$ 90 &   1207 $\pm$ 30 &    163.6 $\pm$ 0.6 &  11.1 $\pm$ 0.1 &  10.73 $\pm$ 0.08 &   9.49 $\pm$ 0.02 &   9.52 $\pm$ 0.02 &   0.36 $\pm$ 0.12 &        2 \\
HAT-P-1 b      &      800 $\pm$ 80 &    1350 $\pm$ 220 &   1306 $\pm$ 33 &    159.0 $\pm$ 1.0 &  10.5 $\pm$ 0.1 &   9.92 $\pm$ 0.18 &   8.73 $\pm$ 0.02 &   8.75 $\pm$ 0.02 &   0.55 $\pm$ 0.21 &        3 \\
WASP-39 b     &     880 $\pm$ 150 &     960 $\pm$ 180 &   1118 $\pm$ 35 &    214.0 $\pm$ 1.7 &  11.1 $\pm$ 0.2 &  11.12 $\pm$ 0.21 &  10.16 $\pm$ 0.02 &  10.22 $\pm$ 0.02 &   0.03 $\pm$ 0.28 &        4 \\
HAT-P-12 b     &     660 $\pm$ 270 &     640 $\pm$ 180 &    958 $\pm$ 18 &    142.8 $\pm$ 0.5 &  12.3 $\pm$ 0.4 &  12.35 $\pm$ 0.31 &  10.08 $\pm$ 0.02 &  10.14 $\pm$ 0.02 &   -0.1 $\pm$ 0.54 &        5 \\
HAT-P-18 b     &     437 $\pm$ 145 &     326 $\pm$ 146 &    847 $\pm$ 26 &    161.4 $\pm$ 0.6 &  12.6 $\pm$ 0.4 &  12.93 $\pm$ 0.49 &   10.2 $\pm$ 0.02 &  10.25 $\pm$ 0.02 &  -0.38 $\pm$ 0.61 &        6 \\
TrES2 b      &    1270 $\pm$ 210 &    2300 $\pm$ 240 &   1498 $\pm$ 32 &    215.3 $\pm$ 1.0 &  10.4 $\pm$ 0.2 &   9.72 $\pm$ 0.12 &   9.78 $\pm$ 0.02 &   9.79 $\pm$ 0.02 &   0.63 $\pm$ 0.21 &        7 \\
WASP-4 b      &    3190 $\pm$ 310 &    3430 $\pm$ 270 &   1651 $\pm$ 27 &    267.2 $\pm$ 3.7 &   9.8 $\pm$ 0.1 &   9.78 $\pm$ 0.09 &  10.68 $\pm$ 0.02 &  10.75 $\pm$ 0.02 &   0.01 $\pm$ 0.15 &        8 \\
XO-2 b        &     810 $\pm$ 170 &     980 $\pm$ 200 &   1322 $\pm$ 23 &    154.3 $\pm$ 1.4 &  11.0 $\pm$ 0.2 &  10.89 $\pm$ 0.22 &   9.24 $\pm$ 0.02 &   9.31 $\pm$ 0.02 &   0.14 $\pm$ 0.32 &        9 \\
GJ3470 b     &      113 $\pm$ 24 &        3 $\pm$ 22 &    662 $\pm$ 45 &     29.4 $\pm$ 0.1 &  15.3 $\pm$ 0.2 &  19.25 $\pm$ 7.96 &   7.81 $\pm$ 0.03 &   7.78 $\pm$ 0.02 &  -3.92 $\pm$ 7.97 &       10 \\
WASP-1 b      &    1170 $\pm$ 160 &    2120 $\pm$ 210 &   1876 $\pm$ 69 &   393.1 $\pm$ 10.5 &   9.6 $\pm$ 0.2 &   8.96 $\pm$ 0.12 &  10.22 $\pm$ 0.02 &  10.25 $\pm$ 0.02 &    0.62 $\pm$ 0.2 &       11 \\
HAT-P-26 b     &        85 $\pm$ 0 &      265 $\pm$ 70 &    994 $\pm$ 48 &    141.8 $\pm$ 1.1 &  14.0 $\pm$ 0.0 &  12.78 $\pm$ 0.29 &   9.54 $\pm$ 0.02 &    9.6 $\pm$ 0.02 &   1.17 $\pm$ 0.29 &        6 \\
WASP-121 b    &    3685 $\pm$ 114 &    4684 $\pm$ 121 &   2359 $\pm$ 61 &    269.9 $\pm$ 1.6 &   8.3 $\pm$ 0.0 &   8.05 $\pm$ 0.04 &   9.36 $\pm$ 0.02 &   9.39 $\pm$ 0.02 &   0.23 $\pm$ 0.06 &       12 \\
WASP-87 b     &    2080 $\pm$ 127 &    2708 $\pm$ 137 &   2343 $\pm$ 68 &    298.4 $\pm$ 3.6 &   8.7 $\pm$ 0.1 &   8.47 $\pm$ 0.06 &    9.4 $\pm$ 0.02 &   9.43 $\pm$ 0.02 &    0.26 $\pm$ 0.1 &       12 \\
WASP-100 b    &     1267 $\pm$ 98 &    1720 $\pm$ 119 &  2200.pm 171 &    364.4 $\pm$ 2.7 &   9.1 $\pm$ 0.1 &   8.74 $\pm$ 0.08 &   9.62 $\pm$ 0.02 &   9.64 $\pm$ 0.02 &   0.31 $\pm$ 0.12 &       12 \\
WASP-78 b     &    2001 $\pm$ 218 &    2013 $\pm$ 351 &  1957.pm 256 &   754.3 $\pm$ 16.7 &   8.3 $\pm$ 0.1 &    8.33 $\pm$ 0.2 &  10.96 $\pm$ 0.02 &  10.98 $\pm$ 0.02 &  -0.01 $\pm$ 0.24 &       12 \\
HAT-P-41 b     &    1842 $\pm$ 319 &    2303 $\pm$ 177 &   1937 $\pm$ 44 &    348.2 $\pm$ 4.5 &   8.7 $\pm$ 0.2 &   8.49 $\pm$ 0.09 &   9.56 $\pm$ 0.02 &    9.6 $\pm$ 0.02 &    0.2 $\pm$ 0.21 &       12 \\
WASP-101 b    &    1161 $\pm$ 111 &    1194 $\pm$ 113 &   1554 $\pm$ 40 &    201.2 $\pm$ 1.1 &   9.9 $\pm$ 0.1 &   9.86 $\pm$ 0.11 &   9.04 $\pm$ 0.02 &   9.07 $\pm$ 0.02 &    0.0 $\pm$ 0.15 &       12 \\
WASP-131 b    &      304 $\pm$ 96 &      289 $\pm$ 80 &   1458 $\pm$ 35 &    200.1 $\pm$ 2.6 &  10.8 $\pm$ 0.3 &   10.91 $\pm$ 0.3 &   8.54 $\pm$ 0.02 &   8.57 $\pm$ 0.02 &  -0.08 $\pm$ 0.46 &       12 \\
WASP-36 b     &     914 $\pm$ 578 &    1953 $\pm$ 544 &   1722 $\pm$ 45 &    386.3 $\pm$ 5.2 &  10.8 $\pm$ 0.7 &   10.04 $\pm$ 0.3 &  11.15 $\pm$ 0.02 &   11.2 $\pm$ 0.02 &   0.77 $\pm$ 0.75 &       12 \\
WASP-63 b     &      486 $\pm$ 96 &     560 $\pm$ 130 &   1531 $\pm$ 45 &    290.7 $\pm$ 2.0 &  10.3 $\pm$ 0.2 &   10.2 $\pm$ 0.25 &   9.34 $\pm$ 0.02 &   9.39 $\pm$ 0.02 &   0.11 $\pm$ 0.33 &       12 \\
WASP-94 A b    &      867 $\pm$ 59 &      995 $\pm$ 93 &   1500 $\pm$ 76 &    211.2 $\pm$ 2.5 &   9.8 $\pm$ 0.1 &   9.72 $\pm$ 0.11 &    8.8 $\pm$ 0.02 &   8.84 $\pm$ 0.02 &   0.11 $\pm$ 0.13 &       12 \\
WASP-62 b     &    1616 $\pm$ 146 &    1359 $\pm$ 130 &   1427 $\pm$ 35 &    175.6 $\pm$ 0.6 &   9.7 $\pm$ 0.1 &   9.86 $\pm$ 0.11 &    8.9 $\pm$ 0.02 &   8.92 $\pm$ 0.02 &  -0.21 $\pm$ 0.15 &       12 \\
CoRoT-1 b   &    4150 $\pm$ 420 &    4820 $\pm$ 420 &   1900 $\pm$ 81 &   787.9 $\pm$ 23.5 &   8.6 $\pm$ 0.1 &   8.45 $\pm$ 0.12 &   12.1 $\pm$ 0.02 &  12.14 $\pm$ 0.02 &   0.12 $\pm$ 0.17 &       13 \\
CoRoT-2 b   &    3550 $\pm$ 200 &    5000 $\pm$ 200 &   1537 $\pm$ 40 &    213.3 $\pm$ 2.5 &   9.5 $\pm$ 0.1 &   9.21 $\pm$ 0.05 &  10.06 $\pm$ 0.02 &   10.1 $\pm$ 0.02 &   0.34 $\pm$ 0.09 &       13 \\
GJ 436 b    &      155 $\pm$ 22 &       34 $\pm$ 20 &    649 $\pm$ 59 &      9.8 $\pm$ 0.0 &  15.6 $\pm$ 0.2 &  16.92 $\pm$ 0.64 &   6.02 $\pm$ 0.11 &   5.69 $\pm$ 0.05 &  -1.32 $\pm$ 0.67 &       14 \\
HAT-P-19 b  &     620 $\pm$ 140 &     620 $\pm$ 140 &   1009 $\pm$ 40 &    202.1 $\pm$ 1.5 &  12.0 $\pm$ 0.2 &  12.05 $\pm$ 0.25 &   10.5 $\pm$ 0.02 &  10.56 $\pm$ 0.02 &  -0.06 $\pm$ 0.35 &        4 \\
HAT-P-2 b   &      996 $\pm$ 72 &     1031 $\pm$ 61 &   1540 $\pm$ 30 &    127.8 $\pm$ 0.4 &   9.5 $\pm$ 0.1 &   9.52 $\pm$ 0.07 &   7.57 $\pm$ 0.02 &   7.58 $\pm$ 0.02 &   0.03 $\pm$ 0.11 &       15 \\
HAT-P-20 b  &      615 $\pm$ 82 &     1096 $\pm$ 77 &    971 $\pm$ 24 &     71.0 $\pm$ 0.2 &  12.3 $\pm$ 0.1 &  11.79 $\pm$ 0.08 &   8.56 $\pm$ 0.03 &   8.65 $\pm$ 0.02 &   0.54 $\pm$ 0.17 &       16 \\
HAT-P-23 b  &    2480 $\pm$ 190 &    3090 $\pm$ 260 &   2051 $\pm$ 71 &    364.8 $\pm$ 4.7 &   9.5 $\pm$ 0.1 &    9.26 $\pm$ 0.1 &  10.75 $\pm$ 0.02 &  10.79 $\pm$ 0.02 &   0.19 $\pm$ 0.13 &       17 \\
HAT-P-26 b  &      -27 $\pm$ 50 &      223 $\pm$ 81 &    994 $\pm$ 48 &    141.8 $\pm$ 1.1 &   - $\pm$ 2.0 &   12.97 $\pm$ 0.4 &   9.54 $\pm$ 0.02 &    9.6 $\pm$ 0.02 &    - $\pm$ 2.05 &        6 \\
HAT-P-3 b   &    1120 $\pm$ 225 &     940 $\pm$ 125 &   1158 $\pm$ 34 &    134.6 $\pm$ 0.5 &  11.1 $\pm$ 0.2 &  11.37 $\pm$ 0.15 &   9.38 $\pm$ 0.02 &   9.45 $\pm$ 0.02 &  -0.26 $\pm$ 0.26 &        5 \\
HAT-P-4 b   &    1420 $\pm$ 160 &    1220 $\pm$ 130 &   1694 $\pm$ 47 &    320.5 $\pm$ 2.8 &   9.3 $\pm$ 0.1 &   9.52 $\pm$ 0.12 &   9.73 $\pm$ 0.02 &   9.77 $\pm$ 0.02 &   -0.2 $\pm$ 0.17 &        5 \\
HAT-P-6 b   &     1170 $\pm$ 80 &     1060 $\pm$ 60 &   1673 $\pm$ 42 &    275.4 $\pm$ 3.6 &   9.4 $\pm$ 0.1 &   9.54 $\pm$ 0.07 &   9.29 $\pm$ 0.02 &    9.3 $\pm$ 0.02 &  -0.12 $\pm$ 0.11 &       18 \\
HAT-P-7 b   &    1560 $\pm$ 130 &    1900 $\pm$ 110 &   2225 $\pm$ 41 &    341.1 $\pm$ 2.4 &   8.6 $\pm$ 0.1 &   8.44 $\pm$ 0.07 &   9.28 $\pm$ 0.02 &    9.3 $\pm$ 0.02 &   0.19 $\pm$ 0.12 &       19 \\
HAT-P-8 b   &     1310 $\pm$ 85 &     1110 $\pm$ 75 &   1772 $\pm$ 48 &    211.6 $\pm$ 1.7 &   9.5 $\pm$ 0.1 &   9.71 $\pm$ 0.08 &   8.93 $\pm$ 0.02 &   8.95 $\pm$ 0.02 &   -0.2 $\pm$ 0.11 &       18 \\
HD 149026 b &      400 $\pm$ 30 &      340 $\pm$ 60 &   1673 $\pm$ 65 &     75.9 $\pm$ 0.2 &  10.9 $\pm$ 0.1 &  11.07 $\pm$ 0.19 &   6.79 $\pm$ 0.07 &    6.8 $\pm$ 0.02 &  -0.19 $\pm$ 0.22 &       20 \\
HD 189733 b &    2560 $\pm$ 140 &    2140 $\pm$ 200 &   1200 $\pm$ 22 &     19.8 $\pm$ 0.0 &  10.3 $\pm$ 0.2 &  10.54 $\pm$ 0.11 &   5.29 $\pm$ 0.15 &   5.34 $\pm$ 0.05 &   -0.25 $\pm$ 0.2 &       21 \\
HD 209458 b &     1190 $\pm$ 70 &     1230 $\pm$ 60 &   1446 $\pm$ 19 &     48.3 $\pm$ 0.1 &  10.2 $\pm$ 0.1 &  10.05 $\pm$ 0.06 &   6.31 $\pm$ 0.09 &   6.19 $\pm$ 0.03 &   0.15 $\pm$ 0.13 &       22 \\
Kepler-12 b &    1370 $\pm$ 200 &    1160 $\pm$ 310 &   1481 $\pm$ 31 &    881.4 $\pm$ 9.7 &   9.5 $\pm$ 0.2 &   9.69 $\pm$ 0.29 &  12.05 $\pm$ 0.02 &  12.08 $\pm$ 0.02 &  -0.21 $\pm$ 0.33 &       23 \\
Kepler-17 b &    2500 $\pm$ 300 &    3100 $\pm$ 350 &   1745 $\pm$ 39 &   720.8 $\pm$ 10.3 &   9.8 $\pm$ 0.1 &   9.58 $\pm$ 0.13 &  12.55 $\pm$ 0.02 &  12.59 $\pm$ 0.02 &   0.19 $\pm$ 0.19 &       24 \\
Kepler-5 b  &    1030 $\pm$ 170 &    1070 $\pm$ 150 &   1807 $\pm$ 35 &   899.8 $\pm$ 16.5 &   9.4 $\pm$ 0.2 &    9.4 $\pm$ 0.16 &  11.68 $\pm$ 0.02 &  11.74 $\pm$ 0.02 &  -0.02 $\pm$ 0.24 &       25 \\
Kepler-6 b  &     690 $\pm$ 270 &    1510 $\pm$ 190 &   1504 $\pm$ 21 &    587.0 $\pm$ 5.0 &  10.6 $\pm$ 0.4 &   9.85 $\pm$ 0.14 &  11.58 $\pm$ 0.02 &  11.64 $\pm$ 0.02 &   0.79 $\pm$ 0.45 &       26 \\
KOI-13 b    &    1560 $\pm$ 310 &    2220 $\pm$ 230 &   2607 $\pm$ 94 &   519.1 $\pm$ 29.1 &   7.8 $\pm$ 0.2 &   7.47 $\pm$ 0.17 &   9.39 $\pm$ 0.02 &   9.41 $\pm$ 0.02 &    0.36 $\pm$ 0.3 &       27 \\
Qatar-1 b   &    1511 $\pm$ 455 &    2907 $\pm$ 415 &   1389 $\pm$ 43 &    185.6 $\pm$ 0.8 &  11.0 $\pm$ 0.3 &  10.39 $\pm$ 0.16 &  10.32 $\pm$ 0.02 &   10.4 $\pm$ 0.02 &   0.64 $\pm$ 0.36 &       28 \\
TrES-3 b    &    3460 $\pm$ 350 &    3720 $\pm$ 540 &   1629 $\pm$ 32 &    231.3 $\pm$ 1.3 &   9.9 $\pm$ 0.1 &   9.86 $\pm$ 0.16 &  10.57 $\pm$ 0.02 &  10.61 $\pm$ 0.02 &    0.04 $\pm$ 0.2 &       29 \\
TrES-4 b    &    1370 $\pm$ 110 &    1480 $\pm$ 160 &   1785 $\pm$ 41 &    516.0 $\pm$ 6.9 &   8.8 $\pm$ 0.1 &   8.79 $\pm$ 0.12 &  10.24 $\pm$ 0.02 &  10.28 $\pm$ 0.02 &   0.05 $\pm$ 0.15 &       30 \\
WASP-10 b   &    1000 $\pm$ 110 &    1460 $\pm$ 160 &    960 $\pm$ 24 &    141.0 $\pm$ 0.7 &  11.7 $\pm$ 0.1 &  11.34 $\pm$ 0.12 &   9.93 $\pm$ 0.02 &   10.0 $\pm$ 0.02 &   0.34 $\pm$ 0.17 &        4 \\
WASP-103 b  &    4458 $\pm$ 383 &    5686 $\pm$ 138 &   2505 $\pm$ 78 &  883.3 $\pm$ 153.1 &   6.9 $\pm$ 0.4 &   6.63 $\pm$ 0.38 &  10.72 $\pm$ 0.02 &  10.75 $\pm$ 0.02 &   0.24 $\pm$ 0.54 &       31 \\
WASP-12 b   &    4210 $\pm$ 110 &    4280 $\pm$ 120 &   2584 $\pm$ 91 &    427.2 $\pm$ 6.0 &   7.9 $\pm$ 0.0 &   7.88 $\pm$ 0.05 &  10.11 $\pm$ 0.02 &  10.11 $\pm$ 0.02 &   0.02 $\pm$ 0.07 &       32 \\
WASP-121 b  &    3150 $\pm$ 103 &    4510 $\pm$ 107 &   2359 $\pm$ 61 &    269.9 $\pm$ 1.6 &   8.5 $\pm$ 0.0 &    8.1 $\pm$ 0.03 &   9.36 $\pm$ 0.02 &   9.39 $\pm$ 0.02 &   0.36 $\pm$ 0.06 &       12 \\
WASP-14 b   &     1870 $\pm$ 70 &    2240 $\pm$ 180 &   1864 $\pm$ 60 &    162.0 $\pm$ 0.8 &   9.3 $\pm$ 0.0 &   9.17 $\pm$ 0.09 &   8.57 $\pm$ 0.02 &    8.6 $\pm$ 0.02 &    0.17 $\pm$ 0.1 &       33 \\
WASP-18 b   &    3000 $\pm$ 200 &    3900 $\pm$ 200 &   2398 $\pm$ 73 &    123.5 $\pm$ 0.4 &   8.9 $\pm$ 0.1 &   8.69 $\pm$ 0.06 &   8.07 $\pm$ 0.02 &   8.12 $\pm$ 0.02 &    0.24 $\pm$ 0.1 &       34 \\
WASP-19 b   &    4830 $\pm$ 250 &    5720 $\pm$ 300 &   2066 $\pm$ 46 &    268.3 $\pm$ 1.7 &   9.1 $\pm$ 0.1 &   8.96 $\pm$ 0.06 &  10.44 $\pm$ 0.02 &  10.49 $\pm$ 0.02 &   0.13 $\pm$ 0.09 &       35 \\
WASP-2 b    &     830 $\pm$ 350 &    1690 $\pm$ 170 &   1300 $\pm$ 71 &    153.2 $\pm$ 1.6 &  11.4 $\pm$ 0.5 &  10.64 $\pm$ 0.11 &   9.58 $\pm$ 0.02 &   9.64 $\pm$ 0.02 &   0.72 $\pm$ 0.47 &       11 \\
WASP-24 b   &    1590 $\pm$ 130 &    2020 $\pm$ 180 &   1769 $\pm$ 39 &    322.1 $\pm$ 4.4 &   9.6 $\pm$ 0.1 &    9.33 $\pm$ 0.1 &   10.1 $\pm$ 0.02 &  10.13 $\pm$ 0.02 &   0.23 $\pm$ 0.14 &       36 \\
WASP-33 b   &    2600 $\pm$ 500 &    4100 $\pm$ 200 &   2694 $\pm$ 53 &    121.9 $\pm$ 1.0 &   8.4 $\pm$ 0.2 &   7.98 $\pm$ 0.06 &   7.38 $\pm$ 0.04 &   7.44 $\pm$ 0.02 &   0.43 $\pm$ 0.22 &       37 \\
WASP-43 b   &    3460 $\pm$ 130 &    3820 $\pm$ 150 &   1375 $\pm$ 79 &     86.7 $\pm$ 0.3 &  10.6 $\pm$ 0.0 &  10.58 $\pm$ 0.05 &   9.15 $\pm$ 0.02 &   9.22 $\pm$ 0.02 &   0.03 $\pm$ 0.07 &       38 \\
WASP-48 b   &    1760 $\pm$ 130 &    2140 $\pm$ 200 &   2033 $\pm$ 68 &    454.1 $\pm$ 4.4 &   8.9 $\pm$ 0.1 &   8.76 $\pm$ 0.11 &  10.33 $\pm$ 0.02 &  10.37 $\pm$ 0.02 &   0.17 $\pm$ 0.14 &       17 \\
WASP-5 b    &    1970 $\pm$ 280 &    2370 $\pm$ 240 &   1742 $\pm$ 68 &    309.1 $\pm$ 3.4 &   9.9 $\pm$ 0.2 &    9.7 $\pm$ 0.11 &  10.54 $\pm$ 0.02 &  10.59 $\pm$ 0.02 &   0.15 $\pm$ 0.19 &       39 \\
WASP-6 b    &     940 $\pm$ 190 &    1150 $\pm$ 220 &   1184 $\pm$ 32 &    197.1 $\pm$ 1.6 &  11.4 $\pm$ 0.2 &  11.21 $\pm$ 0.21 &  10.28 $\pm$ 0.02 &  10.34 $\pm$ 0.02 &    0.16 $\pm$ 0.3 &        4 \\
WASP-67 b   &     220 $\pm$ 130 &     800 $\pm$ 180 &   1028 $\pm$ 32 &    189.5 $\pm$ 1.5 &  12.8 $\pm$ 0.6 &  11.43 $\pm$ 0.25 &  10.03 $\pm$ 0.02 &  10.08 $\pm$ 0.02 &   1.35 $\pm$ 0.69 &        4 \\
WASP-69 b   &      421 $\pm$ 29 &      463 $\pm$ 39 &    961 $\pm$ 21 &     50.0 $\pm$ 0.1 &  12.3 $\pm$ 0.1 &  12.28 $\pm$ 0.09 &   7.32 $\pm$ 0.04 &   7.44 $\pm$ 0.02 &  -0.02 $\pm$ 0.13 &        6 \\
WASP-8 b    &    1130 $\pm$ 180 &      690 $\pm$ 70 &    927 $\pm$ 27 &     90.0 $\pm$ 0.4 &  10.5 $\pm$ 0.2 &  11.05 $\pm$ 0.11 &   7.91 $\pm$ 0.02 &   7.92 $\pm$ 0.02 &  -0.55 $\pm$ 0.21 &       40 \\
WASP-80 b   &     455 $\pm$ 100 &      944 $\pm$ 65 &    775 $\pm$ 25 &     49.8 $\pm$ 0.1 &  13.2 $\pm$ 0.2 &   12.4 $\pm$ 0.08 &    8.3 $\pm$ 0.02 &   8.32 $\pm$ 0.02 &   0.77 $\pm$ 0.25 &       41 \\
XO-3 b      &     1010 $\pm$ 40 &     1580 $\pm$ 36 &   2046 $\pm$ 40 &    213.1 $\pm$ 2.7 &   9.6 $\pm$ 0.1 &   9.13 $\pm$ 0.04 &   8.75 $\pm$ 0.02 &   8.76 $\pm$ 0.02 &   0.47 $\pm$ 0.07 &       42 \\
XO-4 b      &      560 $\pm$ 90 &     1350 $\pm$ 85 &   1639 $\pm$ 35 &    272.7 $\pm$ 2.9 &  10.3 $\pm$ 0.2 &   9.39 $\pm$ 0.08 &   9.37 $\pm$ 0.02 &    9.4 $\pm$ 0.02 &   0.93 $\pm$ 0.19 &       18 \\
HAT-P-13 b  &     851 $\pm$ 107 &    1090 $\pm$ 124 &   1648 $\pm$ 53 &    246.8 $\pm$ 2.2 &   7.2 $\pm$ 0.1 &   6.95 $\pm$ 0.13 &   8.96 $\pm$ 0.02 &   9.01 $\pm$ 0.02 &   0.22 $\pm$ 0.19 &       12 \\
HAT-P-30 b  &    1603 $\pm$ 107 &    1783 $\pm$ 147 &   1637 $\pm$ 43 &    214.0 $\pm$ 2.2 &   6.9 $\pm$ 0.1 &    6.8 $\pm$ 0.09 &   9.04 $\pm$ 0.02 &   9.08 $\pm$ 0.02 &   0.08 $\pm$ 0.12 &       12 \\
HAT-P-33 b  &    1663 $\pm$ 127 &    1896 $\pm$ 199 &   1780 $\pm$ 34 &    396.1 $\pm$ 7.5 &   6.5 $\pm$ 0.1 &   6.33 $\pm$ 0.12 &   10.0 $\pm$ 0.02 &  10.02 $\pm$ 0.02 &   0.13 $\pm$ 0.16 &       12 \\
HAT-P-40 b  &     988 $\pm$ 168 &    1057 $\pm$ 145 &   1765 $\pm$ 66 &    464.5 $\pm$ 6.4 &   6.7 $\pm$ 0.2 &   6.62 $\pm$ 0.15 &   9.98 $\pm$ 0.02 &  10.01 $\pm$ 0.02 &   0.04 $\pm$ 0.24 &       12 \\
KELT-2 A b  &      739 $\pm$ 38 &      761 $\pm$ 47 &   1710 $\pm$ 31 &    134.1 $\pm$ 0.8 &   7.0 $\pm$ 0.1 &   6.99 $\pm$ 0.07 &   7.27 $\pm$ 0.04 &   7.34 $\pm$ 0.02 &   -0.03 $\pm$ 0.1 &       12 \\
KELT-3 b    &     1788 $\pm$ 97 &    1677 $\pm$ 104 &   1822 $\pm$ 44 &    210.3 $\pm$ 5.4 &   6.3 $\pm$ 0.1 &   6.43 $\pm$ 0.09 &   8.57 $\pm$ 0.02 &    8.6 $\pm$ 0.02 &   -0.1 $\pm$ 0.12 &       12 \\
WASP-104 b  &    1709 $\pm$ 195 &    2643 $\pm$ 303 &   1516 $\pm$ 43 &    185.9 $\pm$ 1.5 &   7.9 $\pm$ 0.1 &    7.5 $\pm$ 0.13 &   9.84 $\pm$ 0.02 &   9.91 $\pm$ 0.02 &   0.41 $\pm$ 0.18 &       12 \\
WASP-46 b   &    1360 $\pm$ 701 &    4446 $\pm$ 589 &   1658 $\pm$ 55 &    375.3 $\pm$ 4.4 &   8.1 $\pm$ 0.6 &   6.88 $\pm$ 0.15 &  11.35 $\pm$ 0.02 &  11.37 $\pm$ 0.02 &   1.27 $\pm$ 0.58 &       12 \\
WASP-64 b   &    2859 $\pm$ 270 &    2071 $\pm$ 471 &   1690 $\pm$ 52 &    369.9 $\pm$ 3.0 &   7.0 $\pm$ 0.1 &   7.38 $\pm$ 0.25 &  10.96 $\pm$ 0.02 &  11.01 $\pm$ 0.02 &   -0.4 $\pm$ 0.27 &       12 \\
WASP-65 b   &    1587 $\pm$ 245 &     724 $\pm$ 318 &   1485 $\pm$ 59 &    273.7 $\pm$ 2.7 &   7.6 $\pm$ 0.2 &   8.52 $\pm$ 0.48 &  10.31 $\pm$ 0.02 &  10.35 $\pm$ 0.02 &   -0.9 $\pm$ 0.51 &       12 \\
WASP-76 b   &     2979 $\pm$ 63 &     3762 $\pm$ 82 &   2183 $\pm$ 47 &    194.5 $\pm$ 6.0 &   5.6 $\pm$ 0.1 &   5.35 $\pm$ 0.07 &   8.19 $\pm$ 0.02 &   8.23 $\pm$ 0.02 &    0.22 $\pm$ 0.1 &       12 \\
WASP-77 A b &     2016 $\pm$ 94 &    2487 $\pm$ 127 &   1671 $\pm$ 31 &    105.2 $\pm$ 1.2 &   7.2 $\pm$ 0.1 &   7.06 $\pm$ 0.06 &   8.11 $\pm$ 0.02 &   8.16 $\pm$ 0.02 &   0.17 $\pm$ 0.09 &       12 \\
WASP-97 b   &     1359 $\pm$ 84 &    1534 $\pm$ 101 &   1540 $\pm$ 42 &    151.1 $\pm$ 0.5 &   7.7 $\pm$ 0.1 &   7.65 $\pm$ 0.07 &   8.96 $\pm$ 0.02 &   9.01 $\pm$ 0.02 &    0.08 $\pm$ 0.1 &       12 \\
WASP-74 b     &     1446 $\pm$ 66 &    2075 $\pm$ 100 &   1923 $\pm$ 53 &    149.2 $\pm$ 1.1 &   9.4 $\pm$ 0.1 &   9.03 $\pm$ 0.06 &   8.14 $\pm$ 0.02 &   8.19 $\pm$ 0.02 &   0.34 $\pm$ 0.08 &       12 \\
KELT-7 b      &     1688 $\pm$ 46 &     1896 $\pm$ 57 &   2050 $\pm$ 35 &    136.7 $\pm$ 0.9 &   8.7 $\pm$ 0.0 &   8.65 $\pm$ 0.04 &    7.5 $\pm$ 0.03 &   7.52 $\pm$ 0.02 &    0.1 $\pm$ 0.06 &       12 \\
WASP-79 b     &     1394 $\pm$ 88 &    1783 $\pm$ 106 &   1762 $\pm$ 53 &    246.7 $\pm$ 1.8 &   9.2 $\pm$ 0.1 &   8.96 $\pm$ 0.07 &   9.03 $\pm$ 0.02 &   9.04 $\pm$ 0.02 &    0.25 $\pm$ 0.1 &       12 \\
\hline
\end{longtable}
\tablebib{(1) \citet{Zhao2014};
(2) \citet{Machalek2008};
(3) \citet{Todorov2010};
(4) \citet{Garhart2020};
(5) \citet{Kammer2015};
(6) \citet{Todorov2013};
(7) \citet{Wallack2019};
(8) \citet{ODonovan2010};
(9) \citet{Beerer2011};
(10) \citet{Machalek2009};
(11) \citet{Benneke2019};
(12) \citet{Wheatley2010};
(13) \citet{Deming2011};
(14) \citet{Morley2017};
(15) \citet{Lewis2013};
(16) \citet{Deming2015};
(17) \citet{ORourke2014};
(18) \citet{Todorov2012};
(19) \citet{Christiansen2010};
(20) \citet{Stevenson2012};
(21) \citet{Charbonneau2008};
(22) \citet{Diamond-Lowe2014};
(23) \citet{Fortney2011};
(24) \citet{Desert2011c};
(25) \citet{Desert2011b};
(26) \citet{Desert2011};
(27) \citet{Shporer2014};
(28) \citet{Garhart2020};
(29) \citet{Fressin2010};
(30) \citet{Knutson2009a};
(31) \citet{Kreidberg2018};
(32) \citet{Stevenson2014c};
(33) \citet{Blecic2013};
(34) \citet{Nymeyer2011};
(35) \citet{Anderson2013};
(36) \citet{Smith2012};
(37) \citet{Deming2012};
(38) \citet{Blecic2014};
(39) \citet{Baskin2013};
(40) \citet{Cubillos2013};
(41) \citet{Triaud2015};
(42) \citet{Machalek2010}.}

\end{landscape}

\setlength{\tabcolsep}{3pt}

\begin{longtable}{lllllll}

\caption{\label{P1:tab:pipeline} The optimum parameters used for our pipeline which minimise the $\chi^2$ of a least-squares fit to the systematic errors and the transit. If background method is ``Annulus'', then the two parameters are the radius and width of a circular annulus; if it is ``box'' then the parameter is the size of the box taken in all of the four corners of the image. If the centroiding method is ``baycenter'' then the parameter is the number of pixels over which to create a box over the star for the flux weighting.}\\

\hline\hline
Planet & Channel &  Aperture Size & Background Method &   Background Params & Centroiding Method &  Centroiding Params \\
& & Pixels & & Pixels & & Pixels \\
\hline
\endfirsthead
\caption{continued.} \\
\hline\hline
Planet & Channel &  Aperture Size & Background Method &   Background Params & Centroiding Method &  Centroiding Params \\
& & Pixels & & Pixels & & Pixels \\
\hline
\endhead
\hline
\endfoot
HAT-P-32 b  &     ch1 &           2.50 &           Annulus &         -, 6, 4  &         Barycenter &                 3.0 \\
HAT-P-32 b  &     ch2 &           2.50 &           Annulus &         -, 6, 4  &         Barycenter &                 5.0 \\
XO-1 b     &     ch1 &           2.50 &           Annulus &         -, 6, 4  &             Moffat &                 - \\
XO-1 b     &     ch2 &           2.50 &           Annulus &         -, 6, 4  &           Gaussian &                 - \\
HAT-P-1 b   &     ch1 &           3.00 &               Box &      4, -, -  &         Barycenter &                 5.0 \\
HAT-P-1 b   &     ch2 &           3.50 &               Box &      4, -, -  &         Barycenter &                 5.0 \\
WASP-17 b  &     ch1 &           2.50 &           Annulus &         -, 6, 4  &         Barycenter &                 5.0 \\
WASP-17 b  &     ch2 &           2.50 &           Annulus &         -, 6, 4  &         Barycenter &                 3.0 \\
WASP-39 b  &     ch1 &           2.50 &           Annulus &         -, 6, 4  &         Barycenter &                 3.0 \\
WASP-39 b  &     ch2 &           2.50 &           Annulus &         -, 6, 4  &         Barycenter &                 7.0 \\
HAT-P-12 b  &     ch1 &           2.50 &           Annulus &         -, 6, 4  &             Moffat &                 - \\
HAT-P-12 b  &     ch2 &           2.50 &           Annulus &         -, 6, 4  &             Moffat &                 - \\
HAT-P-18 b  &     ch1 &           2.50 &           Annulus &         -, 6, 4  &         Barycenter &                 5.0 \\
HAT-P-18 b  &     ch2 &           2.50 &           Annulus &         -, 6, 4  &             Moffat &                 - \\
TrES-2 b   &     ch1 &           2.50 &           Annulus &         -, 6, 4  &         Barycenter &                 5.0 \\
TrES-2 b   &     ch2 &           2.50 &         Histogram &   -, -, -  &         Barycenter &                 3.0 \\
WASP-4 b   &     ch1 &           2.50 &           Annulus &         -, 6, 4  &             Moffat &                 - \\
WASP-4 b   &     ch2 &           2.50 &           Annulus &         -, 6, 4  &         Barycenter &                 5.0 \\
XO-2 b      &     ch1 &           2.50 &           Annulus &         -, 6, 4  &             Moffat &                 - \\
XO-2 b      &     ch2 &           2.50 &           Annulus &         -, 6, 4  &          Barycenter &                 7.0 \\
GJ3470 b   &     ch1 &           2.50 &           Annulus &         -, 6, 4  &         Barycenter &                 3.0 \\
GJ3470 b  &     ch2 &           2.50 &           Annulus &         -, 6, 4  &         Barycenter &                 3.0 \\
WASP-21 b  &     ch1 &           2.50 &           Annulus &         -, 6, 4  &         Barycenter &                 5.0 \\
WASP-21 b  &     ch2 &           2.50 &           Annulus &         -, 6, 4  &         Barycenter &                 5.0 \\
WASP-31 b  &     ch1 &           2.50 &           Annulus &         -, 6, 4  &         Barycenter &                 7.0 \\
WASP-31 b  &     ch2 &           2.50 &           Annulus &         -, 6, 4  &         Barycenter &                 7.0 \\
WASP-1 b   &     ch1 &           2.50 &           Annulus &         -, 6, 4  &         Barycenter &                 7.0 \\
WASP-1 b   &     ch2 &           2.50 &           Annulus &         -, 6, 4  &         Barycenter &                 3.0 \\
HAT-P-26 b  &     ch1 &           2.50 &           Annulus &         -, 6, 4  &         Barycenter &                 7.0 \\
HAT-P-26 b  &     ch2 &           2.50 &           Annulus &         -, 6, 4  &         Barycenter &                 5.0 \\
WASP-107 b &     ch1 &           2.25 &         Histogram &   -, -, -  &         Barycenter &                 5.0 \\
WASP-107 b &     ch2 &           2.25 &         Histogram &   -, -, -  &         Barycenter &                 3.0 \\
WASP-13 b  &     ch1 &           2.25 &         Histogram &   -, -, -  &           Gaussian &                 - \\
WASP-13 b  &     ch2 &           2.50 &         Histogram &   -, -, -  &         Barycenter &                 3.0 \\
WASP-121 b &     ch1 &           1.00 &         Histogram &   -, -, -  &         Barycenter &                 3.0 \\
WASP-121 b &     ch2 &           1.00 &               Box &      4, -, -  &         Barycenter &                 3.0 \\
WASP-69 b  &     ch1 &           2.00 &         Histogram &   -, -, -  &           Gaussian &                 - \\
WASP-69 b  &     ch2 &           2.25 &         Histogram &   -, -, -  &         Barycenter &                 3.0 \\
WASP-67 b  &     ch1 &           2.00 &         Histogram &   -, -, -  &             Moffat &                 - \\
WASP-67 b  &     ch2 &           2.00 &         Histogram &   -, -, -  &             Moffat &                 - \\
HATS-7 b   &     ch1 &           2.00 &         Histogram &   -, -, -  &         Barycenter &                 3.0 \\
HATS-7 b   &     ch2 &           2.00 &         Histogram &   -, -, -  &           Gaussian &                 - \\
WASP-29 b  &     ch1 &           2.25 &         Histogram &   -, -, -  &           Gaussian &                 - \\
WASP-29 b  &     ch2 &           2.25 &         Histogram &   -, -, -  &         Barycenter &                 3.0 \\
HAT-P-41 b  &     ch1 &           2.25 &         Histogram &   -, -, -  &         Barycenter &                 3.0 \\
HAT-P-41 b  &     ch2 &           2.25 &         Histogram &   -, -, -  &           Gaussian &                 - \\
WASP-101 b &     ch1 &           2.25 &         Histogram &   -, -, -  &         Barycenter &                 5.0 \\
WASP-101 b &     ch2 &           2.25 &         Histogram &   -, -, -  &         Barycenter &                 3.0 \\
WASP-131 b &     ch1 &           2.25 &         Histogram &   -, -, -  &         Barycenter &                 5.0 \\
WASP-131 b &     ch2 &           2.25 &         Histogram &   -, -, -  &         Barycenter &                 3.0 \\
WASP-36 b  &     ch1 &           2.00 &         Histogram &   -, -, -  &           Gaussian &                 - \\
WASP-36 b  &     ch2 &           2.00 &         Histogram &   -, -, -  &         Barycenter &                 3.0 \\
WASP-63 b  &     ch1 &           2.25 &         Histogram &   -, -, -  &         Barycenter &                 5.0 \\
WASP-63 b  &     ch2 &           2.25 &         Histogram &   -, -, -  &           Gaussian &                 - \\
WASP-79 b  &     ch1 &           2.50 &         Histogram &   -, -, -  &         Barycenter &                 5.0 \\
WASP-79 b  &     ch2 &           2.25 &         Histogram &   -, -, -  &         Barycenter &                 3.0 \\
WASP-94 Ab &     ch1 &           2.50 &         Histogram &   -, -, -  &             Moffat &                 - \\
WASP-94 Ab &     ch2 &           2.25 &         Histogram &   -, -, -  &         Barycenter &                 3.0 \\
WASP-74 b  &     ch1 &           2.25 &         Histogram &   -, -, -  &         Barycenter &                 3.0 \\
WASP-74 b  &     ch2 &           2.00 &         Histogram &   -, -, -  &         Barycenter &                 3.0 \\
WASP-62 b  &     ch1 &           2.25 &         Histogram &   -, -, -  &         Barycenter &                 3.0 \\
WASP-62 b  &     ch2 &           2.25 &               Box &      4, -, -  &         Barycenter &                 3.0 \\
KELT-7 b   &     ch1 &           2.25 &         Histogram &  -, -, - &             Moffat &                 - \\
KELT-7 b   &     ch2 &           2.00 &         Histogram &  -, -, - &         Barycenter &                 3.0 \\
Kepler-45 b &     ch1 &           1.00 &         Histogram &   -, -, -  &         Barycenter &                 3.0 \\
Kepler-45 b &     ch1 &           1.00 &         Histogram &   -, -, -  &           Gaussian &                 - \\
Kepler-45 b &     ch1 &           1.00 &         Histogram &   -, -, -  &           Gaussian &                 - \\
Kepler-45 b &     ch1 &           1.00 &         Histogram &   -, -, -  &           Gaussian &                 - \\
Kepler-45 b &     ch2 &           1.00 &         Histogram &   -, -, -  &           Gaussian &                 - \\
Kepler-45 b &     ch2 &           1.00 &         Histogram &   -, -, -  &         Barycenter &                 3.0 \\
Kepler-45 b &     ch2 &           1.00 &         Histogram &   -, -, -  &             Moffat &                 - \\
Kepler-45 b &     ch2 &           1.00 &               Box &      2, -, -  &           Gaussian &                 - \\
\end{longtable}

\setlength{\tabcolsep}{3pt}

\begin{longtable}[h]{llrrrrrr}

\caption{\label{P1:tab:tests} Statistical tests outputted by our custom-built pipeline. We measure the strength of the dependence on the chosen limb-darkening parameters by varying them within 3$\sigma$ of their error for 500 iterations. For each iteration we perform a least-squares fit and measure the variation on the measured Rp/Rs as a function of the final calculated error on Rp/Rs. `Bad pix' is the number of bad pixels corrected at the beginning of the analysis. Cut time (min) is the number of minutes cut from the beginning of each observation; this value is chosen such that we keep as much baseline as possible while minimizing the chi2 of the different possible baselines. Photon noise is the percentage above pure statistical noise we have for each light curve; typical values for Spitzer are 30-60\% above photon noise.}\\
\hline\hline
Planet & $\lambda (\mu m)$ &  Vary ld 3$\sigma$ &  Bad pix \% &  Cut time (min) &  photon noise \% &  MCMC acceptance fraction \\
\hline
\endfirsthead
\caption{continued.} \\
\hline\hline
Planet & $\lambda (\mu m)$ &  Vary ld 3$\sigma$ &  Bad pix \% &  Cut time (min) &  photon noise \% &  MCMC acceptance fraction \\
\hline
\endhead
\hline
\endfoot

HAT-P-32 b   &               3.6 &           0.804 &      0.306 &      0.0 &          1.31 &                0.384 \\
HAT-P-32 b   &               4.5 &           0.603 &      0.056 &      0.0 &          1.32 &                0.383 \\
XO-1 b      &               3.6 &           0.442 &      0.113 &     10.0 &          1.31 &                0.384 \\
XO-1 b      &               4.5 &           0.194 &      0.061 &     10.0 &          1.23 &                0.385 \\
HAT-P-1 b    &               3.6 &           0.018 &      0.270 &     10.0 &          1.47 &                0.386 \\
HAT-P-1 b    &               4.5 &           0.035 &      0.052 &     10.0 &          1.34 &                0.385 \\
WASP-17 b   &               3.6 &           0.198 &      0.309 &     50.0 &          1.41 &                0.385 \\
WASP-17 b   &               4.5 &           0.106 &      0.058 &     30.0 &          1.37 &                0.386 \\
WASP-39 b   &               3.6 &           1.188 &      0.294 &     20.0 &          1.39 &                0.382 \\
WASP-39 b   &               4.5 &           0.416 &      0.062 &      0.0 &          1.32 &                0.381 \\
HAT-P-12 b   &               3.6 &           0.387 &      0.289 &     10.0 &          1.35 &                0.382 \\
HAT-P-12 b   &               4.5 &           0.273 &      0.062 &     10.0 &          1.34 &                0.383 \\
HAT-P-18 b   &               3.6 &           0.382 &      0.243 &     20.0 &          1.47 &                0.382 \\
HAT-P-18 b   &               4.5 &           0.183 &      0.067 &      0.0 &          1.34 &                0.385 \\
TrES2 b    &               3.6 &           0.332 &      0.275 &     10.0 &          1.33 &                0.383 \\
TrES2 b    &               4.5 &           0.193 &      0.069 &      5.0 &          1.26 &                0.385 \\
WASP-4 b    &               3.6 &           1.424 &      0.295 &     10.0 &          1.53 &                0.384 \\
WASP-4 b    &               4.5 &           0.771 &      0.049 &     30.0 &          1.48 &                0.385 \\
XO-2 b      &               3.6 &           0.156 &      0.250 &     30.0 &          1.27 &                0.385 \\
XO-2 b      &               4.5 &           0.052 &      0.048 &     20.0 &          1.24 &                0.383 \\
GJ3470 b   &               3.6 &           0.523 &      0.007 &      5.0 &          1.22 &                0.382 \\
GJ3470 b   &               4.5 &           0.246 &      0.006 &      5.0 &          1.28 &                0.383 \\
WASP-21 b   &               3.6 &           0.029 &      0.242 &     40.0 &          1.39 &                0.383 \\
WASP-21 b   &               4.5 &           0.020 &      0.070 &     10.0 &          1.32 &                0.383 \\
WASP-31 b   &               3.6 &           0.034 &      0.317 &     30.0 &          1.46 &                0.383 \\
WASP-31 b   &               4.5 &           0.037 &      0.055 &     10.0 &          1.43 &                0.383 \\
WASP-1 b    &               3.6 &           0.742 &      0.331 &     10.0 &          1.43 &                0.384 \\
WASP-1 b    &               4.5 &           0.490 &      0.055 &     40.0 &          1.39 &                0.387 \\
HAT-P-26 b   &               3.6 &           1.501 &      0.265 &     10.0 &          1.35 &                0.384 \\
HAT-P-26 b   &               4.5 &           0.689 &      0.069 &     10.0 &          1.29 &                0.383 \\
WASP-107 b  &               3.6 &           9.628 &      0.010 &     95.0 &          1.30 &                0.384 \\
WASP-107 b  &               4.5 &           4.939 &      0.048 &     15.0 &          1.26 &                0.385 \\
WASP-13 b   &               3.6 &           0.319 &      0.263 &     55.0 &          1.25 &                0.383 \\
WASP-13 b   &               4.5 &           0.137 &      0.060 &     10.0 &          1.26 &                0.384 \\
WASP-121 b  &               3.6 &           0.585 &      0.248 &     45.0 &          1.13 &                0.384 \\
WASP-121 b  &               4.5 &           0.343 &      0.070 &     30.0 &          1.08 &                0.382 \\
WASP-69 b   &               3.6 &           0.078 &      0.008 &     30.0 &          1.25 &                0.383 \\
WASP-69 b   &               4.5 &           0.057 &      0.009 &     25.0 &          1.21 &                0.382 \\
WASP-67 b   &               3.6 &           0.548 &      0.193 &     45.0 &          1.33 &                0.385 \\
WASP-67 b   &               4.5 &           0.196 &      0.075 &     30.0 &          1.24 &                0.385 \\
HATS-7 b    &               3.6 &           0.039 &      0.240 &     35.0 &          1.38 &                0.386 \\
HATS-7 b    &               4.5 &           0.017 &      0.056 &     30.0 &          1.40 &                0.385 \\
WASP-29 b   &               3.6 &           1.719 &      0.274 &     70.0 &          1.35 &                0.384 \\
WASP-29 b   &               4.5 &           0.822 &      0.054 &     10.0 &          1.22 &                0.381 \\
HAT-P-41 b   &               3.6 &           0.403 &      0.221 &     30.0 &          1.31 &                0.382 \\
HAT-P-41 b   &               4.5 &           0.201 &      0.061 &    150.0 &          1.27 &                0.384 \\
WASP-101 b  &               3.6 &           0.231 &      0.261 &     30.0 &          1.33 &                0.385 \\
WASP-101 b  &               4.5 &           0.140 &      0.064 &     30.0 &          1.21 &                0.384 \\
WASP-131 b  &               3.6 &           0.027 &      0.304 &     30.0 &          1.26 &                0.383 \\
WASP-131 b  &               4.5 &           0.033 &      0.068 &     35.0 &          1.25 &                0.383 \\
WASP-36 b   &               3.6 &           0.001 &      0.320 &    110.0 &          1.49 &                0.384 \\
WASP-36 b   &               4.5 &           0.003 &      0.075 &     40.0 &          1.42 &                0.383 \\
WASP-63 b   &               3.6 &           0.298 &      0.307 &     60.0 &          1.30 &                0.385 \\
WASP-63 b   &               4.5 &           0.107 &      0.058 &     10.0 &          1.23 &                0.384 \\
WASP-79 b   &               3.6 &           0.169 &      0.291 &     35.0 &          1.45 &                0.385 \\
WASP-79 b   &               4.5 &           0.158 &      0.062 &     10.0 &          1.22 &                0.385 \\
WASP-94 Ab  &               3.6 &           0.591 &      0.268 &    105.0 &          1.35 &                0.384 \\
WASP-94 Ab  &               4.5 &           0.348 &      0.065 &     30.0 &          1.21 &                0.385 \\
WASP-74 b   &               3.6 &           0.894 &      0.008 &     90.0 &          1.26 &                0.384 \\
WASP-74 b   &               4.5 &           0.560 &      0.007 &     70.0 &          1.24 &                0.384 \\
WASP-62 b   &               3.6 &           0.373 &      0.255 &     85.0 &          1.47 &                0.384 \\
WASP-62 b   &               4.5 &           0.239 &      0.064 &     40.0 &          1.21 &                0.384 \\
KELT-7 b    &               3.6 &           0.248 &      0.009 &     60.0 &          1.29 &                0.384 \\
KELT-7 b    &               4.5 &           0.182 &      0.007 &     50.0 &          1.21 &                0.382 \\
Kepler-45 b &               3.6 &           0.022 &      0.277 &     30.0 &          1.50 &                0.383 \\
Kepler-45 b &               3.6 &           0.021 &      0.280 &     10.0 &          1.55 &                0.384 \\
Kepler-45 b &               3.6 &           0.026 &      0.269 &     40.0 &          1.71 &                0.384 \\
Kepler-45 b &               3.6 &           0.024 &      0.265 &     20.0 &          1.78 &                0.384 \\
Kepler-45 b &               4.5 &           0.005 &      0.047 &     20.0 &          1.44 &                0.384 \\
Kepler-45 b &               4.5 &           0.006 &      0.065 &     20.0 &          1.51 &                0.383 \\
Kepler-45 b &               4.5 &           0.008 &      0.059 &     20.0 &          1.41 &                0.383 \\
Kepler-45 b &               4.5 &           0.005 &      0.052 &     20.0 &          1.46 &                0.385 \\
\hline
\end{longtable}

\end{appendix}

\clearpage

\setcounter{table}{3}
\setlength{\tabcolsep}{6pt}

\begin{longtable}[h]{lllllll}

\caption{\label{P1:tab:results} Results from the MCMC analysis. We show the semi-major axis (a/R*), the inclination (degrees), the percentage transit depth $(R_p/R_s)^2$, the corresponding impact parameter (b) and the mid-transit time in BJD\_UTC. Values for the semi-major axis and the inclination are from the initial MCMC fits and then these are fixed with Gaussian priors for a second MCMC run where the final values for the transit depths are determined.}\\

\hline\hline
Planet & $\lambda$  & a/R* & inc  & depth  &     b & T0 \\
& $(\mu m)$ & & degrees & $(R_p/R_s)^2$ (\%) & &  BJD\_UTC \\ 
\hline
\endfirsthead
\caption{continued.} \\
\hline\hline
Planet & $\lambda$  & a/R* & inc  & depth  &     b & T0 \\
& $(\mu m)$ & & degrees & $(R_p/R_s)^2$ (\%) & & BJD\_UTC \\ 
\hline
\endhead
\hline
\endfoot

HAT-P-32 b   &               3.6 &   ${6.13}^{+0.02}_{-0.02}$ &  ${89.5}^{+0.3}_{-0.5}$ &  ${2.15}^{+0.01}_{-0.01}$ &    ${0.05}^{+2.06}_{-2.90}$ &  ${2456250.103520}^{+0.000112}_{-0.000120}$ \\
HAT-P-32 b   &               4.5 &   ${6.13}^{+0.02}_{-0.04}$ &  ${89.4}^{+0.4}_{-0.6}$ &  ${2.21}^{+0.02}_{-0.02}$ &    ${0.06}^{+2.62}_{-3.95}$ &  ${2456370.504208}^{+0.000154}_{-0.000152}$ \\
XO-1 b      &               3.6 &  ${11.46}^{+0.05}_{-0.11}$ &  ${89.5}^{+0.3}_{-0.4}$ &  ${1.67}^{+0.01}_{-0.01}$ &    ${0.09}^{+3.56}_{-4.43}$ &  ${2456426.076095}^{+0.000115}_{-0.000120}$ \\
XO-1 b      &               4.5 &  ${11.24}^{+0.20}_{-0.23}$ &  ${88.8}^{+0.5}_{-0.4}$ &  ${1.72}^{+0.01}_{-0.01}$ &    ${0.24}^{+5.40}_{-4.49}$ &  ${2456437.900819}^{+0.000158}_{-0.000157}$ \\
HAT-P-1 b    &               3.6 &   ${9.91}^{+0.13}_{-0.13}$ &  ${85.7}^{+0.1}_{-0.1}$ &  ${1.40}^{+0.01}_{-0.01}$ &    ${0.74}^{+0.99}_{-0.99}$ &  ${2456547.478364}^{+0.000156}_{-0.000150}$ \\
HAT-P-1 b    &               4.5 &  ${10.07}^{+0.16}_{-0.15}$ &  ${85.8}^{+0.1}_{-0.1}$ &  ${1.39}^{+0.01}_{-0.01}$ &    ${0.73}^{+1.25}_{-1.17}$ &  ${2456556.409109}^{+0.000187}_{-0.000190}$ \\
WASP-17 b   &               3.6 &   ${7.16}^{+0.14}_{-0.16}$ &  ${88.1}^{+0.8}_{-0.7}$ &  ${1.52}^{+0.01}_{-0.01}$ &    ${0.24}^{+6.05}_{-4.97}$ &  ${2456423.188874}^{+0.000233}_{-0.000221}$ \\
WASP-17 b   &               4.5 &   ${7.24}^{+0.09}_{-0.15}$ &  ${88.6}^{+0.8}_{-0.8}$ &  ${1.57}^{+0.02}_{-0.02}$ &    ${0.17}^{+6.13}_{-5.89}$ &  ${2456426.923243}^{+0.000288}_{-0.000285}$ \\
WASP-39 b   &               3.6 &  ${10.47}^{+0.19}_{-0.17}$ &  ${87.0}^{+0.2}_{-0.2}$ &  ${2.15}^{+0.02}_{-0.02}$ &    ${0.56}^{+1.83}_{-1.68}$ &  ${2456401.396438}^{+0.000159}_{-0.000176}$ \\
WASP-39 b   &               4.5 &  ${11.38}^{+0.28}_{-0.25}$ &  ${87.7}^{+0.3}_{-0.2}$ &  ${2.16}^{+0.02}_{-0.02}$ &    ${0.45}^{+3.14}_{-2.66}$ &  ${2456575.774315}^{+0.000200}_{-0.000194}$ \\
HAT-P-12 b   &               3.6 &  ${11.23}^{+0.26}_{-0.26}$ &  ${88.0}^{+0.3}_{-0.3}$ &  ${1.89}^{+0.01}_{-0.01}$ &    ${0.39}^{+3.64}_{-3.30}$ &  ${2456359.882148}^{+0.000131}_{-0.000138}$ \\
HAT-P-12 b   &               4.5 &  ${10.90}^{+0.35}_{-0.30}$ &  ${87.8}^{+0.4}_{-0.3}$ &  ${1.93}^{+0.03}_{-0.03}$ &    ${0.41}^{+4.47}_{-3.60}$ &  ${2456363.095398}^{+0.000197}_{-0.000187}$ \\
HAT-P-18 b   &               3.6 &  ${15.28}^{+0.47}_{-0.41}$ &  ${88.5}^{+0.3}_{-0.3}$ &  ${1.77}^{+0.02}_{-0.02}$ &    ${0.41}^{+5.19}_{-4.07}$ &  ${2456461.067141}^{+0.000195}_{-0.000197}$ \\
HAT-P-18 b   &               4.5 &  ${15.48}^{+0.45}_{-0.44}$ &  ${88.5}^{+0.3}_{-0.3}$ &  ${1.93}^{+0.02}_{-0.02}$ &    ${0.41}^{+5.07}_{-4.47}$ &  ${2456483.099518}^{+0.000215}_{-0.000215}$ \\
TrES-2 b    &               3.6 &   ${7.96}^{+0.16}_{-0.15}$ &  ${83.9}^{+0.2}_{-0.2}$ &  ${1.37}^{+0.02}_{-0.02}$ &    ${0.84}^{+1.33}_{-1.34}$ &  ${2456252.834601}^{+0.000203}_{-0.000196}$ \\
TrES-2 b    &               4.5 &   ${8.20}^{+0.25}_{-0.23}$ &  ${84.2}^{+0.3}_{-0.2}$ &  ${1.40}^{+0.02}_{-0.02}$ &    ${0.83}^{+2.09}_{-1.92}$ &  ${2456257.775215}^{+0.000268}_{-0.000267}$ \\
WASP-4 b    &               3.6 &   ${5.58}^{+0.03}_{-0.04}$ &  ${89.3}^{+0.5}_{-0.7}$ &  ${2.28}^{+0.02}_{-0.02}$ &    ${0.07}^{+2.71}_{-4.14}$ &  ${2456288.955465}^{+0.000137}_{-0.000142}$ \\
WASP-4 b    &               4.5 &   ${5.46}^{+0.05}_{-0.11}$ &  ${88.7}^{+0.9}_{-1.3}$ &  ${2.34}^{+0.03}_{-0.03}$ &    ${0.12}^{+4.80}_{-7.20}$ &  ${2456292.969500}^{+0.000208}_{-0.000212}$ \\
XO-2 b      &               3.6 &   ${8.17}^{+0.09}_{-0.17}$ &  ${88.9}^{+0.7}_{-0.8}$ &  ${1.07}^{+0.01}_{-0.01}$ &    ${0.15}^{+5.81}_{-6.13}$ &  ${2456295.370617}^{+0.000139}_{-0.000140}$ \\
XO-2 b      &               4.5 &   ${7.77}^{+0.22}_{-0.22}$ &  ${87.6}^{+0.7}_{-0.6}$ &  ${1.07}^{+0.01}_{-0.01}$ &    ${0.33}^{+5.43}_{-4.50}$ &  ${2456292.754728}^{+0.000198}_{-0.000191}$ \\
GJ3470 b   &               3.6 &  ${14.63}^{+0.52}_{-0.50}$ &  ${88.4}^{+0.3}_{-0.3}$ &  ${0.57}^{+0.01}_{-0.01}$ &    ${0.42}^{+4.13}_{-3.73}$ &  ${2456284.001794}^{+0.000118}_{-0.000115}$ \\
GJ3470 b   &               4.5 &  ${14.41}^{+0.65}_{-0.54}$ &  ${88.4}^{+0.4}_{-0.3}$ &  ${0.61}^{+0.01}_{-0.01}$ &    ${0.41}^{+5.23}_{-4.09}$ &  ${2456294.011801}^{+0.000151}_{-0.000150}$ \\
WASP-21 b   &               3.6 &   ${9.55}^{+0.30}_{-0.28}$ &  ${87.1}^{+0.4}_{-0.4}$ &  ${1.08}^{+0.01}_{-0.01}$ &    ${0.49}^{+3.75}_{-3.34}$ &  ${2456532.561048}^{+0.000261}_{-0.000260}$ \\
WASP-21 b   &               4.5 &   ${9.61}^{+0.40}_{-0.34}$ &  ${87.1}^{+0.5}_{-0.4}$ &  ${1.14}^{+0.02}_{-0.02}$ &    ${0.48}^{+5.28}_{-4.23}$ &  ${2456536.882998}^{+0.000308}_{-0.000322}$ \\
WASP-31 b   &               3.6 &   ${8.06}^{+0.20}_{-0.18}$ &  ${84.5}^{+0.2}_{-0.2}$ &  ${1.54}^{+0.02}_{-0.02}$ &    ${0.77}^{+1.81}_{-1.66}$ &  ${2456360.907660}^{+0.000317}_{-0.000328}$ \\
WASP-31 b   &               4.5 &   ${8.86}^{+0.34}_{-0.32}$ &  ${85.2}^{+0.3}_{-0.3}$ &  ${1.50}^{+0.03}_{-0.03}$ &    ${0.74}^{+2.86}_{-2.77}$ &  ${2456371.125690}^{+0.000407}_{-0.000412}$ \\
WASP-1 b    &               3.6 &   ${5.72}^{+0.03}_{-0.05}$ &  ${89.3}^{+0.5}_{-0.8}$ &  ${1.07}^{+0.01}_{-0.01}$ &    ${0.07}^{+2.91}_{-4.60}$ &  ${2456361.902274}^{+0.000263}_{-0.000250}$ \\
WASP-1 b    &               4.5 &   ${5.41}^{+0.17}_{-0.20}$ &  ${86.5}^{+1.2}_{-1.1}$ &  ${1.09}^{+0.02}_{-0.02}$ &    ${0.33}^{+6.58}_{-6.12}$ &  ${2456371.982150}^{+0.000354}_{-0.000364}$ \\
HAT-P-26 b   &               3.6 &  ${13.22}^{+0.75}_{-0.94}$ &  ${88.3}^{+0.8}_{-0.7}$ &  ${0.53}^{+0.01}_{-0.01}$ &   ${0.39}^{+10.08}_{-9.85}$ &  ${2456545.361384}^{+0.000296}_{-0.000288}$ \\
HAT-P-26 b   &               4.5 &  ${13.92}^{+0.16}_{-0.32}$ &  ${89.5}^{+0.3}_{-0.6}$ &  ${0.55}^{+0.01}_{-0.01}$ &    ${0.11}^{+4.33}_{-8.72}$ &  ${2456405.622835}^{+0.000356}_{-0.000364}$ \\
WASP-107 b  &               3.6 &  ${18.19}^{+0.03}_{-0.04}$ &  ${89.9}^{+0.1}_{-0.1}$ &  ${1.96}^{+0.01}_{-0.01}$ &    ${0.04}^{+1.45}_{-2.26}$ &  ${2457876.124941}^{+0.000060}_{-0.000064}$ \\
WASP-107 b  &               4.5 &  ${18.09}^{+0.05}_{-0.09}$ &  ${89.8}^{+0.1}_{-0.2}$ &  ${2.06}^{+0.01}_{-0.01}$ &    ${0.06}^{+2.54}_{-3.09}$ &  ${2457870.403743}^{+0.000081}_{-0.000077}$ \\
WASP-13 b   &               3.6 &   ${7.64}^{+0.20}_{-0.19}$ &  ${85.6}^{+0.4}_{-0.3}$ &  ${0.86}^{+0.01}_{-0.01}$ &    ${0.58}^{+2.70}_{-2.50}$ &  ${2456480.940869}^{+0.000231}_{-0.000246}$ \\
WASP-13 b   &               4.5 &   ${7.78}^{+0.27}_{-0.23}$ &  ${85.7}^{+0.4}_{-0.4}$ &  ${0.87}^{+0.01}_{-0.01}$ &    ${0.59}^{+3.43}_{-2.85}$ &  ${2456315.526437}^{+0.000293}_{-0.000303}$ \\
WASP-121 b  &               3.6 &   ${3.84}^{+0.02}_{-0.03}$ &  ${88.9}^{+0.8}_{-1.1}$ &  ${1.47}^{+0.01}_{-0.01}$ &    ${0.07}^{+2.95}_{-4.10}$ &  ${2457906.807311}^{+0.000148}_{-0.000144}$ \\
WASP-121 b  &               4.5 &   ${3.82}^{+0.02}_{-0.03}$ &  ${89.0}^{+0.8}_{-1.3}$ &  ${1.49}^{+0.01}_{-0.01}$ &    ${0.07}^{+3.02}_{-4.80}$ &  ${2457910.632374}^{+0.000183}_{-0.000171}$ \\
WASP-69 b   &               3.6 &  ${12.26}^{+0.09}_{-0.08}$ &  ${86.8}^{+0.0}_{-0.0}$ &  ${1.60}^{+0.00}_{-0.00}$ &    ${0.68}^{+0.60}_{-0.58}$ &  ${2457992.354188}^{+0.000053}_{-0.000054}$ \\
WASP-69 b   &               4.5 &  ${12.30}^{+0.11}_{-0.10}$ &  ${86.8}^{+0.1}_{-0.1}$ &  ${1.67}^{+0.01}_{-0.01}$ &    ${0.68}^{+0.77}_{-0.71}$ &  ${2457996.222243}^{+0.000066}_{-0.000069}$ \\
WASP-67 b   &               3.6 &  ${13.50}^{+0.39}_{-0.33}$ &  ${86.2}^{+0.2}_{-0.2}$ &  ${1.97}^{+0.03}_{-0.03}$ &    ${0.91}^{+2.28}_{-2.72}$ &  ${2457776.271136}^{+0.000219}_{-0.000220}$ \\
WASP-67 b   &               4.5 &  ${13.90}^{+0.31}_{-0.39}$ &  ${86.3}^{+0.1}_{-0.1}$ &  ${1.92}^{+0.03}_{-0.03}$ &    ${0.89}^{+1.84}_{-1.93}$ &  ${2457979.305753}^{+0.000282}_{-0.000276}$ \\
HATS-7 b    &               3.6 &  ${11.09}^{+0.62}_{-1.07}$ &  ${88.2}^{+1.0}_{-1.3}$ &  ${0.38}^{+0.02}_{-0.02}$ &  ${0.35}^{+11.03}_{-14.14}$ &  ${2457694.120917}^{+0.000590}_{-0.000538}$ \\
HATS-7 b    &               4.5 &  ${10.80}^{+0.48}_{-0.91}$ &  ${88.4}^{+1.1}_{-1.2}$ &  ${0.40}^{+0.03}_{-0.03}$ &  ${0.30}^{+12.05}_{-13.12}$ &  ${2457697.305788}^{+0.000797}_{-0.000795}$ \\
WASP-29 b   &               3.6 &  ${12.58}^{+0.05}_{-0.11}$ &  ${89.7}^{+0.2}_{-0.4}$ &  ${0.95}^{+0.01}_{-0.01}$ &    ${0.07}^{+2.95}_{-4.91}$ &  ${2457807.234478}^{+0.000115}_{-0.000120}$ \\
WASP-29 b   &               4.5 &  ${12.53}^{+0.05}_{-0.08}$ &  ${89.7}^{+0.2}_{-0.3}$ &  ${0.93}^{+0.01}_{-0.01}$ &    ${0.06}^{+2.59}_{-4.13}$ &  ${2457826.848225}^{+0.000150}_{-0.000149}$ \\
HAT-P-41 b   &               3.6 &   ${5.53}^{+0.03}_{-0.06}$ &  ${89.0}^{+0.7}_{-0.9}$ &  ${1.00}^{+0.01}_{-0.01}$ &    ${0.10}^{+3.76}_{-4.92}$ &  ${2457772.203860}^{+0.000220}_{-0.000217}$ \\
HAT-P-41 b   &               4.5 &   ${5.55}^{+0.03}_{-0.04}$ &  ${89.3}^{+0.5}_{-0.8}$ &  ${1.09}^{+0.01}_{-0.01}$ &    ${0.07}^{+2.78}_{-4.43}$ &  ${2457788.367795}^{+0.000274}_{-0.000263}$ \\
WASP-101 b  &               3.6 &   ${8.60}^{+0.17}_{-0.16}$ &  ${85.2}^{+0.2}_{-0.2}$ &  ${1.18}^{+0.01}_{-0.01}$ &    ${0.73}^{+1.55}_{-1.54}$ &  ${2457760.332526}^{+0.000170}_{-0.000175}$ \\
WASP-101 b  &               4.5 &   ${8.51}^{+0.19}_{-0.18}$ &  ${85.0}^{+0.2}_{-0.2}$ &  ${1.14}^{+0.01}_{-0.01}$ &    ${0.74}^{+1.72}_{-1.64}$ &  ${2457771.089626}^{+0.000225}_{-0.000227}$ \\
WASP-131 b  &               3.6 &   ${8.34}^{+0.20}_{-0.19}$ &  ${85.0}^{+0.2}_{-0.2}$ &  ${0.61}^{+0.01}_{-0.01}$ &    ${0.73}^{+1.88}_{-1.80}$ &  ${2457696.837080}^{+0.000253}_{-0.000256}$ \\
WASP-131 b  &               4.5 &   ${8.41}^{+0.28}_{-0.26}$ &  ${85.0}^{+0.3}_{-0.3}$ &  ${0.61}^{+0.01}_{-0.01}$ &    ${0.73}^{+2.63}_{-2.53}$ &  ${2457909.718452}^{+0.000336}_{-0.000332}$ \\
WASP-36 b   &               3.6 &   ${6.06}^{+0.25}_{-0.22}$ &  ${83.7}^{+0.6}_{-0.5}$ &  ${1.78}^{+0.03}_{-0.03}$ &    ${0.67}^{+3.75}_{-3.19}$ &  ${2457805.166629}^{+0.000262}_{-0.000262}$ \\
WASP-36 b   &               4.5 &   ${6.19}^{+0.44}_{-0.39}$ &  ${84.4}^{+1.1}_{-1.0}$ &  ${1.82}^{+0.03}_{-0.03}$ &    ${0.60}^{+6.70}_{-6.13}$ &  ${2457975.813843}^{+0.000346}_{-0.000353}$ \\
WASP-63 b   &               3.6 &   ${6.26}^{+0.23}_{-0.21}$ &  ${86.6}^{+1.0}_{-0.8}$ &  ${0.61}^{+0.01}_{-0.01}$ &    ${0.37}^{+6.38}_{-5.07}$ &  ${2457865.520635}^{+0.000353}_{-0.000330}$ \\
WASP-63 b   &               4.5 &   ${6.52}^{+0.16}_{-0.21}$ &  ${87.7}^{+1.1}_{-1.0}$ &  ${0.55}^{+0.01}_{-0.01}$ &    ${0.26}^{+7.01}_{-6.45}$ &  ${2457922.436423}^{+0.000470}_{-0.000461}$ \\
WASP-79 b   &               3.6 &   ${7.31}^{+0.15}_{-0.14}$ &  ${85.9}^{+0.3}_{-0.3}$ &  ${1.20}^{+0.01}_{-0.01}$ &    ${0.52}^{+2.36}_{-2.05}$ &  ${2457713.374126}^{+0.000167}_{-0.000167}$ \\
WASP-79 b   &               4.5 &   ${7.12}^{+0.15}_{-0.14}$ &  ${85.6}^{+0.3}_{-0.3}$ &  ${1.17}^{+0.01}_{-0.01}$ &    ${0.55}^{+2.29}_{-2.02}$ &  ${2457720.699409}^{+0.000215}_{-0.000213}$ \\
WASP-94 Ab  &               3.6 &   ${7.34}^{+0.02}_{-0.04}$ &  ${89.5}^{+0.3}_{-0.5}$ &  ${1.12}^{+0.01}_{-0.01}$ &    ${0.06}^{+2.43}_{-3.56}$ &  ${2457795.021530}^{+0.000147}_{-0.000147}$ \\
WASP-94 Ab  &               4.5 &   ${7.34}^{+0.02}_{-0.04}$ &  ${89.5}^{+0.4}_{-0.5}$ &  ${1.13}^{+0.01}_{-0.01}$ &    ${0.07}^{+2.62}_{-3.61}$ &  ${2457972.780291}^{+0.000187}_{-0.000180}$ \\
WASP-74 b   &               3.6 &   ${4.75}^{+0.08}_{-0.07}$ &  ${79.7}^{+0.3}_{-0.2}$ &  ${0.87}^{+0.01}_{-0.01}$ &    ${0.85}^{+1.17}_{-1.01}$ &  ${2457768.164558}^{+0.000178}_{-0.000176}$ \\
WASP-74 b   &               4.5 &   ${5.13}^{+0.11}_{-0.10}$ &  ${80.8}^{+0.3}_{-0.3}$ &  ${0.86}^{+0.01}_{-0.01}$ &    ${0.82}^{+1.44}_{-1.42}$ &  ${2457770.304101}^{+0.000228}_{-0.000230}$ \\
WASP-62 b   &               3.6 &   ${9.47}^{+0.16}_{-0.16}$ &  ${88.2}^{+0.4}_{-0.3}$ &  ${1.29}^{+0.01}_{-0.01}$ &    ${0.30}^{+3.57}_{-3.02}$ &  ${2457717.229937}^{+0.000138}_{-0.000138}$ \\
WASP-62 b   &               4.5 &   ${9.32}^{+0.20}_{-0.18}$ &  ${87.9}^{+0.4}_{-0.3}$ &  ${1.20}^{+0.01}_{-0.01}$ &    ${0.35}^{+3.88}_{-3.05}$ &  ${2457730.466206}^{+0.000165}_{-0.000167}$ \\
Kepler-45 b &               3.6 &      - &   - &  ${3.37}^{+0.13}_{-0.13}$ &       - &  - \\
Kepler-45 b &               4.5 &      - &   - &  ${3.50}^{+0.14}_{-0.14}$ &       - &  - \\
\end{longtable}

\end{document}